\documentclass[numberedappendix,onecolumn]{emulateapj} 
\usepackage{rotating} 
\usepackage{amsmath,amssymb,bbold,mathrsfs}
\usepackage{subfigure}
\usepackage{bm}

\begin{document}

\title{The magnetic sensitivity of the resonance and subordinate lines of
       \ion{Mg}{2} in the solar chromosphere}


 \author{T.\ del Pino Alem\'an$^{1,2}$, J.\ Trujillo Bueno$^{1,2,3,a}$,
         R.\ Casini$^{4}$, and R.\ Manso Sainz$^{5}$}

 \affil{$^1$Instituto de Astrof\'{\i}sica de Canarias, E-38205 La Laguna,
        Tenerife, Spain}
 \affil{$^2$Departamento de Astrof\'\i sica, Universidad de La Laguna, E-38206
        La Laguna, Tenerife, Spain}
 \affil{$^3$Consejo Superior de Investigaciones Cient\'{\i}ficas, Spain}
 \affil{$^4$High Altitude Observatory, National Center for Atmospheric
 Research${}^{b}$.\break
 P.O.~Box 3000, Boulder, CO 80307-3000, U.S.A.\break}
  \affil{$^5$Max-Planck-Institut f\"ur Sonnensystemforschung,\break
  Justus-von-Liebig-Weg 3, 37077 G\"ottingen, Germany}
 \footnotetext[a]{Affiliate scientist of the National Center for Atmospheric
                  Research, Boulder, U.S.A.}
 \footnotetext[b]{The National Center for Atmospheric Research is sponsored by
                  the National Science Foundation.}

\begin{abstract}
We carry out a theoretical study of the polarization of the solar \ion{Mg}{2} h-k
doublet (including its extended wings) and the subordinate UV triplet around
280\,nm. These lines are of great diagnostic interest, as they encode information
on the physical properties of the solar atmosphere from the upper photosphere to the
chromosphere-corona transition region. We base our study on radiative transfer
calculations of spectral line polarization in one-dimensional models of quiet
and plage regions of the solar atmosphere. Our calculations take into
account the combined action of atomic polarization, quantum level interference,
frequency redistribution, and magnetic fields of arbitrary strength. In particular,
we study the sensitivity of the emergent Stokes profiles to changes in the magnetic
field through the Zeeman and Hanle effects.
We also study the impact of the chromospheric plasma dynamics on the
emergent Stokes profiles, taking into account the angle-dependent
frequency redistribution in the h-k resonance transitions. The results
presented here are of interest for the interpretation of spectropolarimetric
observations in this important region of the solar ultraviolet spectrum.
\end{abstract}



\section{Introduction}\label{Sintro}

The magnetic field plays a key role in determining the behavior of the solar
plasma. In the outer layers of the solar atmosphere, i.e., the chromosphere,
the transition region, and the corona, the fast decrease in density is such that
the magnetic field dominates the structuring of the low-$\beta$ plasma (e.g.,
\citealt{BPriest2014}). The determination of
the magnetic fields in these regions of the solar atmosphere is one of the main
challenges faced in solar physics nowadays.

The polarization profiles of spectral lines carry information on many
physical properties of the emitting plasma, in particular, of the
magnetic field strength and geometry. Therefore, by studying the
polarization of the electromagnetic radiation emerging from
the solar atmosphere we can aspire to determine its magnetic
properties.

Most of the atomic lines in the visible solar spectrum are formed in the lower
layers of the solar atmosphere. In order to study the magnetic fields in the
outer atmospheric regions, we have to look for spectral lines that are formed
there, many of which are strong resonance lines in the
ultraviolet (UV) region of the solar spectrum. The challenge of diagnosing the
magnetic field is then twofold.

First, UV observations are challenging or simply impossible from ground-based observing
facilities due to the absorption of the Earth atmosphere. Therefore, space
telescopes are preferable or necessary to observe
many UV atomic lines of diagnostic relevance. In 2013, NASA
launched the Interface Region Imaging Spectrograph (IRIS;
see \citealt{DePontieuetal2014}), which continues to provide excellent
spectroscopic observations of UV lines such as \ion{Mg}{2} h-k (e.g.,
\citealt{Pereiraetal2014}). Two years later, the Chromospheric Lyman
Alpha Spectro-Polarimeter sounding rocket experiment (CLASP;
\citealt{Kanoetal2012,Kobayashi2012})
provided the first on-disk observations of the linear polarization
spectrum of the hydrogen Lyman-$\alpha$ line (\citealt{Kanoetal2017}).
Theoretical modeling of those observations has allowed to constrain
the geometric complexity of the
chromosphere-corona transition region (\citealt{Trujilloetal2018}). Very
recently, while this paper was being written, NASA launched with success the
Chromospheric LAyer Spectro-Polarimeter (CLASP-2;
\citealt{Narukageetal2016}) to
measure the intensity and polarization across the \ion{Mg}{2} h-k lines with high
polarimetric sensitivity and spectral resolution, in both quiet and active
regions of the solar disk. More than three decades ago, the Ultra-Violet
Spectro-Polarimeter (UVSP; \citealt{Calvertetal1979,Woodgateetal1980}) on board
the Solar Maximum Mission
({\emph SMM}; \citealt{Bohlinetal1980,BStrongetal1999})
satellite was able to detect $Q/I$ scattering
signals of the order of a few percent in the far wings of the \ion{Mg}{2} h-k lines
\citep{HenzeStenflo1987}. Such large $Q/I$ wing signals, as well as
the negative (radial) scattering polarization in the region between the h
and k lines, had been theoretically predicted by \cite{Aueretal1980} using
the approximation of coherent scattering in the observer's frame.
\cite{Stenflo1980} observed such $Q/I$ pattern across the Ca {\sc ii} H
and K resonance lines and pointed out that it is due to quantum mechanical
interference between the sublevels pertaining to the upper $J$ levels of the
H and K lines. Evidence for the expected negative polarization at wing
wavelengths between the Mg {\sc ii} resonance lines has indeed been found in 
a reanalysis of the observations from SMM \citep{Mansoetal2019}.

Second, because of the rapid decrease of the density with height in the
solar atmosphere, collisional processes become less important in
the statistical equilibrium of the ion compared to radiative
processes. Moreover, as photons can more easily escape
from the plasma, the radiation field becomes more and more anisotropic. This
anisotropic radiation field produces atomic polarization, that is, population
imbalances and coherence among the magnetic sublevels of the atom.
{\emph The differential absorption and emission of radiation in the presence of
atomic polarization is able to produce linear polarization---the so-called
line scattering polarization---even in the absence of magnetic fields.}
A magnetic field splits the energy of the
magnetic sublevels (Zeeman effect) and modifies the scattering polarization
by relaxing the atomic level coherence (Hanle effect). The lower
rate of collisional processes also implies that the frequency correlation
between absorbed and re-emitted photons in the scattering events becomes
relevant in the line formation process, giving rise to the so-called
partial frequency redistribution (PRD) regime. Both lines observed by the
CLASP experiments are affected by PRD effects and by quantum mechanical
interference between the two upper levels of
those transitions (\citealt{BelluzziTrujillo2012,Belluzzietal2012,
Kanoetal2017}).

Clearly, the interpretation of spectropolarimetric observations of strong
UV resonance lines requires a thorough theoretical
understanding---and a comparatively adequate numerical modeling---of the
generation and transfer of spectral line polarized radiation
in magnetized plasmas,
(see the review by \citealt{Trujilloetal2017}). For instance,
the Lyman-$\alpha$ and \ion{Mg}{2} h-k lines result
from transitions between the ground term, composed of the single level
${}^2$S${}_{1/2}$ and the first excited term, composed of the
${}^2$P${}_{1/2}$ and ${}^2$P${}_{3/2}$ levels. A rigorous radiative
transfer modeling requires taking into account correlation effects between
the incoming and outgoing photons in the scattering events (PRD effects),
as well as the effects of quantum interference between the
two upper levels with $J=3/2$ and $J=1/2$. This
radiative transfer problem has been solved by
\cite{BelluzziTrujillo2012,BelluzziTrujillo2014} for the \ion{Mg}{2} h-k lines
and by \cite{Belluzzietal2012} for the Lyman-$\alpha$ line using one-dimensional
(1D) model atmospheres without magnetic fields. These authors demonstrated that
PRD effects and $J$-state interference
have a very important impact outside
the center of the lines, and that the approximation of coherent scattering in
the observer's frame used by \cite{Aueretal1980} to investigate the scattering
polarization in the wings of \ion{Mg}{2} h-k produces a significant
overestimation of the polarization amplitudes. They also showed that the
approximation of complete frequency redistribution (CRD) without $J$-state
interference used by \cite{Trujilloetal2011} is suitable for
estimating the line-center scattering polarization where the Hanle effect
operates. The impact of the joint action of the Hanle, Zeeman, and PRD effects
has been investigated by \cite{Alsinaetal2016} for the \ion{Mg}{2} k line
(using a two-level atomic model, therefore without $J$-state interference) and
by \cite{delPinoetal2016} for the \ion{Mg}{2} h-k lines (using a two-term
atomic model, therefore with $J$-state interference). These papers highlighted
that the $\rho_VQ$ and $\rho_VU$ magneto-optical (M-O) terms of the transfer
equations for Stokes $U$ and $Q$, respectively, give rise to very significant
signals in the wings of $U/I$.
The two-term atom investigation by \cite{delPinoetal2016}
showed that the linear polarization is sensitive to the magnetic field
all over the wings of the \ion{Mg}{2} resonance lines.

\begin{figure}[!t]
\centering
\includegraphics[width=.75\hsize]{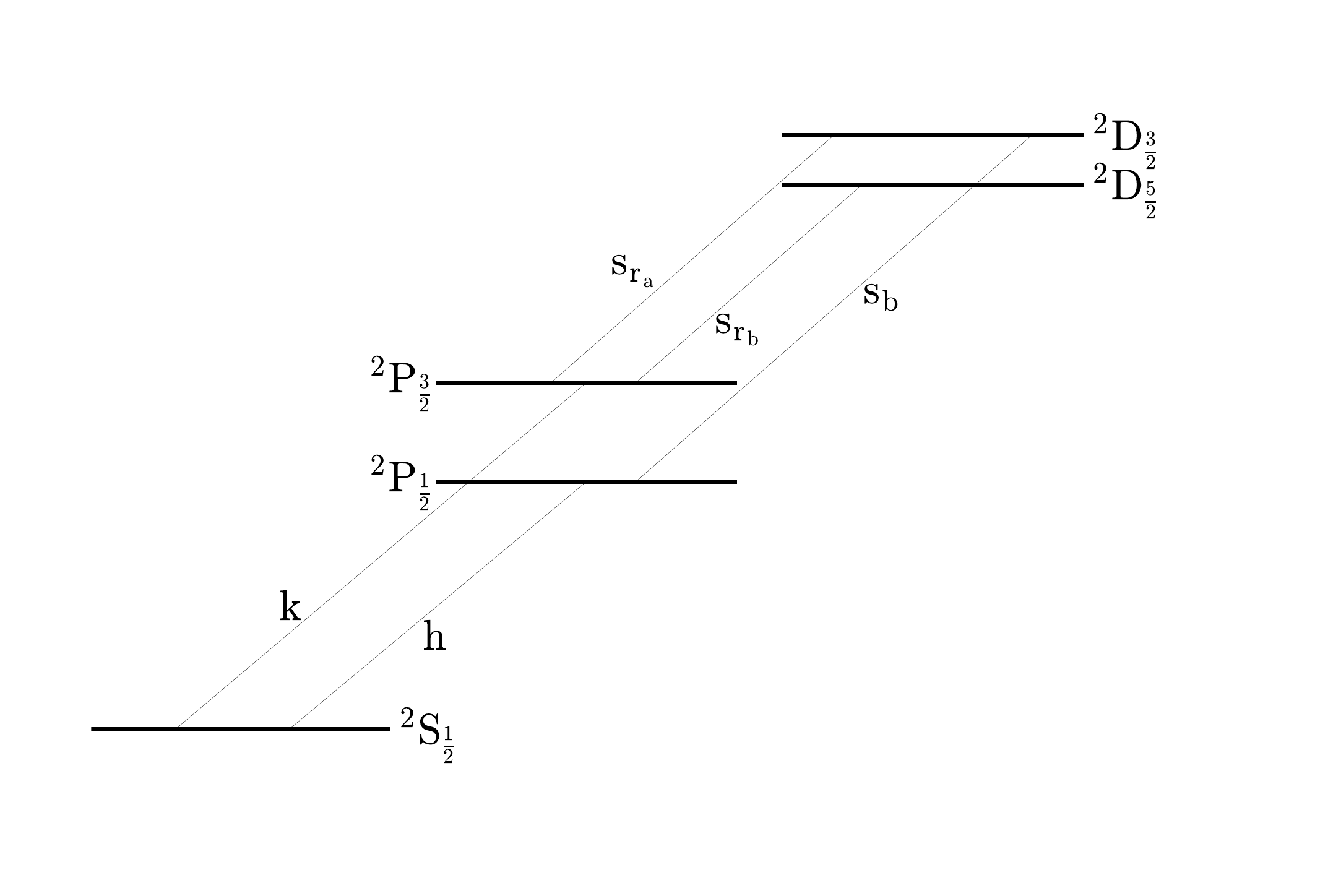}
\caption{Energy diagram of the three-term (S,P,D) atomic model of \ion{Mg}{2} 
adopted for this work. The ground state of \ion{Mg}{3} is also taken into
account for solving the radiative and collisional statistical equilibrium
of the system of transitions shown in the figure. The corresponding
four visible spectral lines are: k (279.64\,nm), h (280.35\,nm),
$\rm s_b$ (279.16\,nm), and $\rm s_{r_a}+s_{r_b}$ (279.88\,nm).
Our numerical model fully takes into account quantum interference in the P and
D terms due to their fine structure, as well as magnetic-induced, level-crossing
interference.}
\label{fig:Grotrian}
\vspace{3.1ex}
\end{figure}

These studies on the polarization of the \ion{Mg}{2}
resonance lines accounting for PRD effects and scattering polarization were
carried out using atomic models that only included either or both the h and k
lines. However, the same spectral region also contains a triplet of subordinate
lines of \ion{Mg}{2} (see Fig.~\ref{fig:Grotrian}), which also are of
diagnostic interest (see \citealt{BelluzziTrujillo2012,Pereiraetal2015}). One of
the lines in such triplet is located in the blue wing of the k line, while the
other two transitions are blended and located in the red wing of the k line.
Interestingly, the heights of line-center optical depth unity in such
subordinate lines are located about $600\,$km below those corresponding to
the h and k line centers (see Fig. 1 of \citealt{BelluzziTrujillo2012}).
As seen in Fig.~\ref{fig:Grotrian}, the lower levels of this triplet around
279.16 and 279.88\,nm are the upper levels of the h-k doublet.
Because this is a cascade transition, rather than a three-term
transition of the $\Lambda$-type, the theoretical approach outlined in
\cite{CasiniManso2016} (see also \citealt{Casinietal2017,Casinietal2017b}) can
only be applied neglecting PRD effects in the subordinate lines.
Nevertheless, the CRD approximation is valid for modeling the intensity profiles
of this triplet (\citealt{Pereiraetal2015}), so we are confident that this
approximation is also suitable for modeling their polarization given that such
lines are weaker than the doublet and form lower in the atmosphere.

After presenting the formulation of the problem in \S\ref{Sproblem}, including
information about the numerical approach and the atomic and atmospheric models
adopted, in \S\ref{SResults} we focus on a detailed presentation of the
results. 
In particular, by using semi-empirical models of quiet and plage regions of the
solar disk, we study in depth the thermal and magnetic sensitivity of the
polarization of the emergent spectral line radiation. Since the solar
chromosphere is highly dynamic, we also pay particular attention to
the sensitivity of the intensity and polarization across the \ion{Mg}{2} lines
to the dynamical state of the plasma. To this end, we have carried out additional
radiative transfer calculations in a 1D model of chromospheric dynamics 
developed by \cite{CarlssonStein1997}. In particular, we solved the
non-equilibrium polarization transfer problem at each time step of the hydrodynamic
simulation, taking into account the effects of angle-dependent frequency
redistribution. Finally, in \S\ref{Sconclusions} we present our main
conclusions, including an outlook of future research.     

\section{Formulation of the Problem}\label{Sproblem}
We solve the radiation transfer problem with polarization for the
three-term model
atom (Fig.~\ref{fig:Grotrian}) underlying the formation of the \ion{Mg}{2} h-k
doublet at 279.64 and 280.35\,nm and the UV triplet at 279.16 and 
279.88\,nm. Our objective is to study the sensitivity of these lines to the
magnetic field and to the plasma velocity gradients. We consider several 1D
atmospheric models (semi-empirical models and a hydrodynamical time-dependent
model) without assuming local thermodynamical equilibrium (hereafter, NLTE),
and taking into account scattering polarization (both in the lines and in the
continuum), partial frequency redistribution effects, and the Zeeman and Hanle
effects.

\subsection{Solution Method}\label{SSsol}
In order to solve the NLTE problem of the generation and transfer of polarized
radiation in an optically thick and magnetized plasma, we must solve,
simultaneously, the radiation transfer (RT) equations and the statistical
equilibrium (SE) equations. The former determine how the polarized radiation is
absorbed and emitted at every point in the plasma, while the latter determine the
excitation state of the atom, that is, the populations of the atomic
levels and the quantum coherence among them.

We follow the same approach as in \cite{delPinoetal2016}.
Therefore, we consider the SE equations to perturbative first order in the
atom-photon interaction (see \citealt{BLandiLandolfi2004}), and we solve the
RT equations taking into account PRD effects in the transitions ${\rm h}$ and
${\rm k}$. Because the lower term of the transitions ${\rm s}_{\rm b}$,
${\rm s}_{\rm r_{\rm a}}$, and ${\rm s}_{\rm r_{\rm b}}$ is also the upper
term of ${\rm h}$ and ${\rm k}$, our modeling framework cannot account for
coherent scattering effects in their formation (\citealt{CasiniManso2016b}).
However, because PRD effects are typically only important for resonance
transitions (in the case of \ion{Mg}{2}, the ${\rm h}$ and ${\rm k}$ lines), it
is resonable to assume that complete complete redistribution (CRD) is
a suitable approximation for the subordinate transitions.
Regardless, this approximation needs to be justified once a self-consistent
theory becames available to handle this type of atomic system.

Concerning the iterative method of solution, we first solve the unpolarized 
radiative transfer problem applying the accelerated Lambda iteration method
(e.g., \citealt{RybickiHummer1991} and references therein). Once the self-consistent
solution is obtained for the populations and intensity, we then apply
$\Lambda$-iteration (e.g., \citealt{BMihalas1970}) for obtaining the density
matrix elements and the polarization profiles, initializing the problem with the
previously obtained unpolarized solution. This strategy allows to obtain the
self-consistent solution for the Stokes profiles after only a few
$\Lambda$-iterations \citep{TrujilloManso1999}. In a standard quad-core laptop
using 5 processes a typical solution of the non-magnetic problem with
$\sim 10^3$ frequency nodes and $8$ quadrature directions takes of the order
of $2.5$ minutes, while a typical magnetic solution with $\sim 10^3$ frequency
nodes and $64$ quadrature directions takes of the order of $10^2$ minutes, both
of them with the angle-averaged approximation. With $32$ processes in a cluster,
a typical solution of the non-magnetic dynamic case with angle-dependent
redistribution took of the order of $1$ hour, although some snapshots could take
up to $3$ hours depending on the particular stratification\footnote{Because the
velocity in the time series we use is vertical, we can take advantage of the
axial symmetry of the problem, which decreases significantly the computational
cost. If this was not the case, the computing time would be at least two orders
of magnitude larger.}.

\subsection{The Atomic and Atmospheric Models}\label{SSatom}
\begin{figure}[htp!]
\centering
\includegraphics[width=.55\hsize]{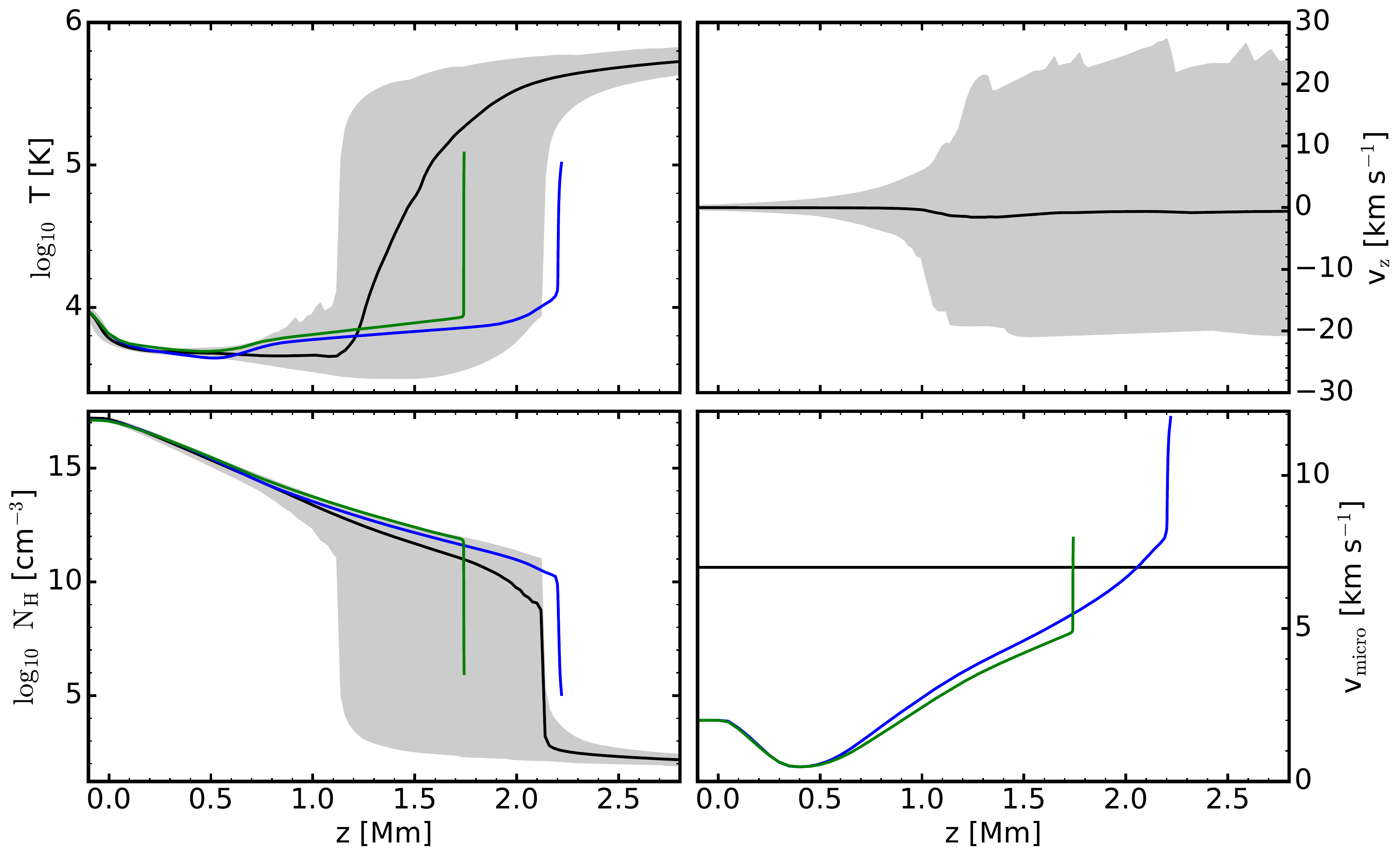}
\caption{Variation with height of the temperature (top left), vertical
velocity (top right), \ion{H}{1} number density (bottom left),
and microtubulent velocity (bottom right), for the model atmospheres used in this
work. The black curve is the time-averaged value of the CS model,
and the shaded areas represent the span of all the possible variations
for the full time series over the full range of heights.
The colored curves are for the static models FAL-C (blue) and FAL-P (green).}
\label{fig:Atmos}
\end{figure}

We solve numerically the NLTE problem of the generation and transfer of
polarized radiation in several 1D plane-parallel models of the solar atmosphere.
We chose the static models C and P of \citeauthor{Fontenlaetal1993}
(\citeyear{Fontenlaetal1993}; hereafter FAL-C and FAL-P models, respectively),
representative of the average quiet Sun and a plage region, respectively. We also
calculate the polarized spectrum in the strongly dynamic time-dependent
hydrodynamical model of \citeauthor{CarlssonStein1997} 
(\citeyear{CarlssonStein1997}; hereafter CS model). Fig.~\ref{fig:Atmos} shows
the variation with height of the temperature, neutral hydrogen number
density, vertical velocity, and microturbulent velocity in the mentioned
atmospheric models. In the radiative transfer calculations in the static
FAL-C and FAL-P models we have taken into account the spectral line broadening
produced by the model's microturbulent velocity. No microturbulent velocity is
specified for the dynamic CS model, but in our calculations we have chosen a
value of $v_{\rm micro} = $ 7\,km/s, intermediate between the microturbulence
at the two formation heights of the doublet and triplet in the FAL-C model.

We used a model atom that includes the three terms of \ion{Mg}{2} relevant
for the transitions in the spectral range of interest (see Fig.~
\ref{fig:Grotrian}) and the ground term of \ion{Mg}{3}. We verified that
the emergent Stokes profiles are practically indistinguishable whether we
include just these four terms or use a more complex atomic model with eight
terms, considering CRD for all the transitions but the h-k doublet.
Consequently, we have adopted the smallest atomic model to reduce
the computational cost of this investigation.

The energy values of the atomic levels are taken from the NIST database 
(\citealt{NIST}). The Einstein $A$-coefficient for spontaneous emission of
the transition between two terms of orbital angular momenta $L_u$ and
$L_\ell$ is the weighted average of the corresponding coefficients of
the transitions between the atomic $J$-levels within the terms, which are also
taken from the NIST database:
\begin{equation} 
A_{L_uL_\ell} = \frac{1}{(2L_u+1)(2S+1)}\,\sum_{J_u J_\ell}(2J_u+1)
A_{J_uJ_\ell}\;,
\label{eq:Aul}
\end{equation}
where $J_\ell$ and $J_u$ are the total angular momenta of the lower and upper
levels, respectively, and $S$ is the spin angular momentum.
The photoionization cross-sections for the ${}^2{\rm S}$ and ${}^2{\rm P}$ terms
are taken from the TOPbase database \citep{Cuntoetal1993} and the one
corresponding to the ${}^2{\rm D}$ term is taken as hydrogenic
(e.g., \citealt{BMihalas1978}). The collisional
excitation rates are taken from \cite{SigutPradhan1995}. Inelastic and
super-elastic collisions in the multi-term atom are implemented similarly to
the formalism of \cite{Belluzzietal2013}. Because \cite{SigutPradhan1995}
provide the collisional rates between fine structure levels, we compute the
collisional rate between terms by applying Eq. \eqref{eq:Aul} with the
substitution $A\rightarrow C$. The bound-free inelastic collisional rates with
electrons are computed using the approximation given in \cite{BAllen1963}.
The rates of depolarizing collisions with neutral hydrogen are taken from
\cite{Mansoetal2014}. 

\section{Results}\label{SResults}
Due to the computational cost of calculating the redistribution function for
every pair of virtually absorbed-emitted photon frequencies and propagation
directions, the synthesis problem with PRD effects is usually solved under the
angle-average approximation (e.g., \citealt{BMihalas1970,BelluzziTrujillo2012}).
The angle-average approximation is a good one for unmagnetized models, but 
this is not generally the case for magnetized model atmospheres (see
\citealt{Sampoornaetal2017} for a recent study based on academic lines in
isothermal model atmospheres).
For simplicity and computational time reasons we have used the angle-average
approximation for our investigation with magnetic fields in the FAL-C and FAL-P
models. This is justified because accurate angle-dependent PRD numerical
calculations in the presence of inclined fields are very costly, and the aim of
this paper is not to model spectropolarimetric observations. However, we relaxed
it to derive the results of \S\ref{SSDynamics}, where we consider a time-dependent
unmagnetized model of solar chromospheric dynamics.

\subsection{Thermal sensitivity}\label{SSAtomAtmo}

\begin{figure}[htp!]
\centering
\includegraphics[width=.32\hsize]{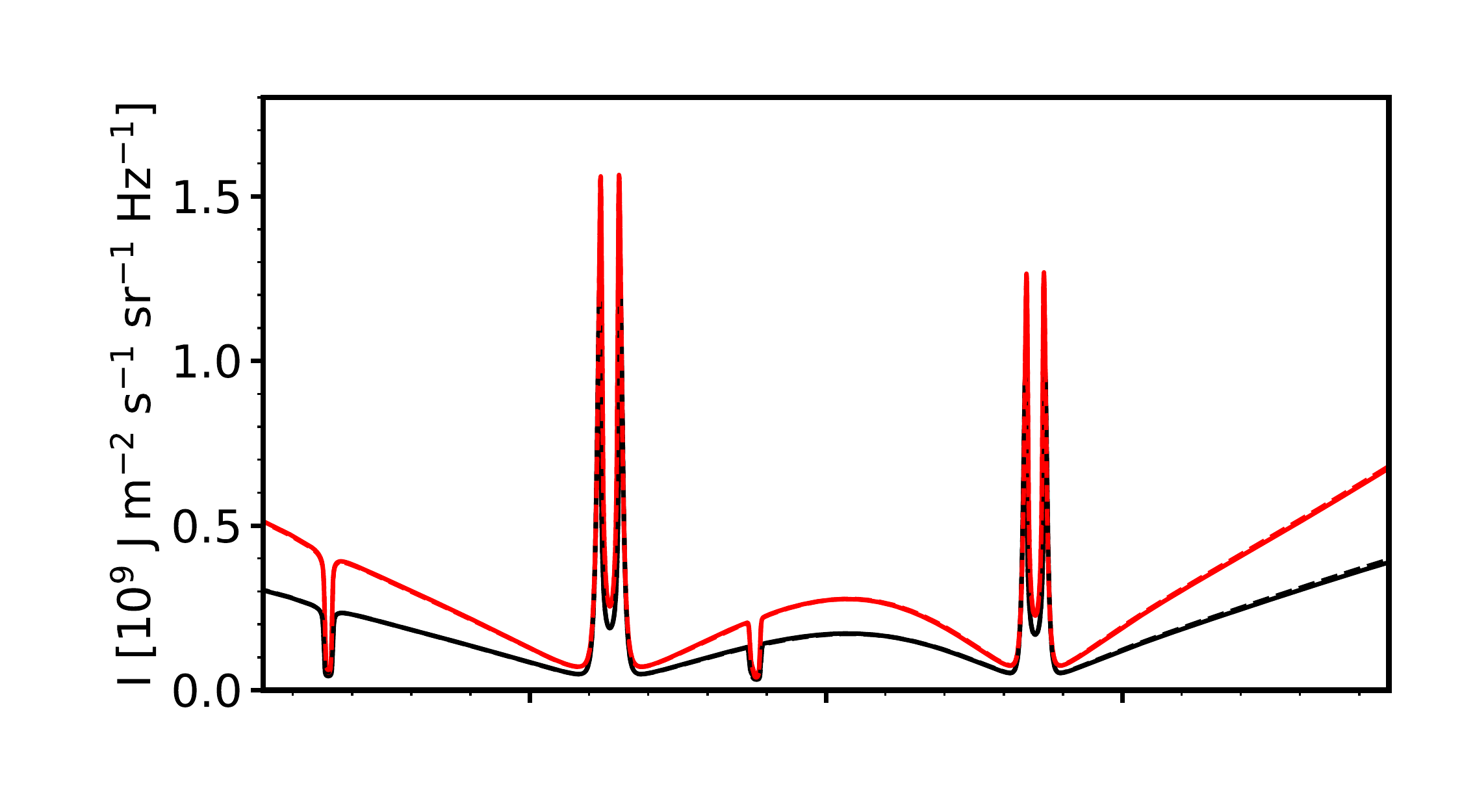}
\includegraphics[width=.32\hsize]{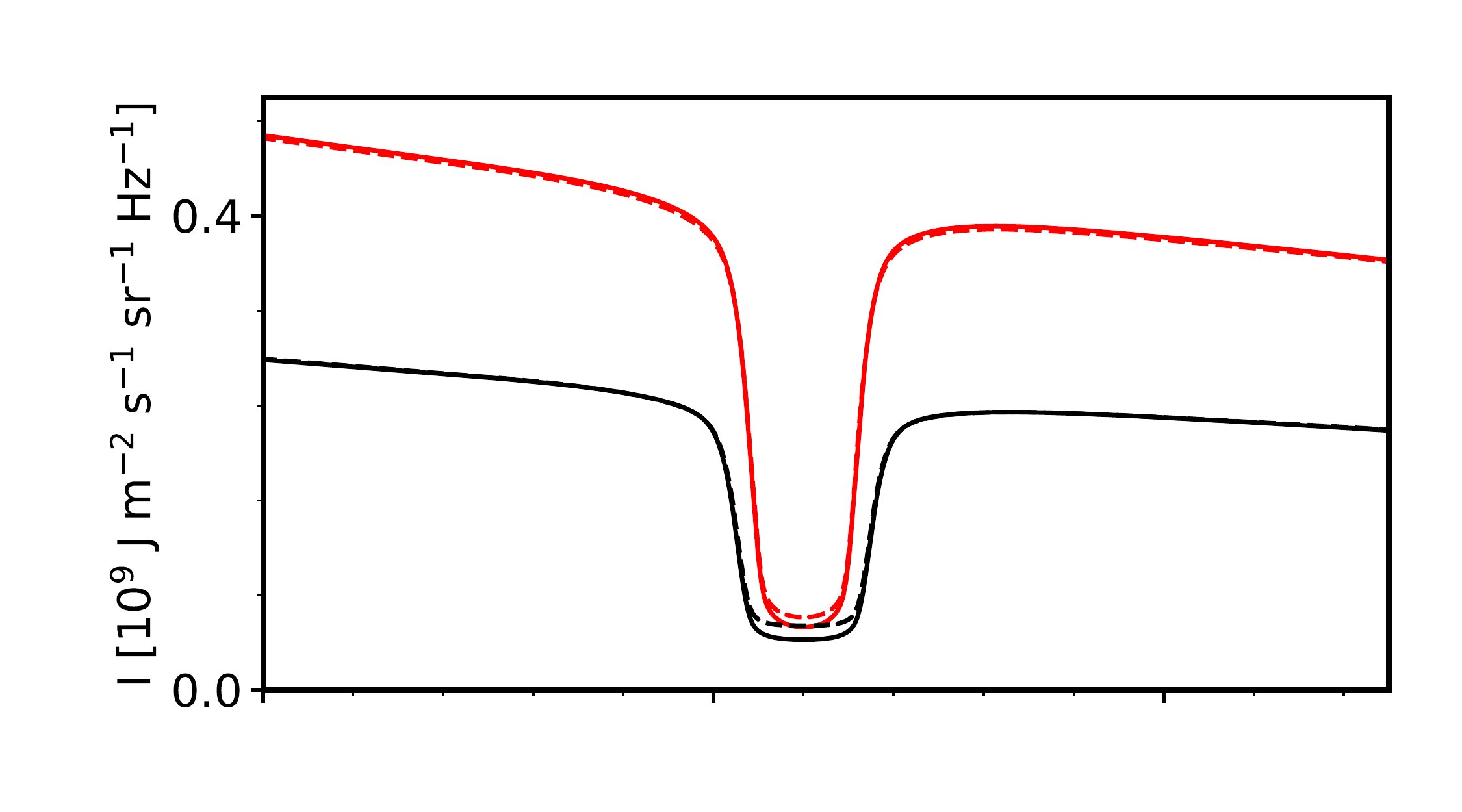}
\includegraphics[width=.32\hsize]{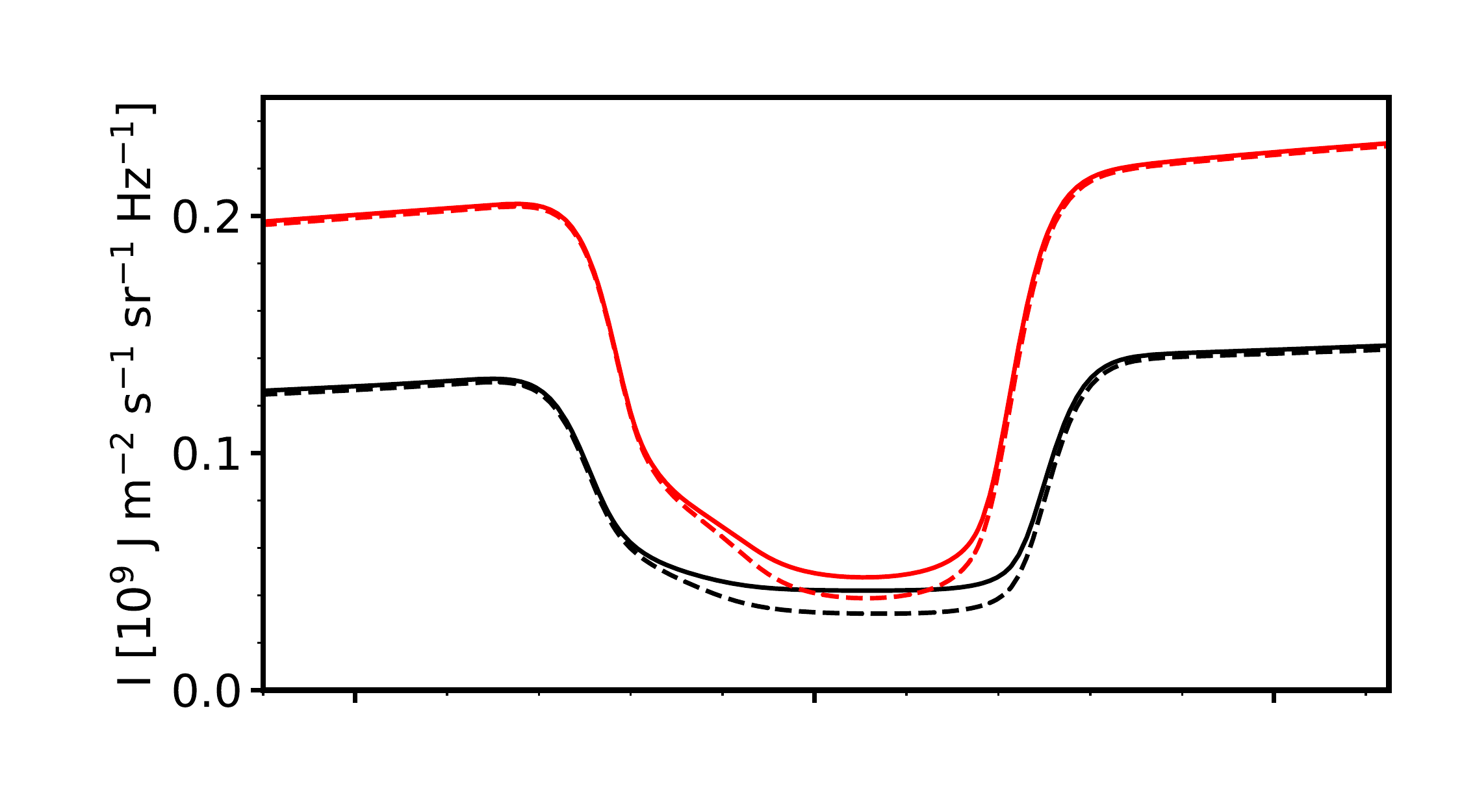} \\ \vspace{-1.5em}
\includegraphics[width=.32\hsize]{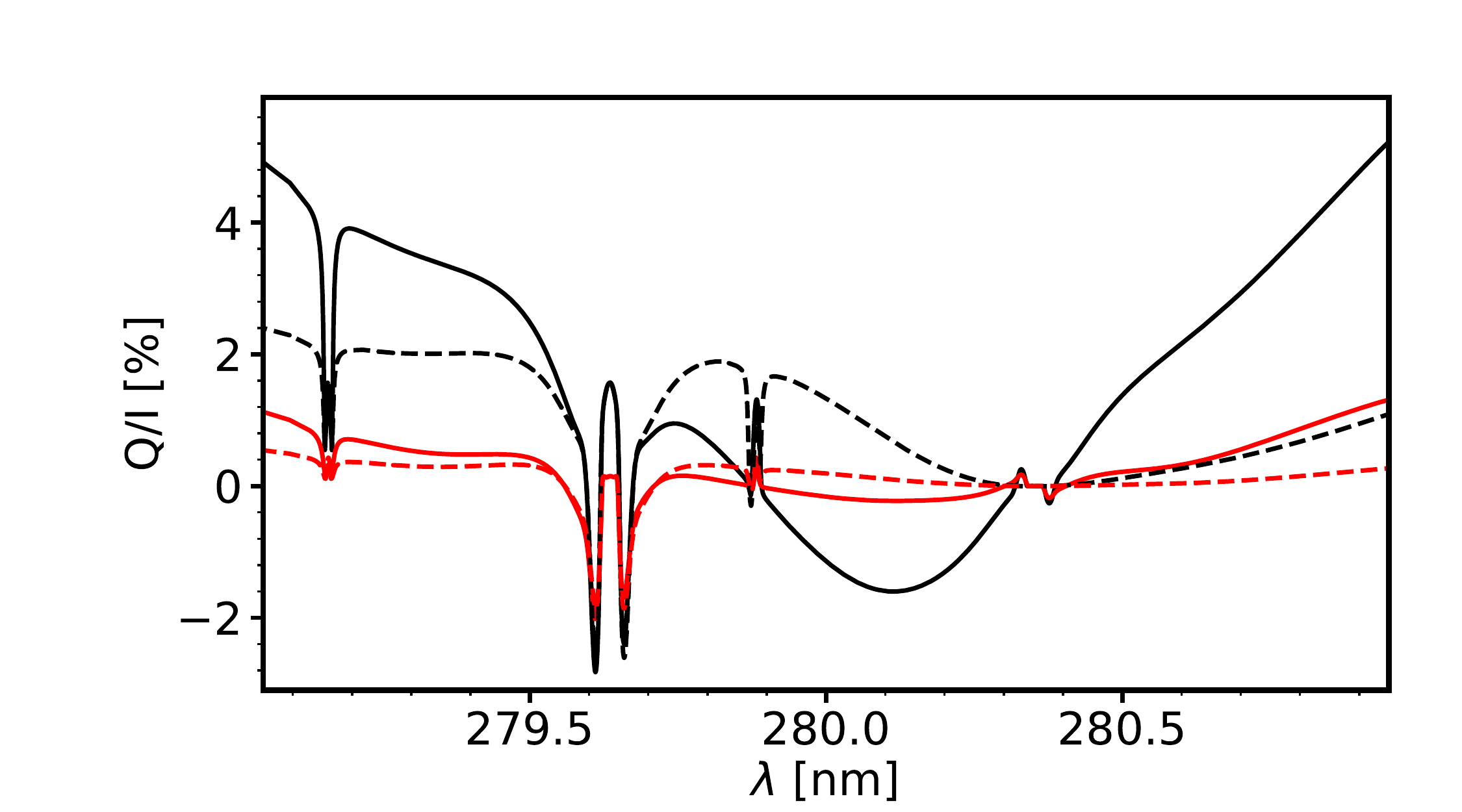}
\includegraphics[width=.32\hsize]{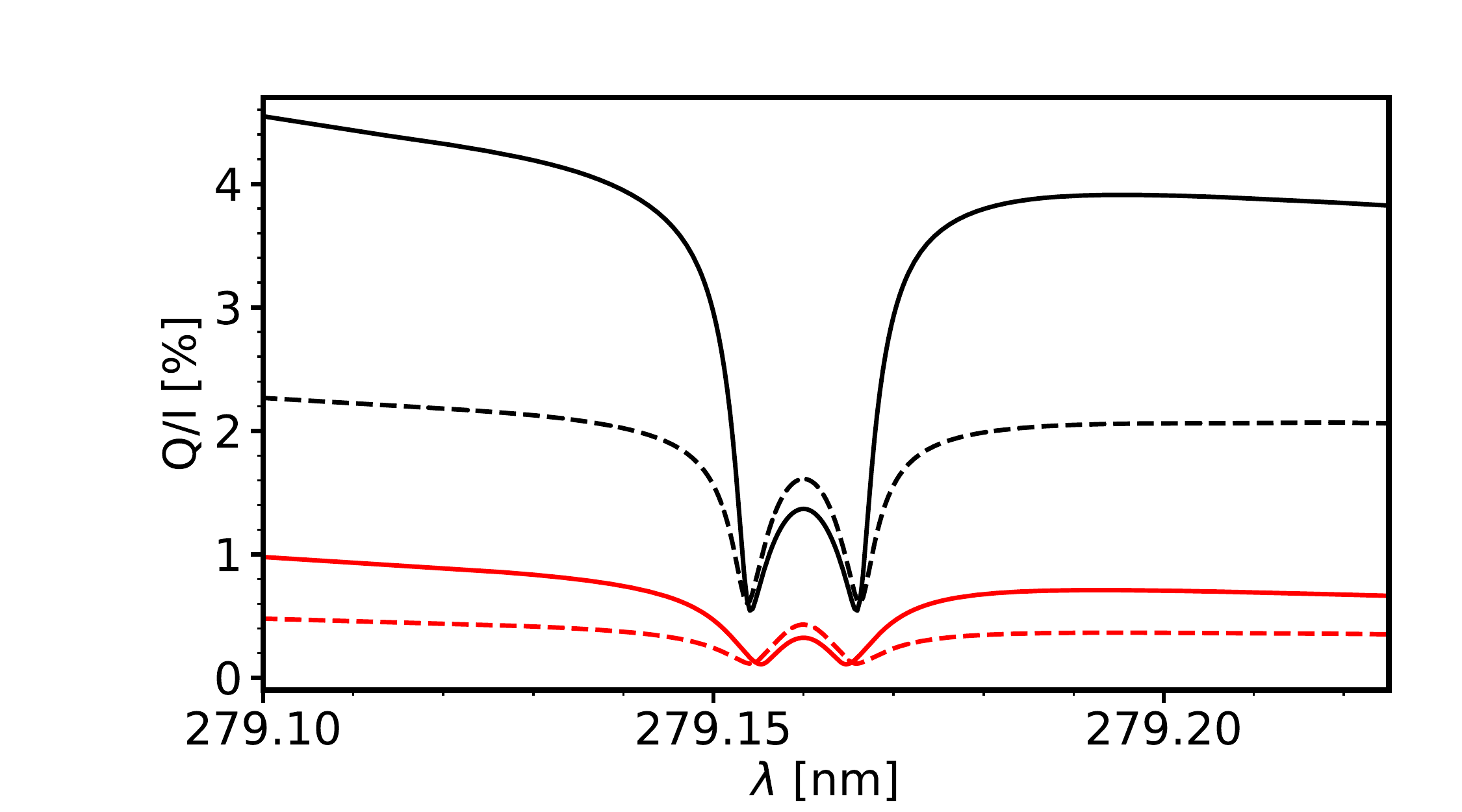}
\includegraphics[width=.32\hsize]{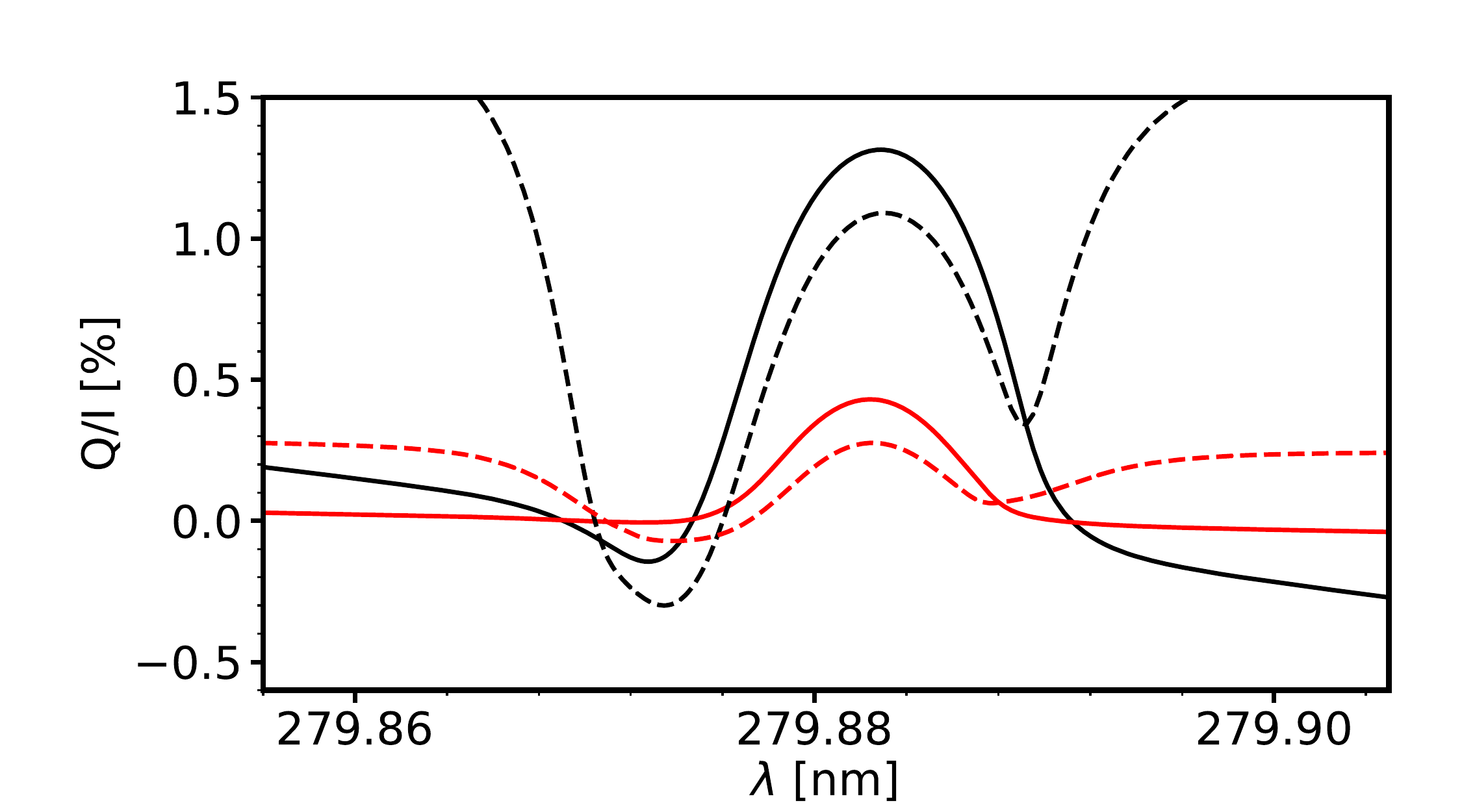}
\caption{Profiles of Stokes $I$ (top row) and $Q/I$ (bottom row) of the
\ion{Mg}{2} h-k doublet and UV triplet in the FAL-C model. The first column
shows the full spectral range, while the second and third columns show the spectral
regions around the subordinate lines. The solid (dashed) curves represent the
multi-term (multi-level) solution. The color of the curves indicates the $\mu$
value of the LOS: $0.1$ (black) and $0.5$ (red).}
\label{fig:MTvsML}
\end{figure}

As shown by \cite{BelluzziTrujillo2012}, PRD effects and quantum
interference in the upper term of the \ion{Mg}{2} h-k doublet is
fundamental for the correct modeling of the linear polarization produced by
scattering processes in these lines (see also \citealt{delPinoetal2016}). In
order to take into account this physical ingredient, we must solve the SE
equations for the multi-term atom. The bottom-left panel of Fig.~\ref{fig:MTvsML}
shows the impact of the upper-term quantum interference on the shape of
the broadband polarization pattern around the h-k resonances (solid curves)
for two lines of sight with different values of $\mu=\cos\theta$, where
$\theta$ is the heliocentric angle.\footnote{It is well known that,
for symmetry reasons,
the emergent polarization along the vertical direction of an axially symmetric
1D model cannot be linearly polarized. Larger heliocentric angles favor
scattering polarization and thus we choose values of $\mu = 0.1$ and $0.5$ in
our figures} For comparison, we also show the $Q/I$ profile obtained by
solving the SE equations for the multi-level atom, i.e., neglecting the
upper-term quantum interference (dashed curves).  With regard to the
broadband $Q/I$ pattern across the h-k doublet, the differences
between the multi-term and multi-level solutions are significant, and
characteristic of the ${}^2{\rm P}$--${}^2{\rm S}$ resonant transitions.

We point out that in our
multi-term approach we assume that every transition pertaining to the same multiplet
shares the same average radiation field (i.e., a ``flat'' spectrum
over the frequency range spanned by the fine-structure transitions within a
given multiplet). While this assumption
ensures the internal consistency of the SE equations in the CRD limit
(\citealt{BLandiLandolfi2004}), it can have an impact on the emergent spectral
profiles. The \ion{Mg}{2} transitions in this study are all
very close in wavelength (within $\sim 7\,$\AA), and the spectral modulation of
the emergent intensity profile is not significant enough to warrant relaxing the CRD
approximation for the solution of the SE problem. Although there are measurable
differences between the radiation field tensors of the h and k lines, those
differences around the height of formation are not significant, and have
a minimal impact on the line core of the emergent Stokes profiles.
On the other hand, the impact of the flat-spectrum approximation is more
significant for the subordinate lines, and the differences in the emergent
linear polarization follow closely the behavior of the radiation anisotropy:
${\rm s}_{\rm b}$ has larger anisotropy and linear polarization,
while ${\rm s}_{\rm r}$ shows less anisotropy and linear polarization.

Nevertheless, because the quantum interference between the $J$-levels
of the ${}^2{\rm P}$ term is a necessary physical ingredient to realistically
model the linear polarization of the \ion{Mg}{2} system, all the synthetic
profiles shown in the rest of this paper have been computed using the multi-term
model.

\begin{figure}[htp!]
\centering
\includegraphics[width=.32\hsize]{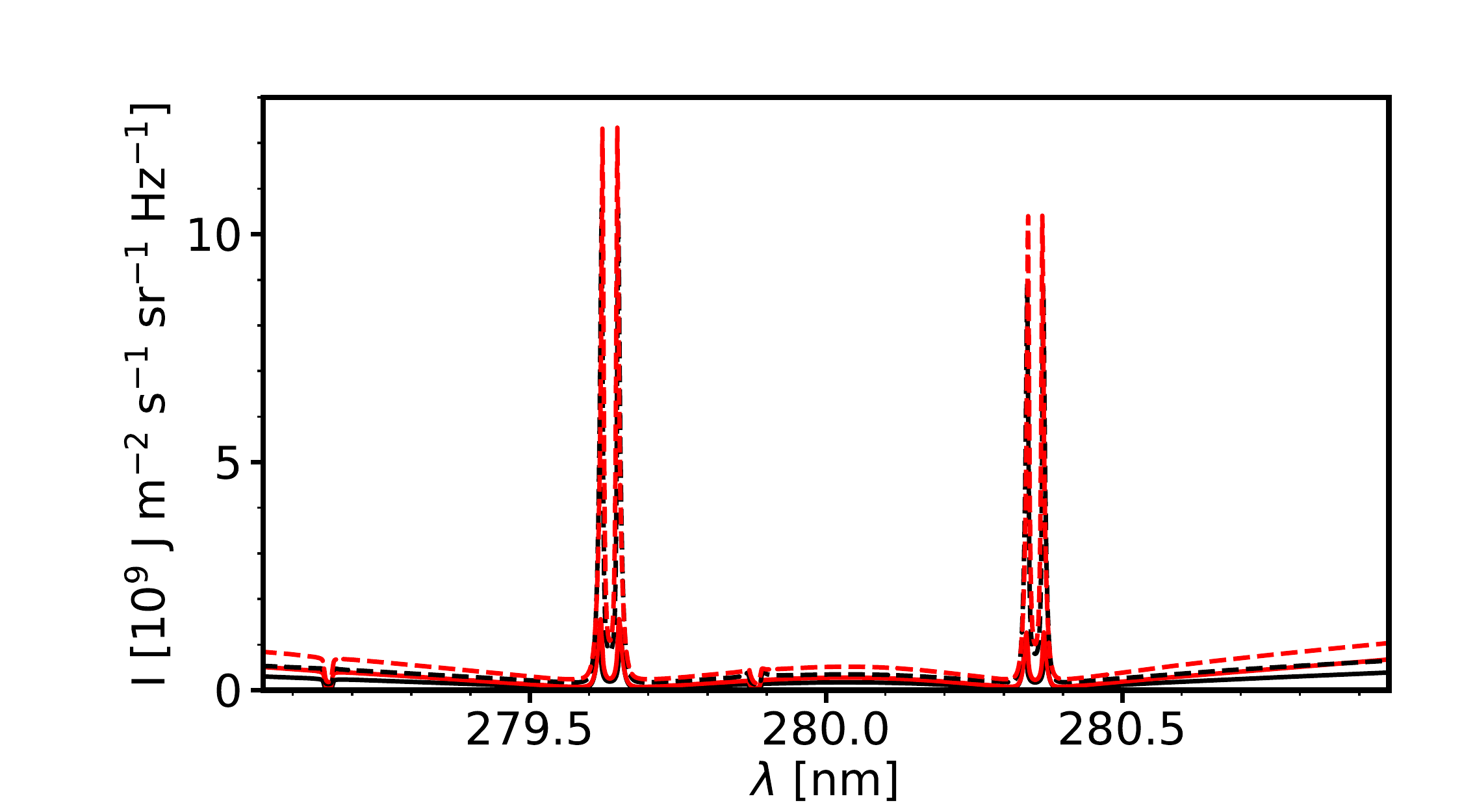}
\includegraphics[width=.32\hsize]{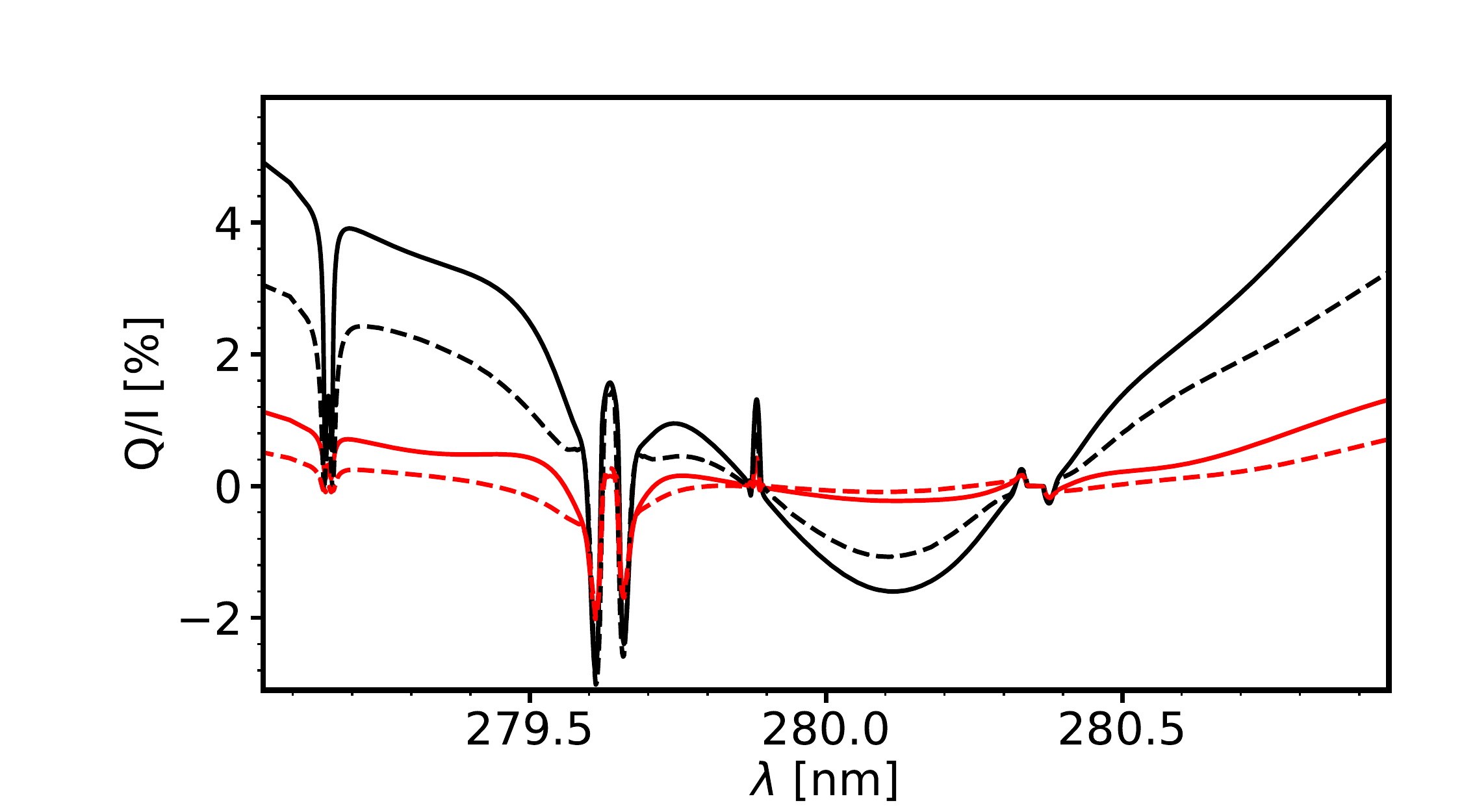} \\
\includegraphics[width=.32\hsize]{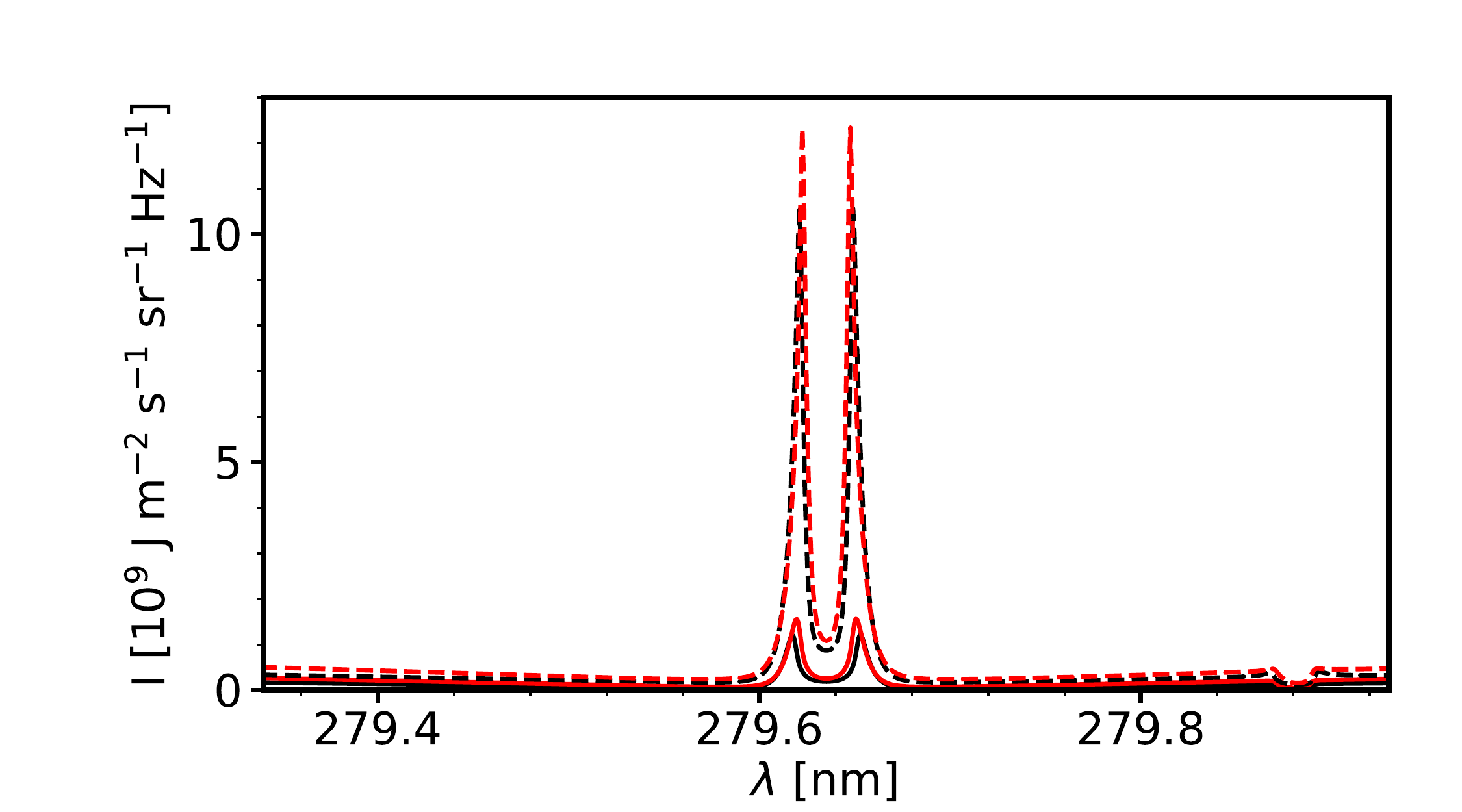}
\includegraphics[width=.32\hsize]{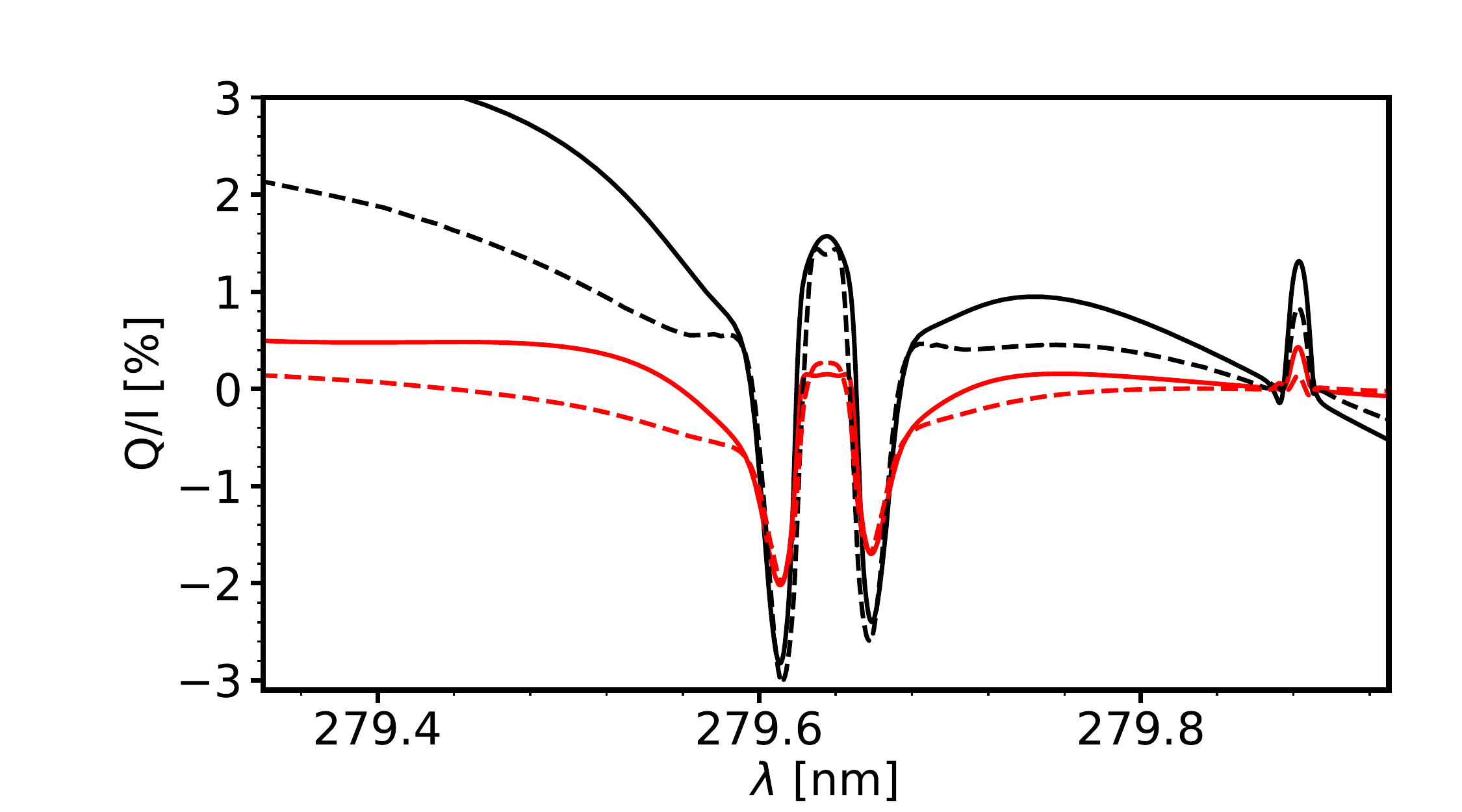} \\
\includegraphics[width=.32\hsize]{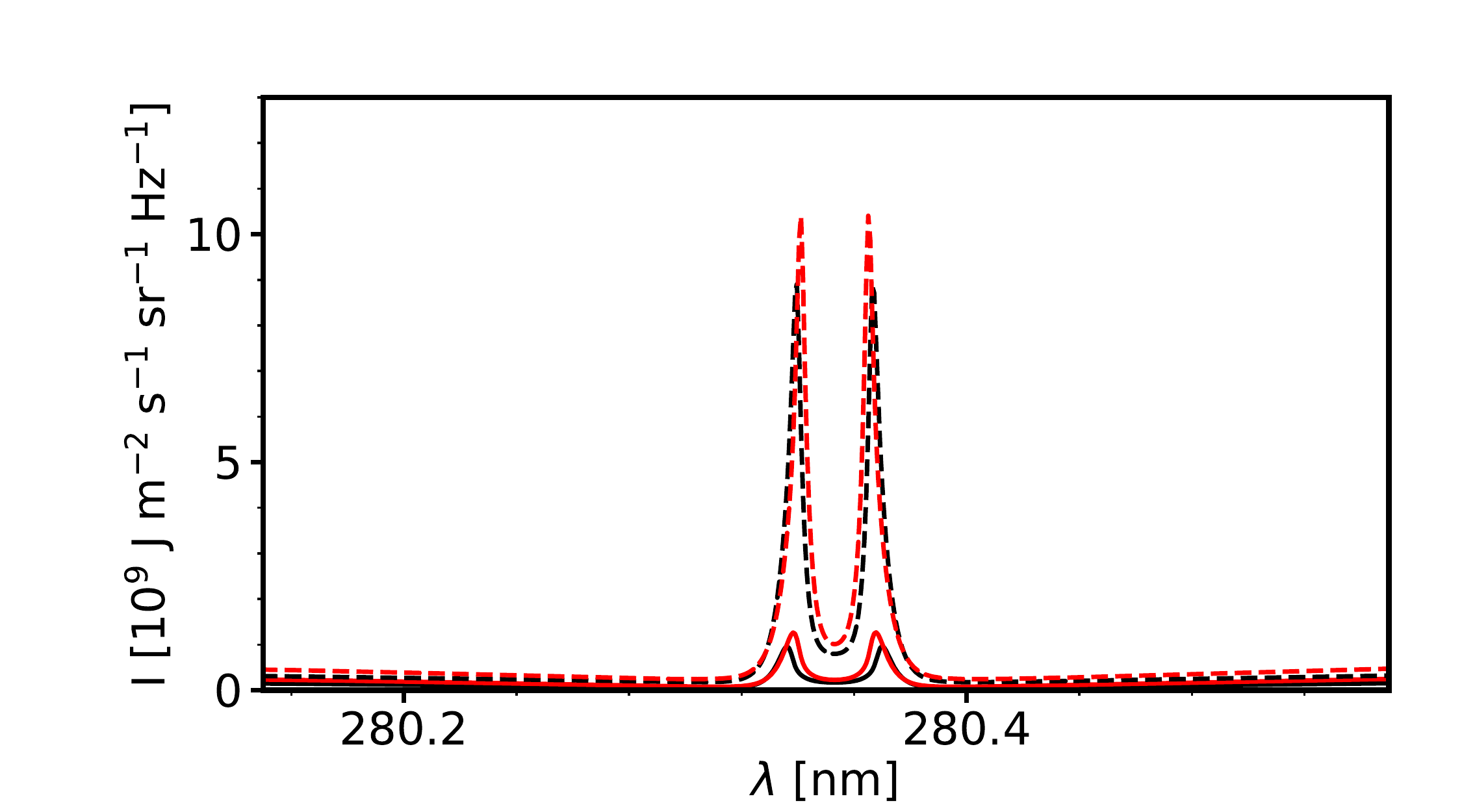}
\includegraphics[width=.32\hsize]{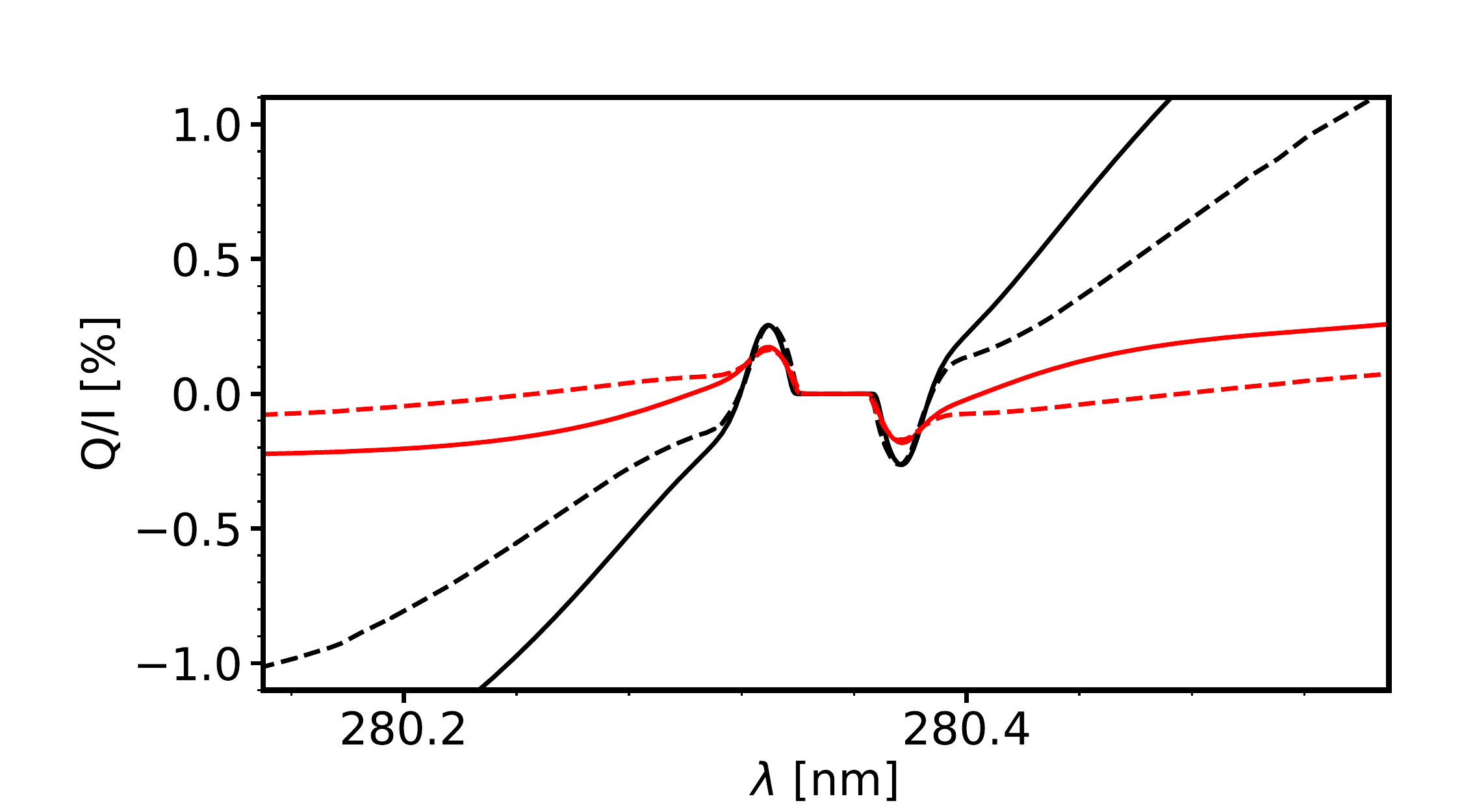} \\
\includegraphics[width=.32\hsize]{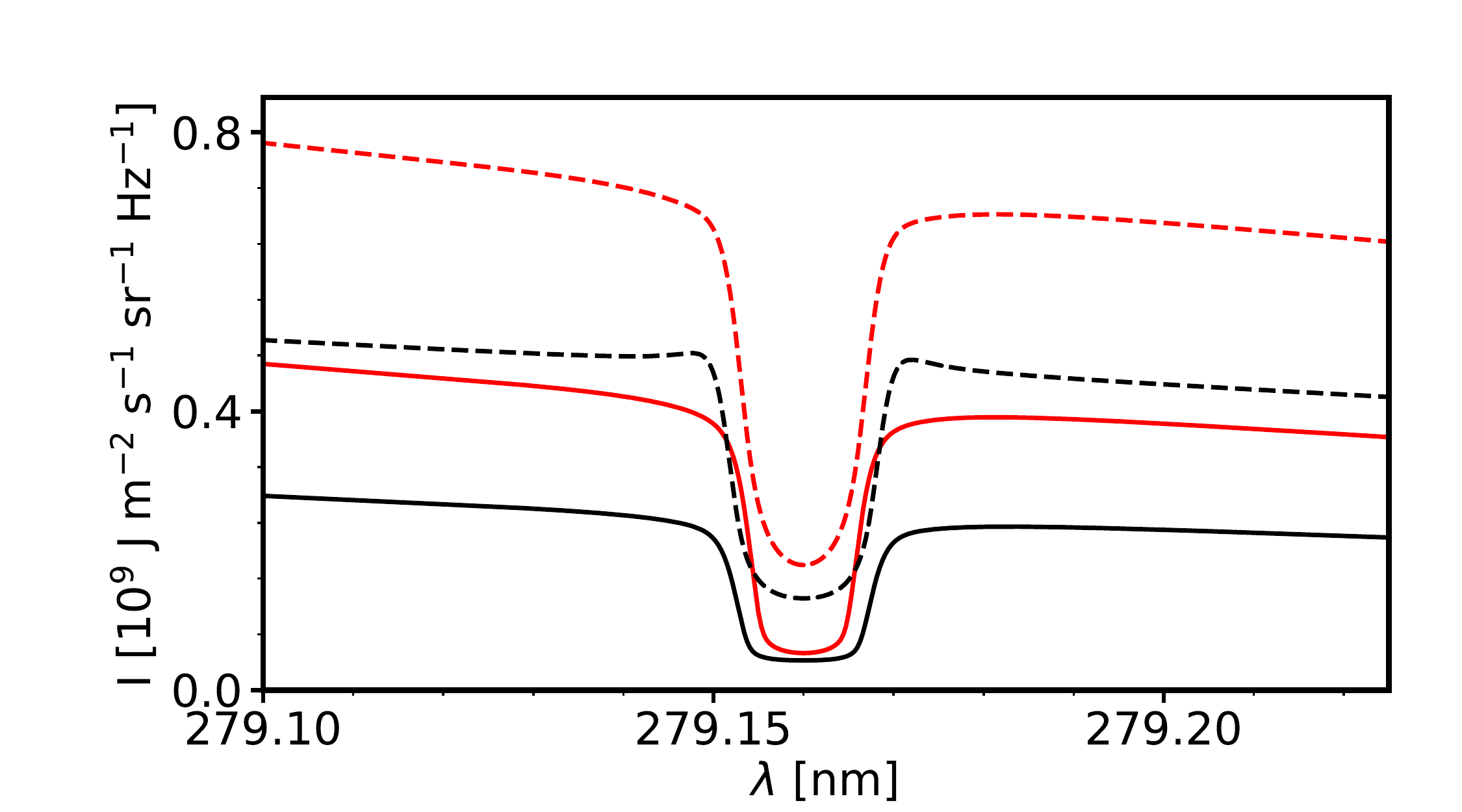}
\includegraphics[width=.32\hsize]{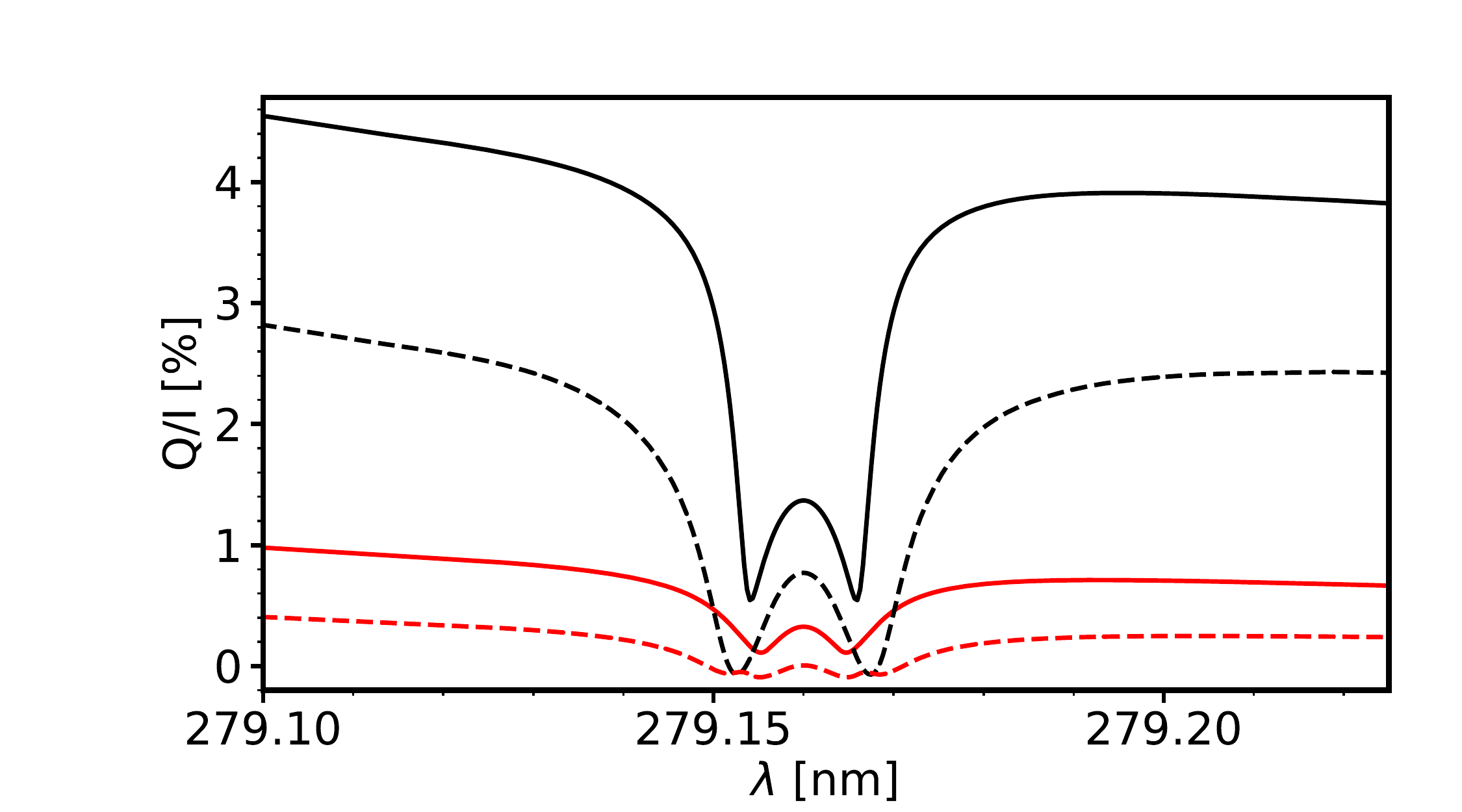} \\
\includegraphics[width=.32\hsize]{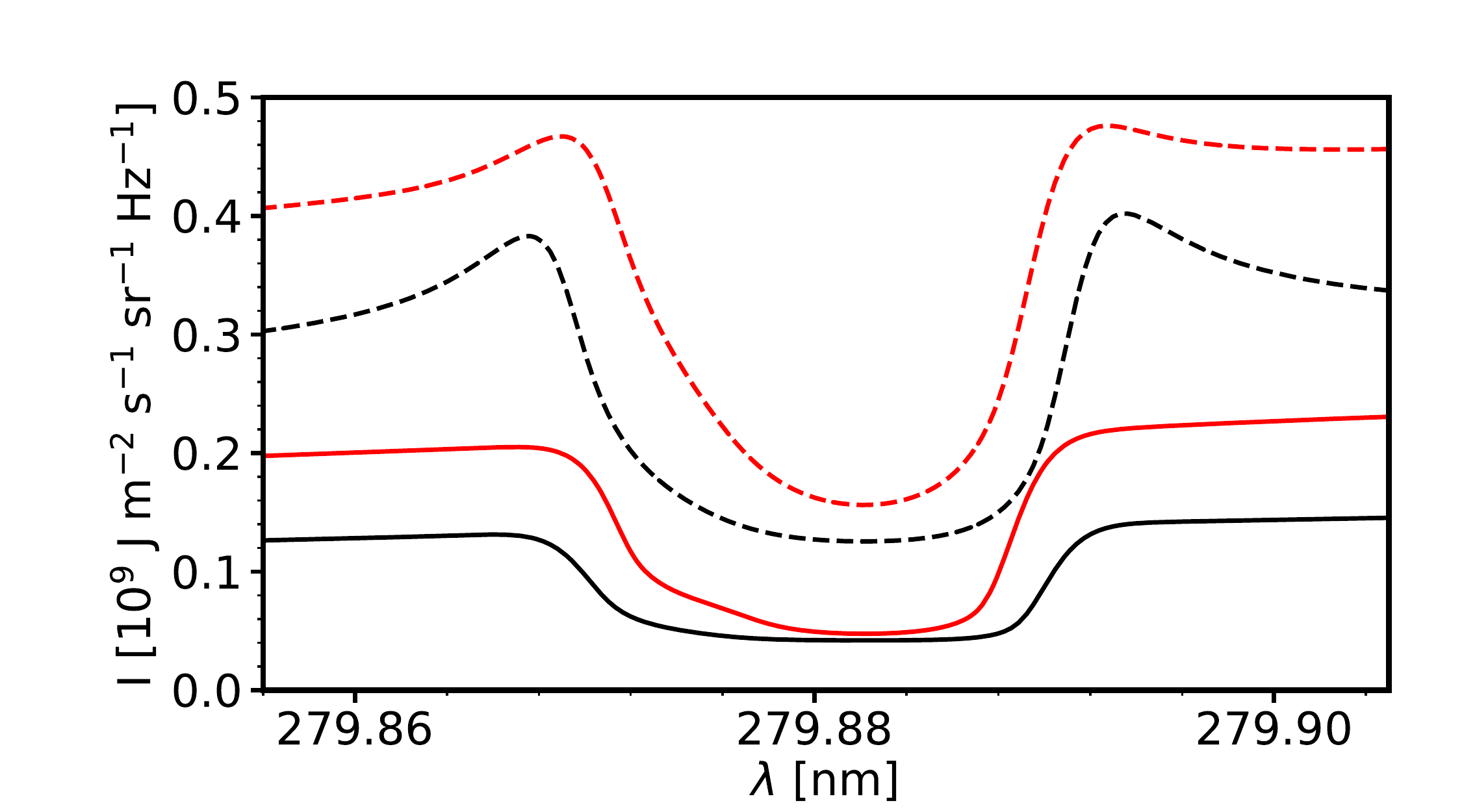}
\includegraphics[width=.32\hsize]{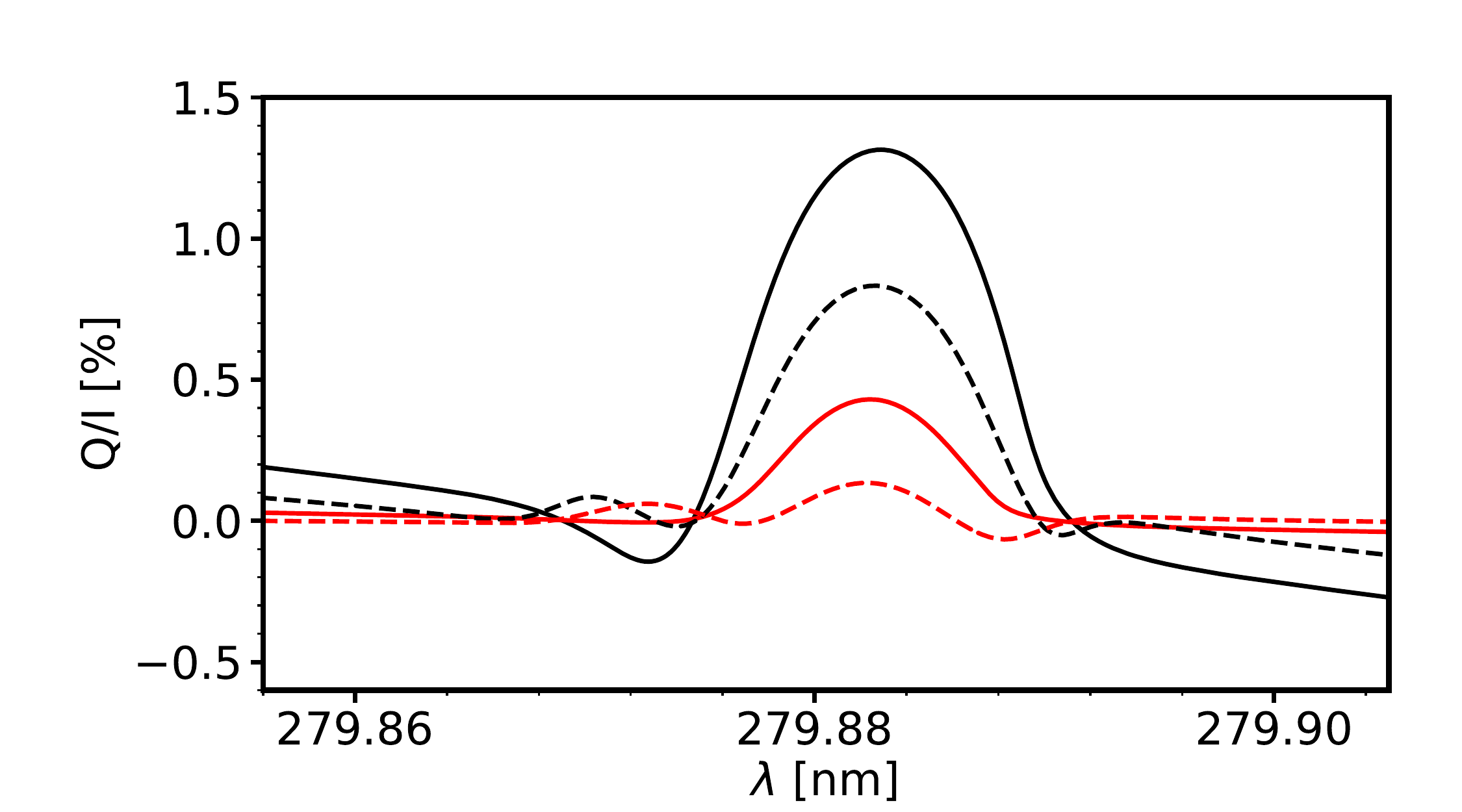}
\caption{Profiles of Stokes $I$ (left column) and $Q/I$ (right column) of
the \ion{Mg}{2} h-k doublet and UV triplet. The first row shows the full spectral
range, while the other rows show, in order, the spectral details of k, h,
$\rm s_b$, and $\rm s_{r_a}+s_{r_b}$. The solid (dashed) curves correspond to the
calculations using the FAL-C (FAL-P) model atmosphere. The color of the curves
indicates the $\mu$ value of the LOS: $0.1$ (black) and $0.5$ (red).}
\label{fig:Thermal}
\end{figure}

\begin{figure}[htp!]
\centering
\includegraphics[width=.45\hsize]{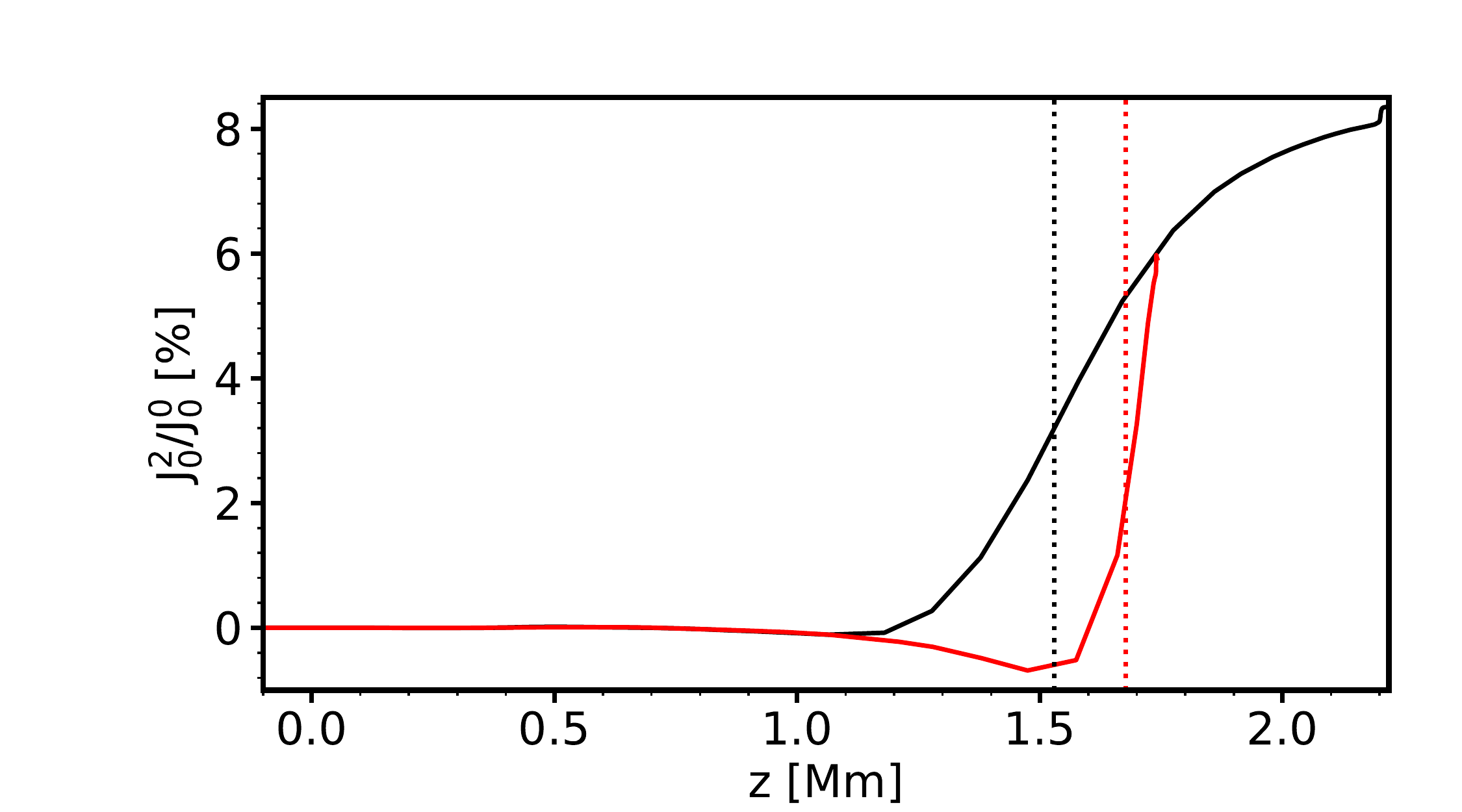}
\caption{Variation with height of the fractional anisotropy 
$J^2_0/J^0_0$ for the \ion{Mg}{2} UV triplet in the FAL-C (black)
and FAL-P (red) models. The vertical dotted lines show the largest height where
the optical depth is unity within the range of the subordinate lines for
a LOS with $\mu = 0.1$}
\label{fig:J20sub}
\end{figure}

Figure \ref{fig:Thermal} shows the intensity and fractional linear polarization
profiles for the FAL-C (average atmosphere) and FAL-P (plage)
semi-empirical models.
Because these models are
static and non-magnetic, all the differences between the two calculations are due
to the thermodynamic structure of the model atmosphere, that is, how the
atmospheric temperature and density change with height. The FAL-P emergent
intensity is larger than its FAL-C counterpart across the whole spectral
region.
Such an increase in intensity is a consequence of the larger temperature (and
consequently larger electron density) below the chromosphere-corona transition
region in the case of the FAL-P model. The effect on the fractional
linear polarization is the opposite, being smaller in the FAL-P model.
In fact, FAL-P produces a lower degree of anisotropy in the subordinate lines
than FAL-C (see Fig.~\ref{fig:J20sub}), leading to a smaller degree of
atomic alignment and, consequently, a smaller fractional linear
polarization in the plage model.

\subsection{Magnetic Sensitivity}\label{SSMagnetic}

\begin{figure}[htp!]
\centering
\includegraphics[width=.4\hsize]{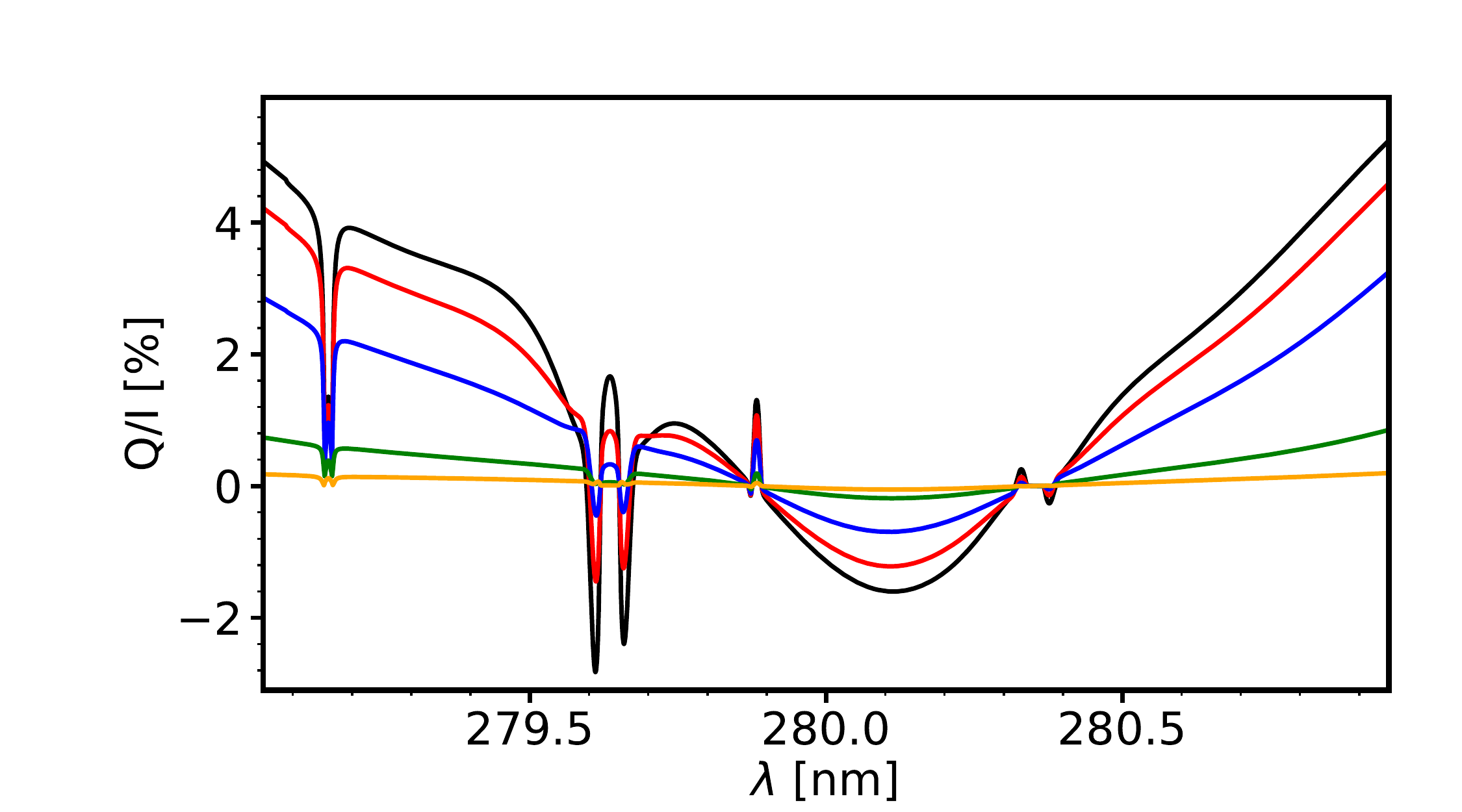}
\includegraphics[width=.4\hsize]{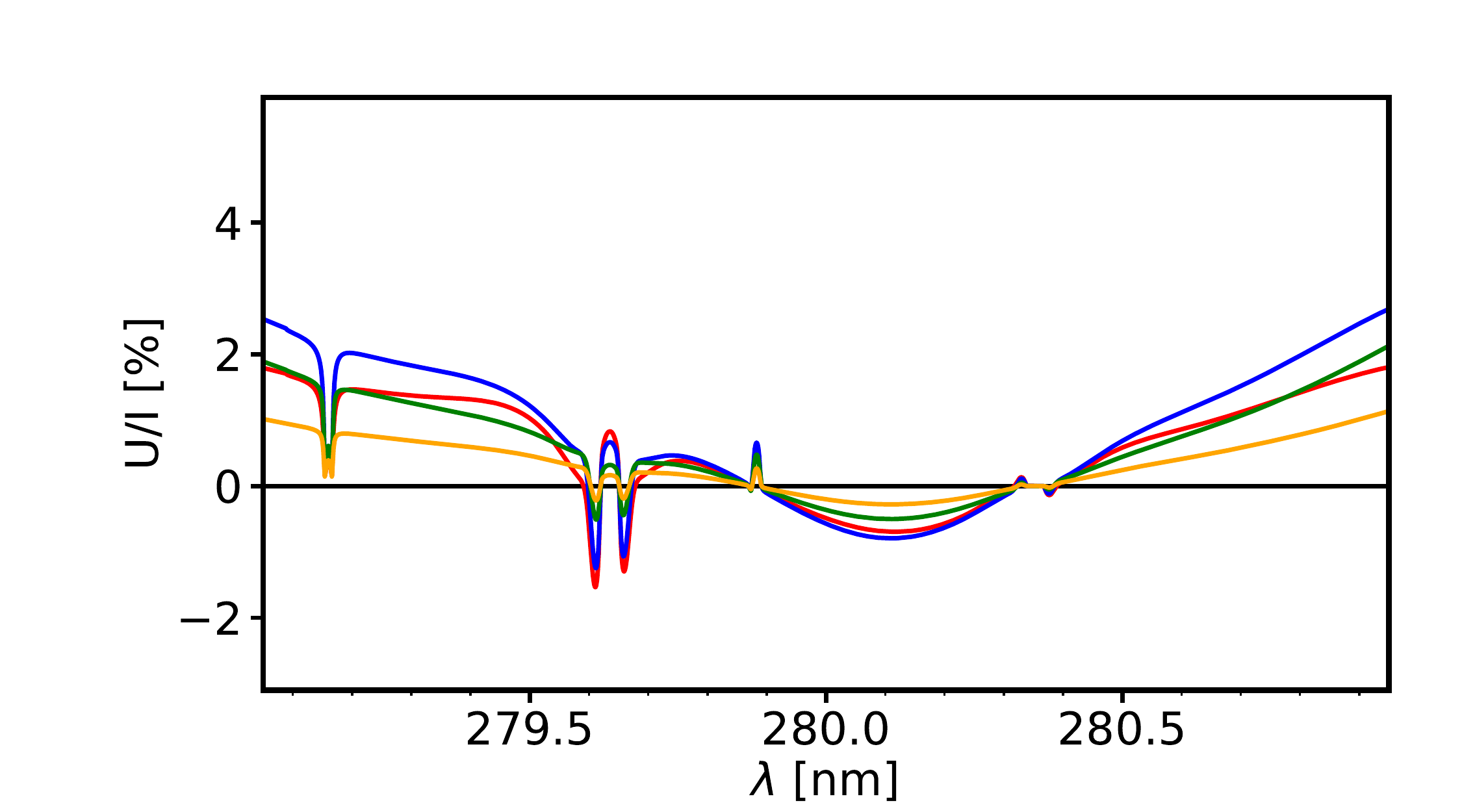} \\
\includegraphics[width=.4\hsize]{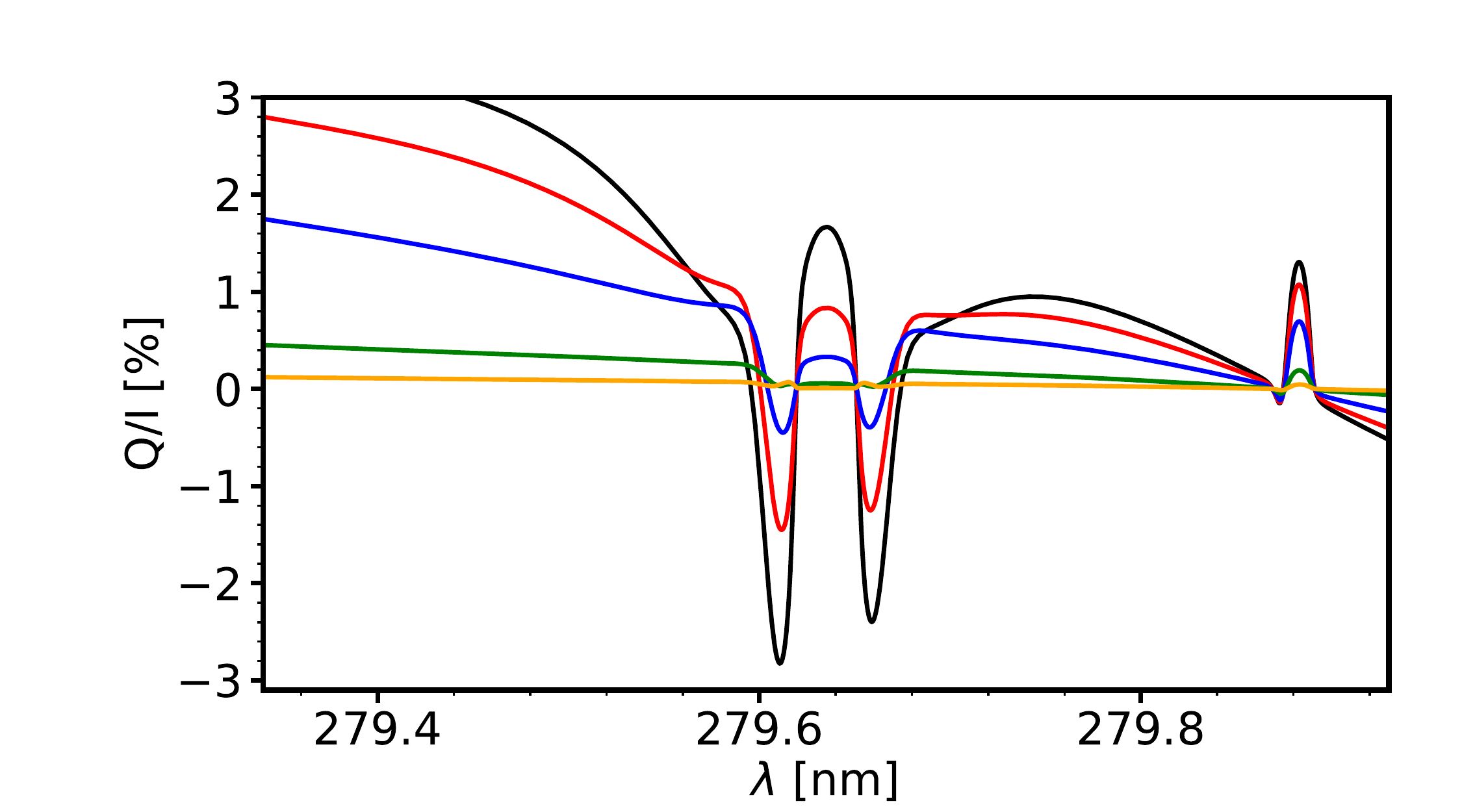}
\includegraphics[width=.4\hsize]{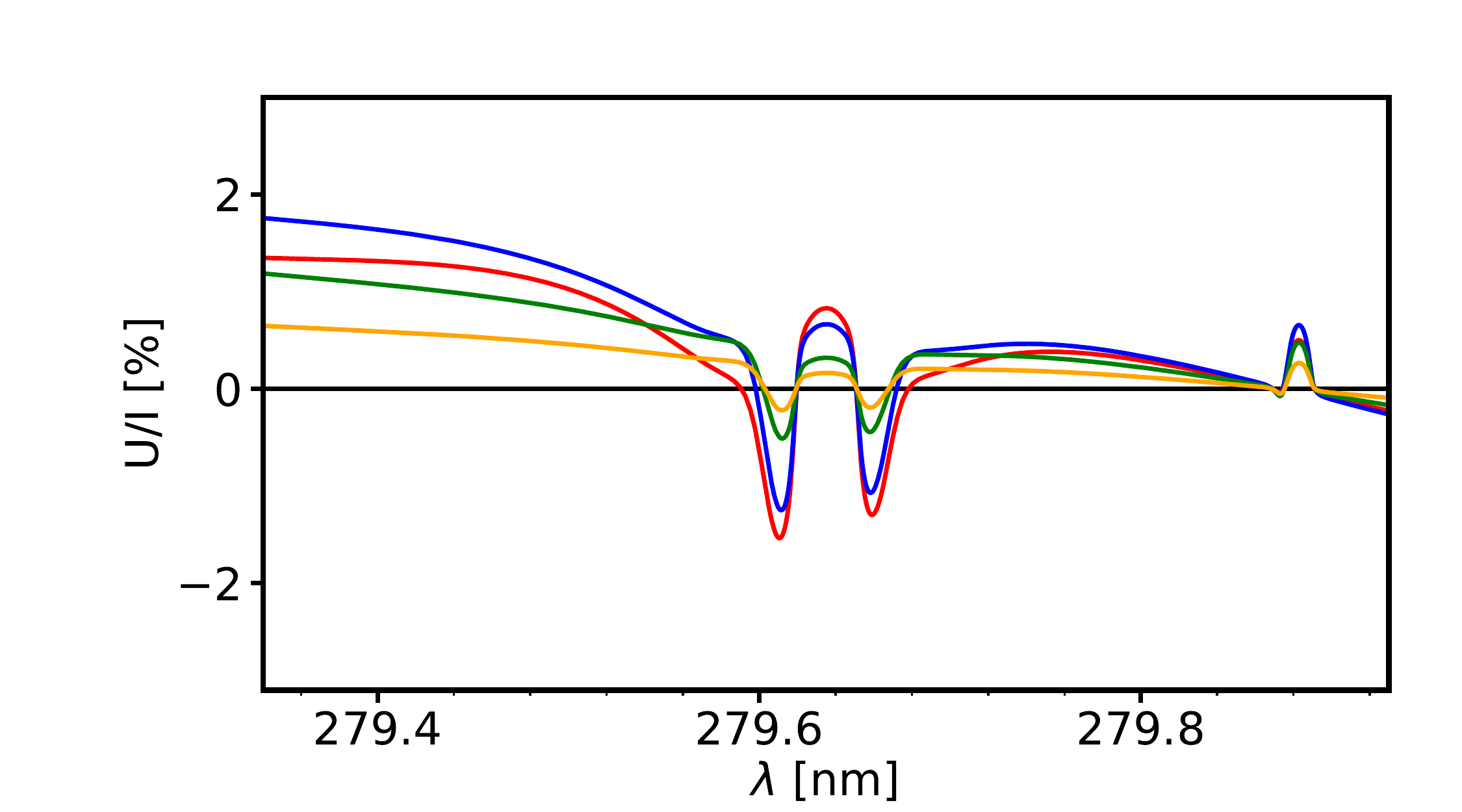} \\
\includegraphics[width=.4\hsize]{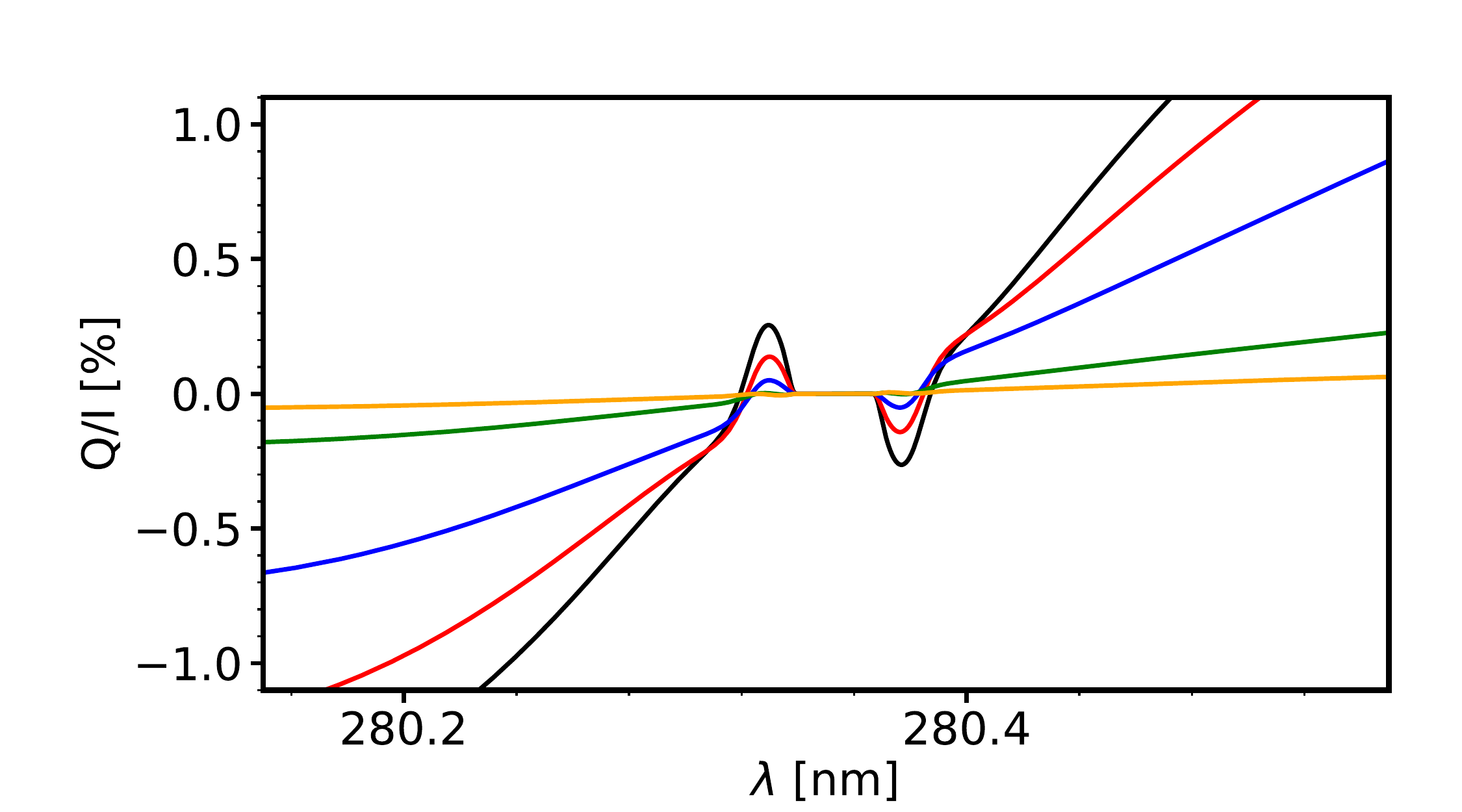}
\includegraphics[width=.4\hsize]{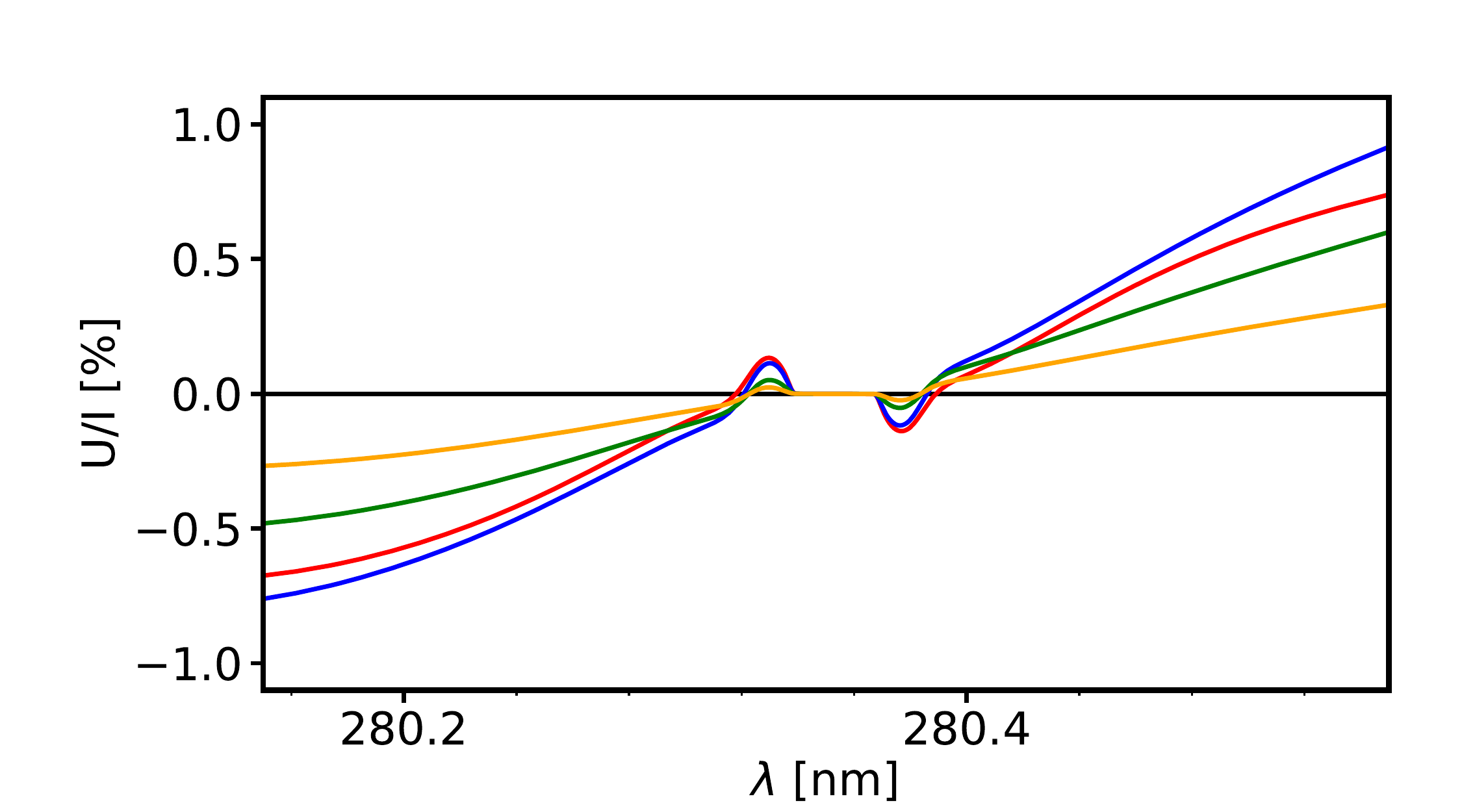} \\
\includegraphics[width=.4\hsize]{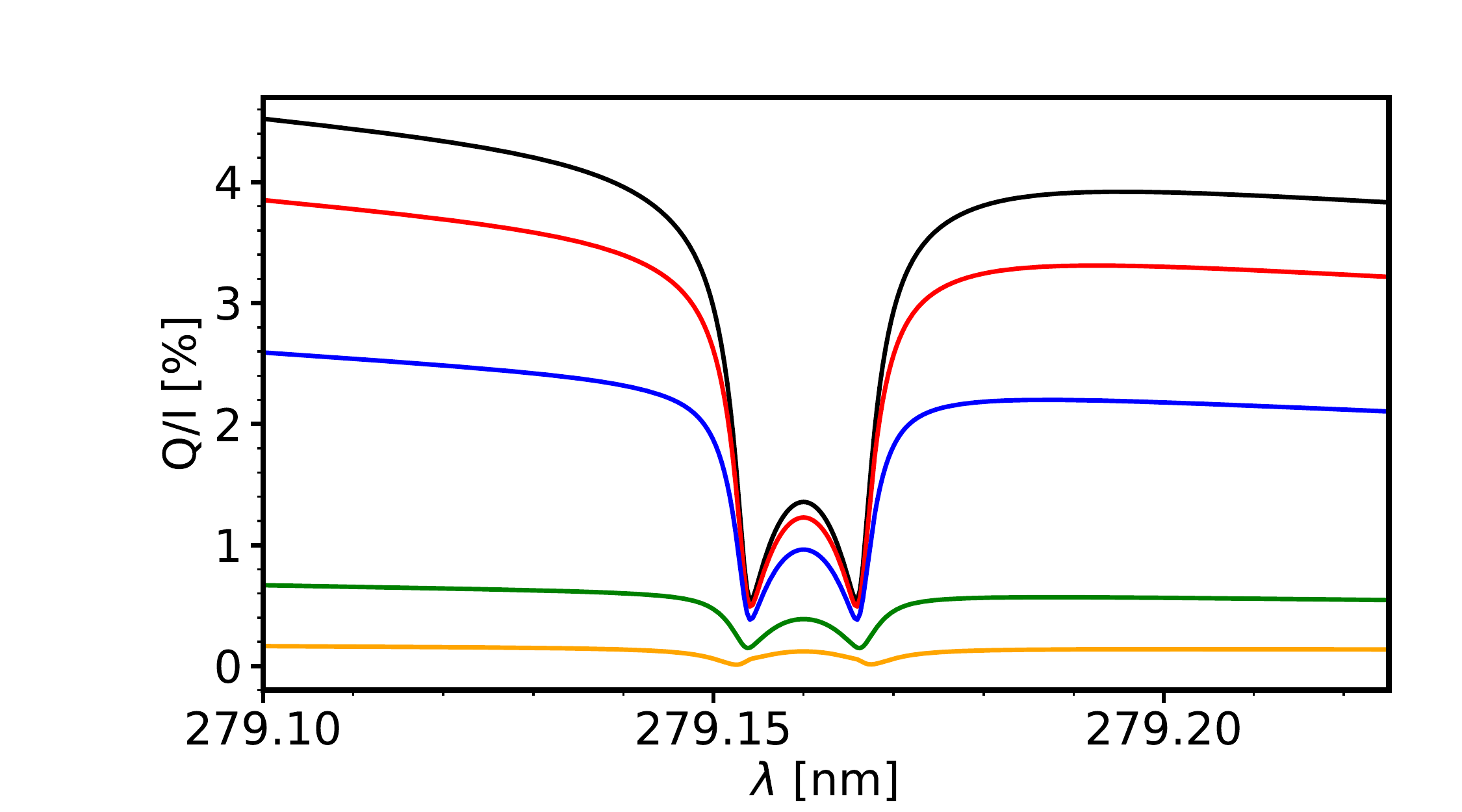}
\includegraphics[width=.4\hsize]{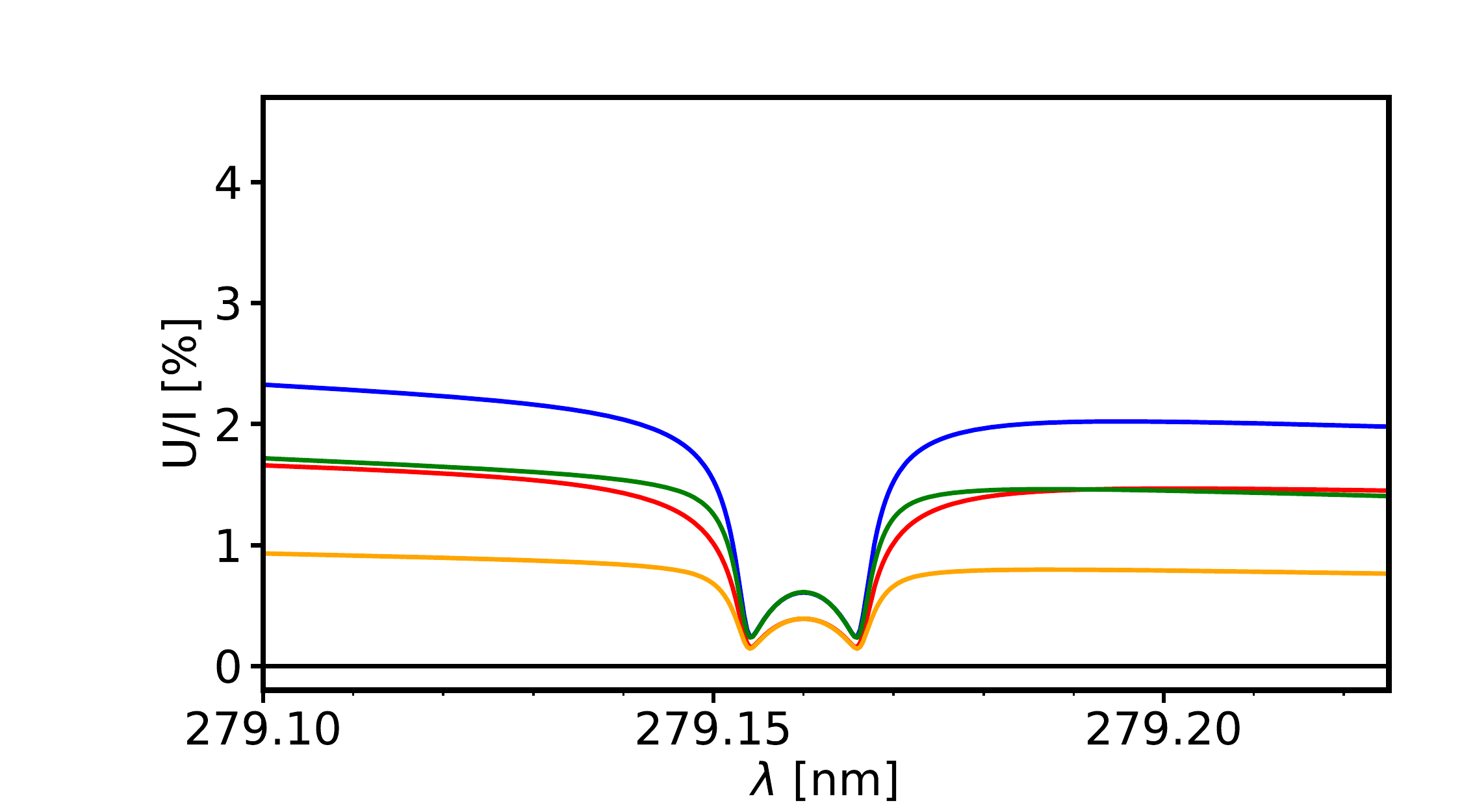} \\
\includegraphics[width=.4\hsize]{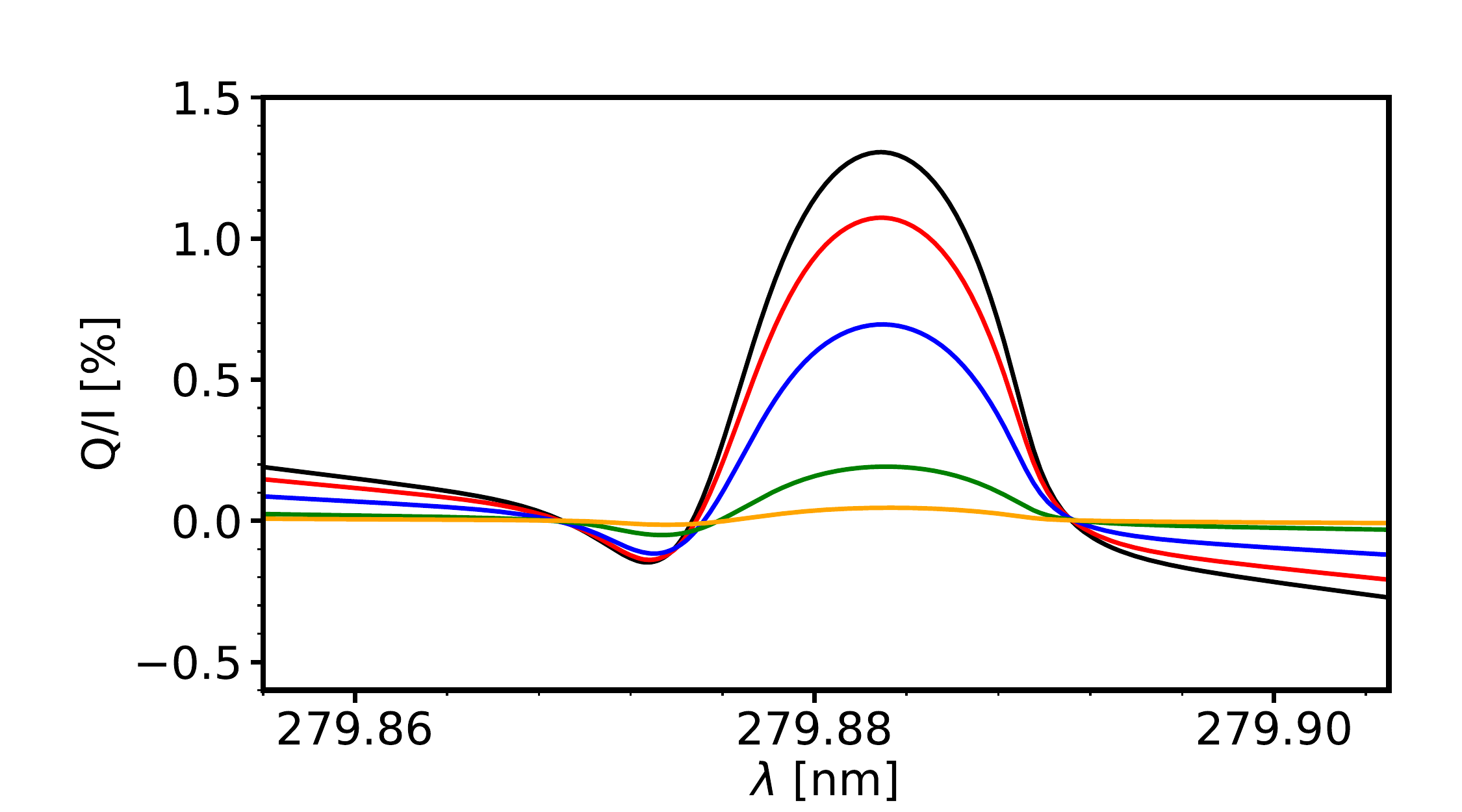}
\includegraphics[width=.4\hsize]{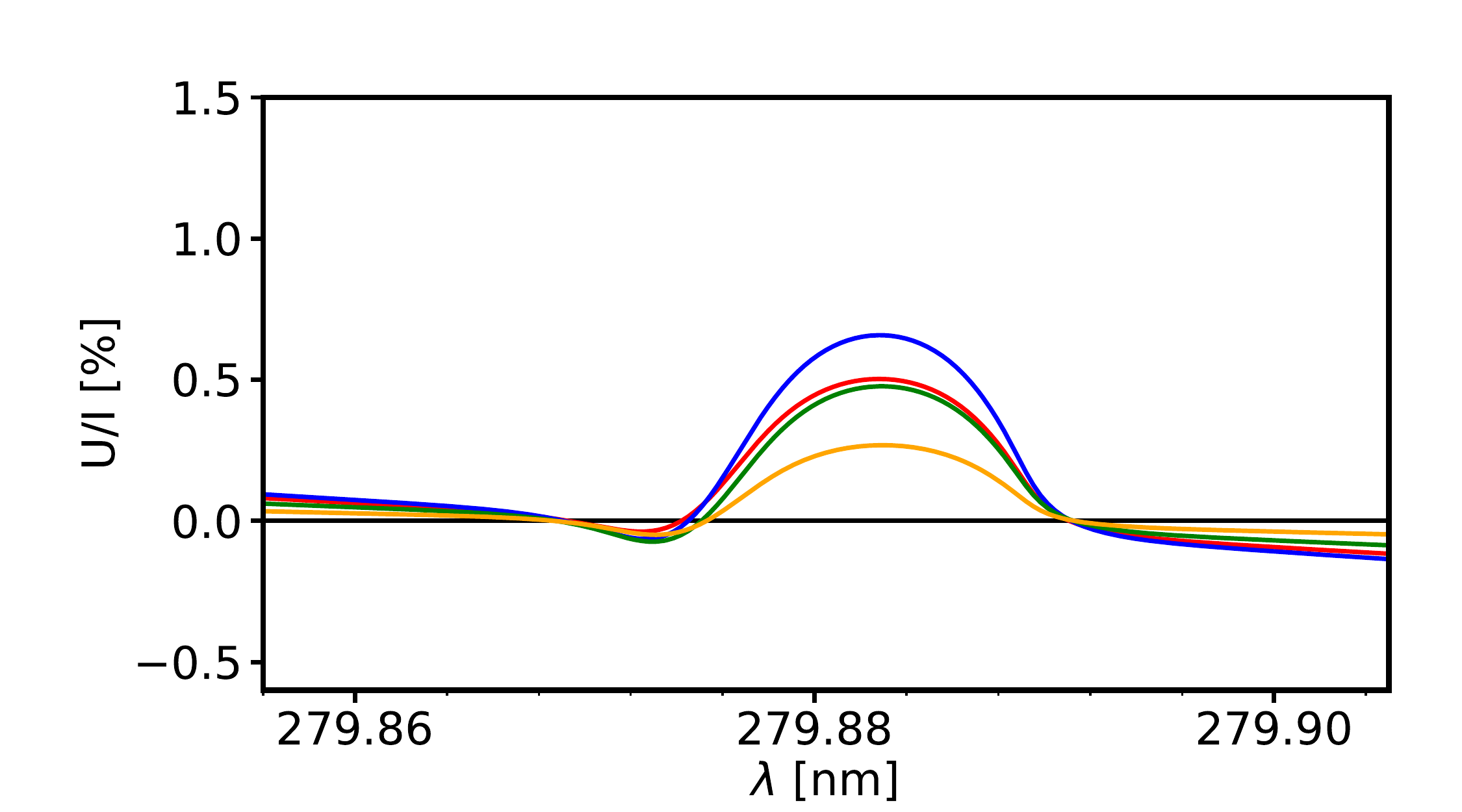} \\
\caption{Fractional linear polarization profiles $Q/I$ (left column) and $U/I$
(right column) for the \ion{Mg}{2} h-k doublet and UV triplet in the FAL-C model,
for a LOS with $\mu=0.1$. The first row shows the full spectral range, while the
second to fifth rows show the spectral regions around the transition lines. The
color of the curves indicates the strength of the horizontal magnetic field
pointing towards the observer: $0\,$G (black), $10\,$G (red),
$20\,$G (blue), $50\,$G (green), and $100\,$G (orange).}
\label{fig:MagnProfC}
\end{figure}

\begin{figure}[htp!]
\centering
\includegraphics[width=.4\hsize]{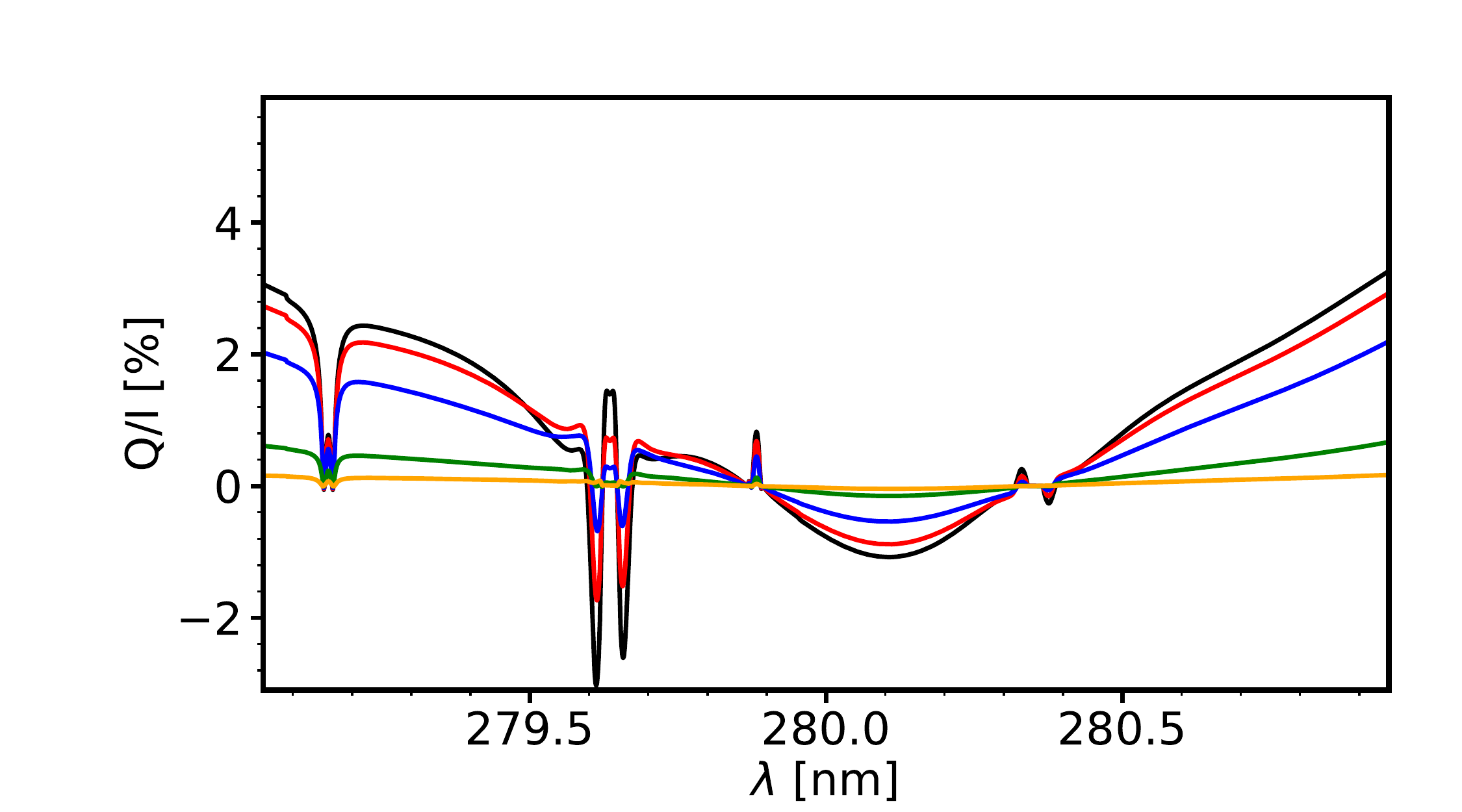} 
\includegraphics[width=.4\hsize]{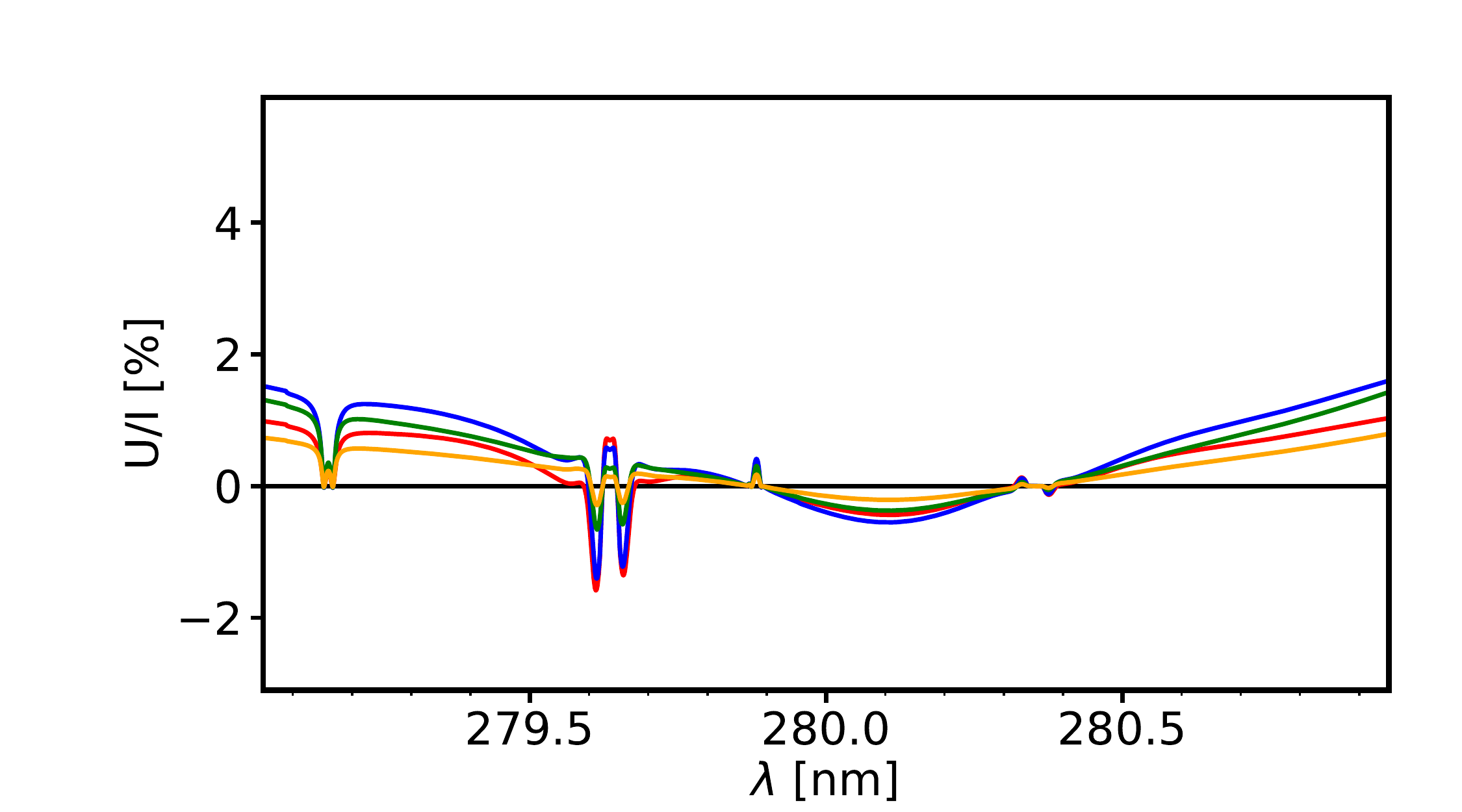} \\
\includegraphics[width=.4\hsize]{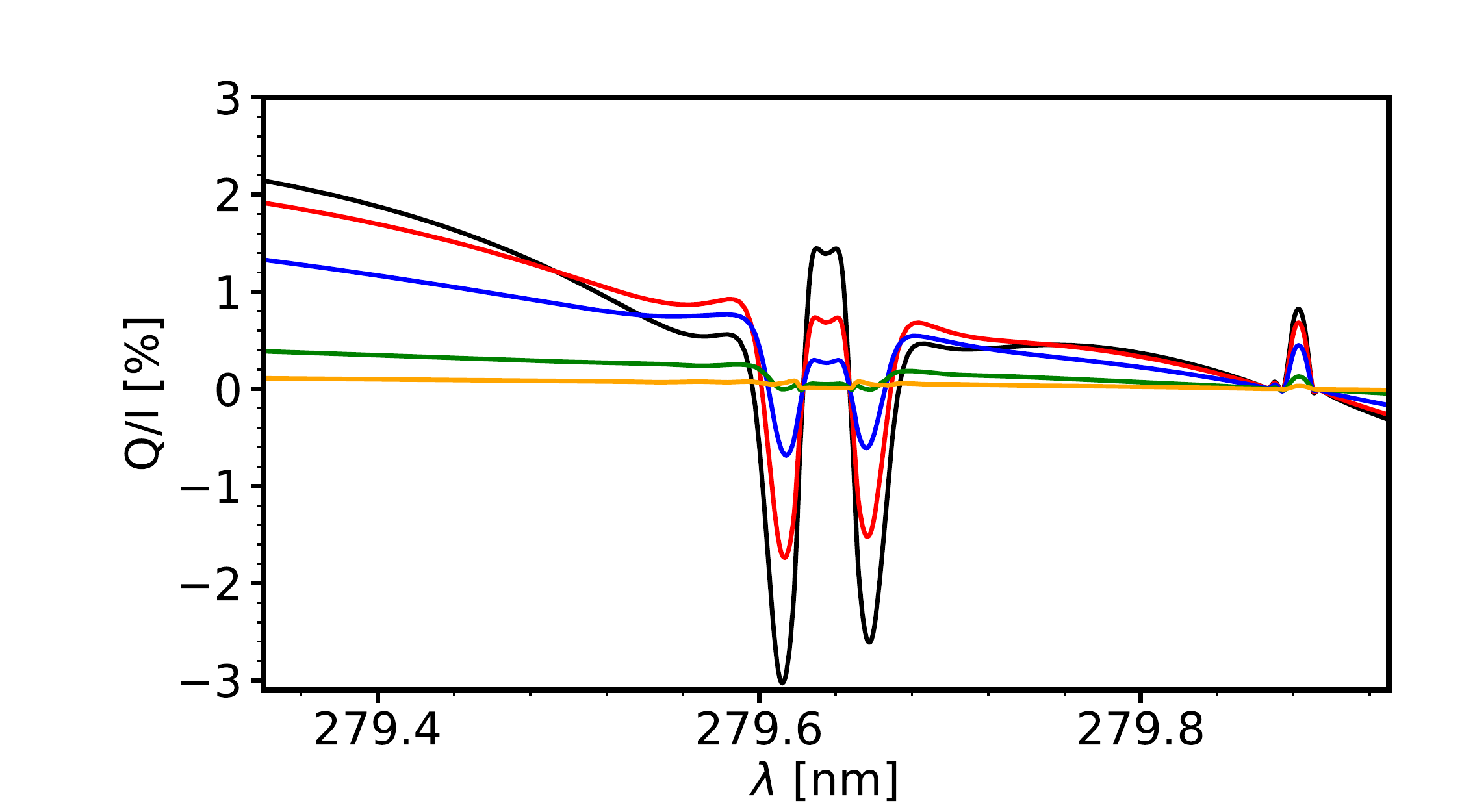}
\includegraphics[width=.4\hsize]{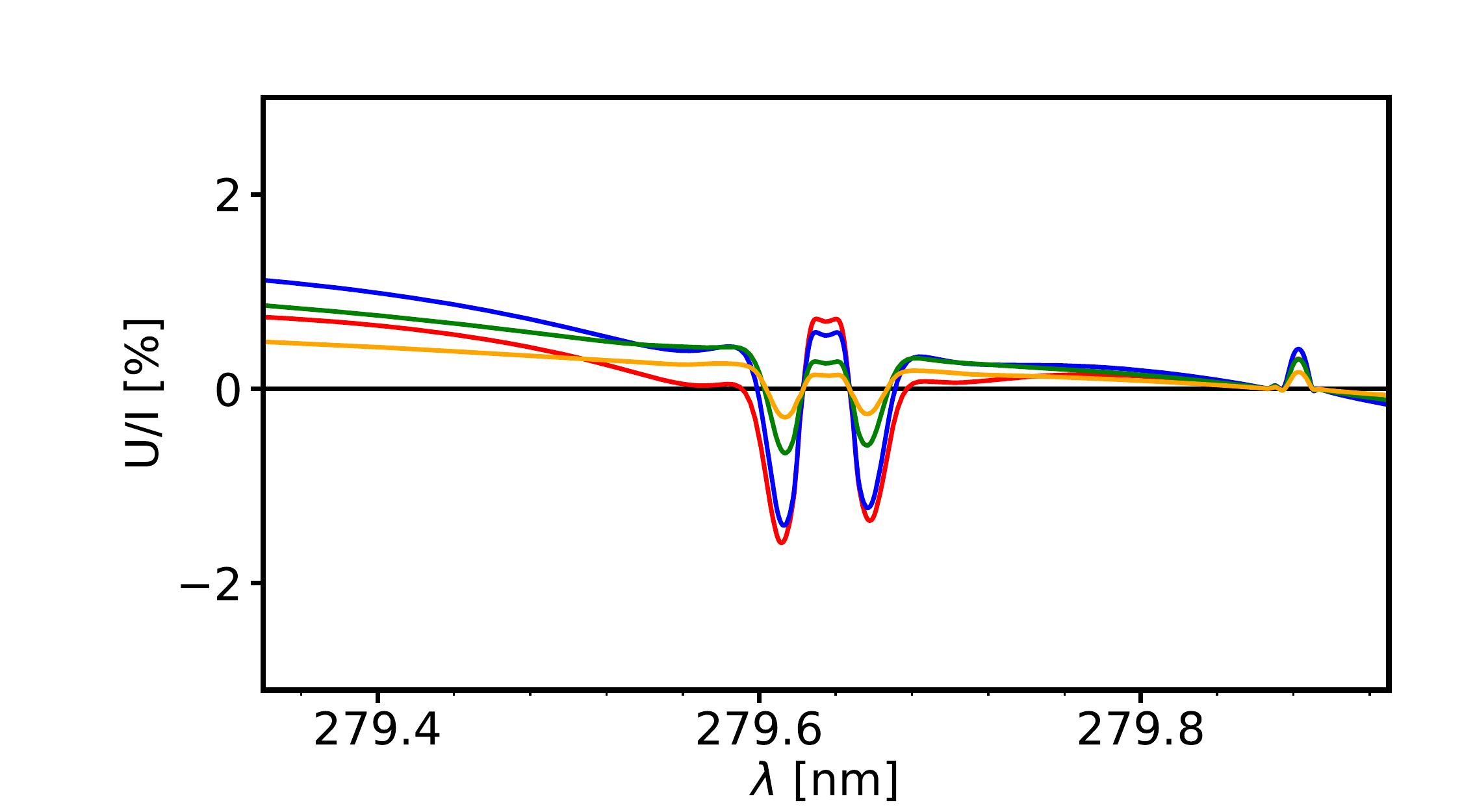} \\
\includegraphics[width=.4\hsize]{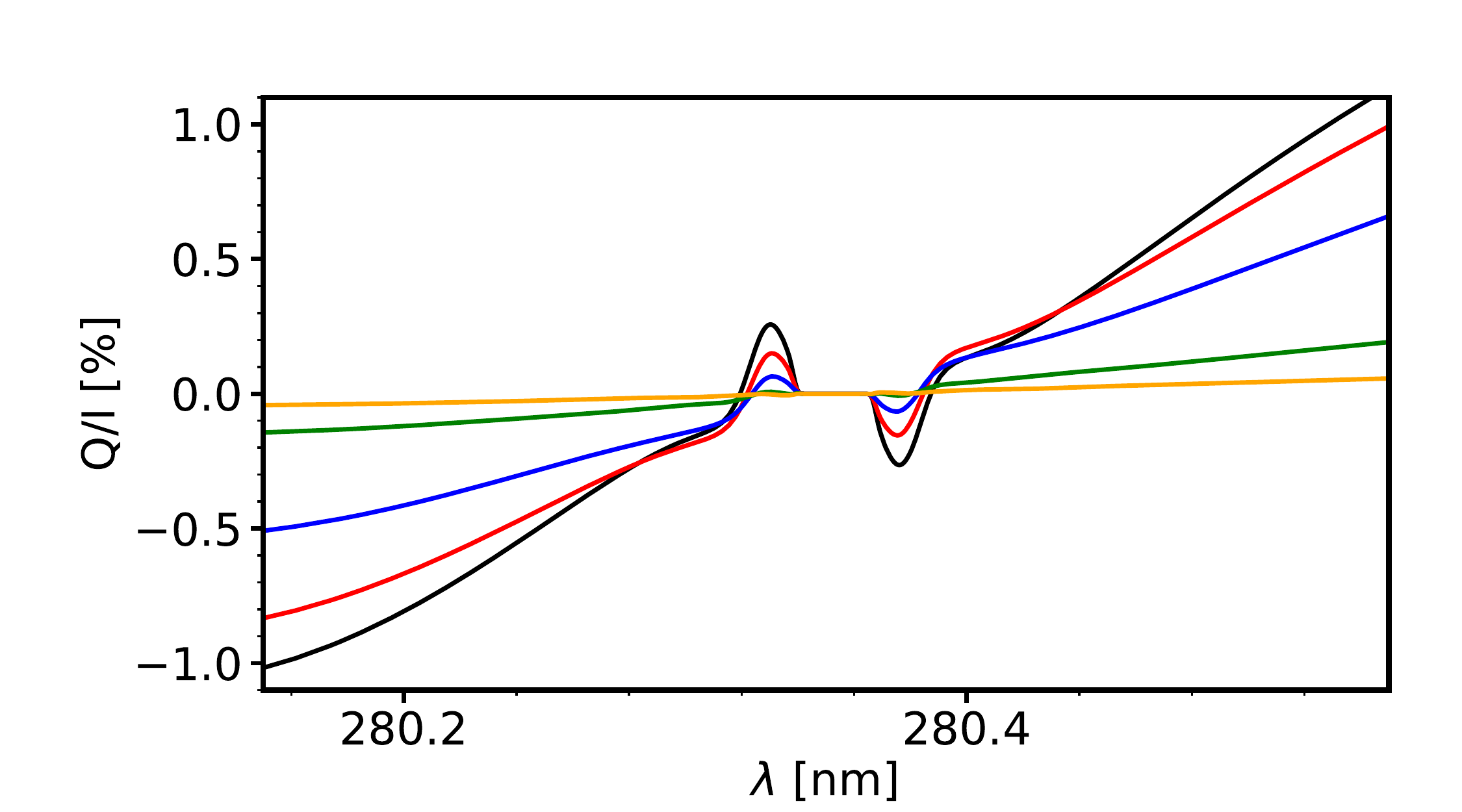}
\includegraphics[width=.4\hsize]{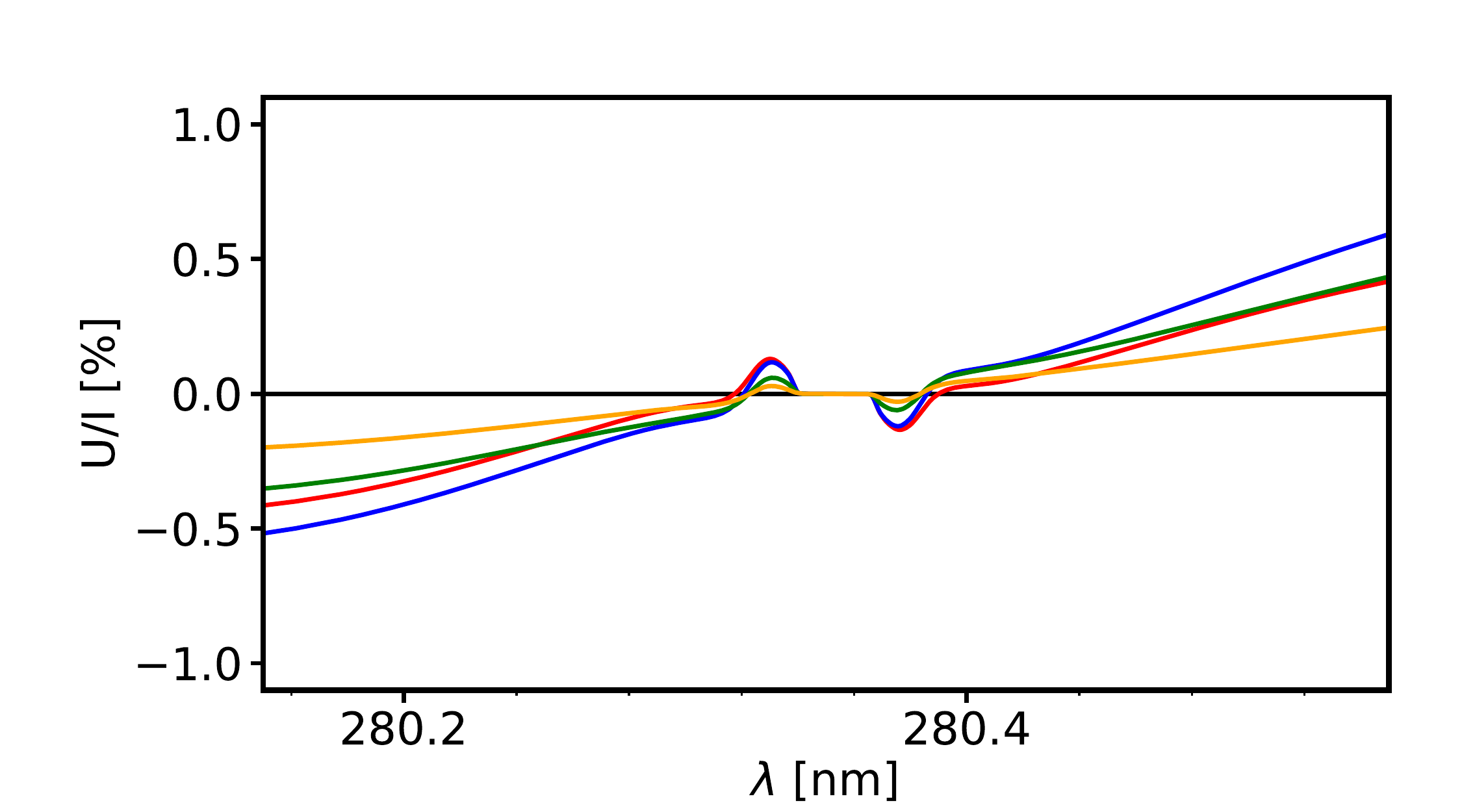} \\
\includegraphics[width=.4\hsize]{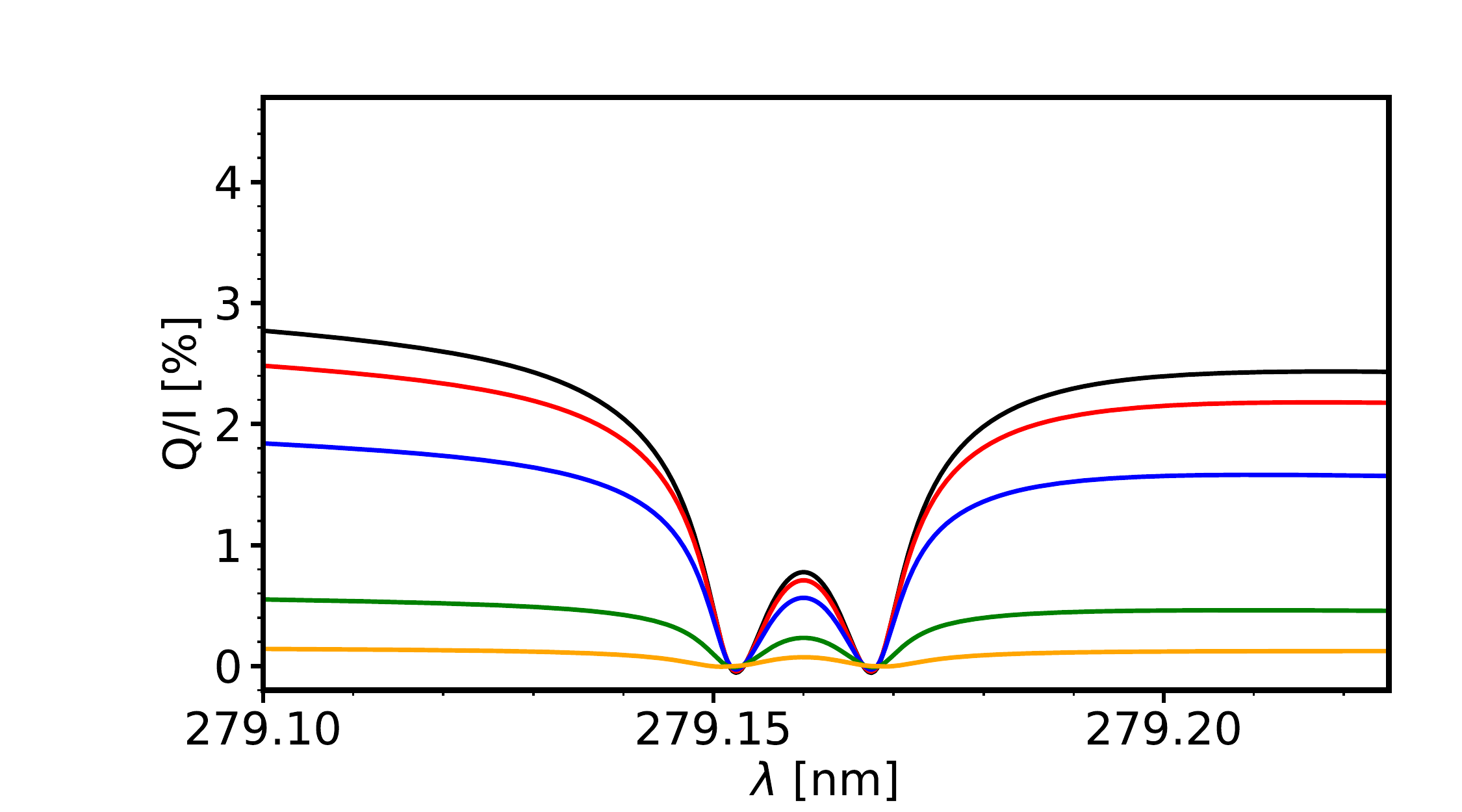}
\includegraphics[width=.4\hsize]{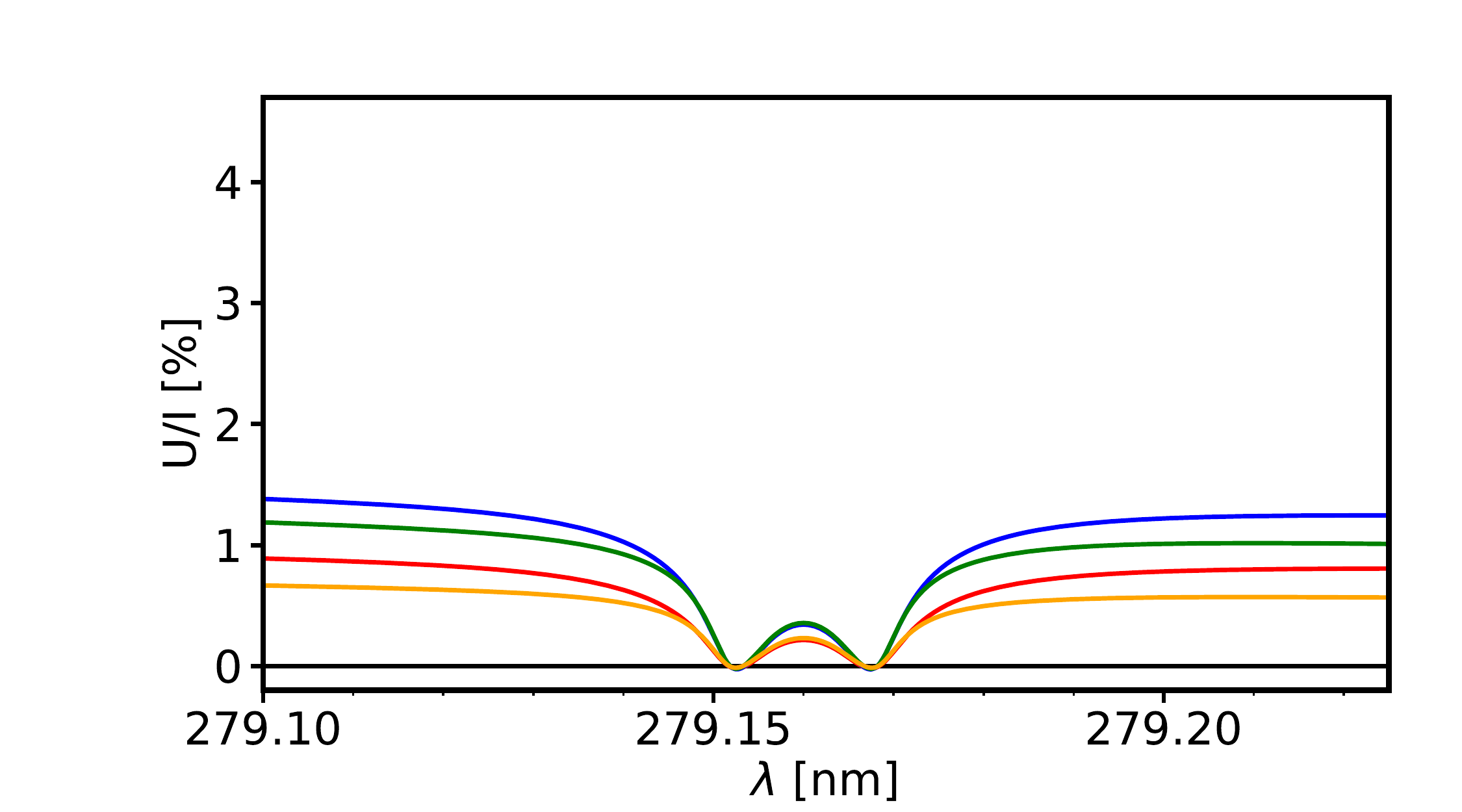} \\
\includegraphics[width=.4\hsize]{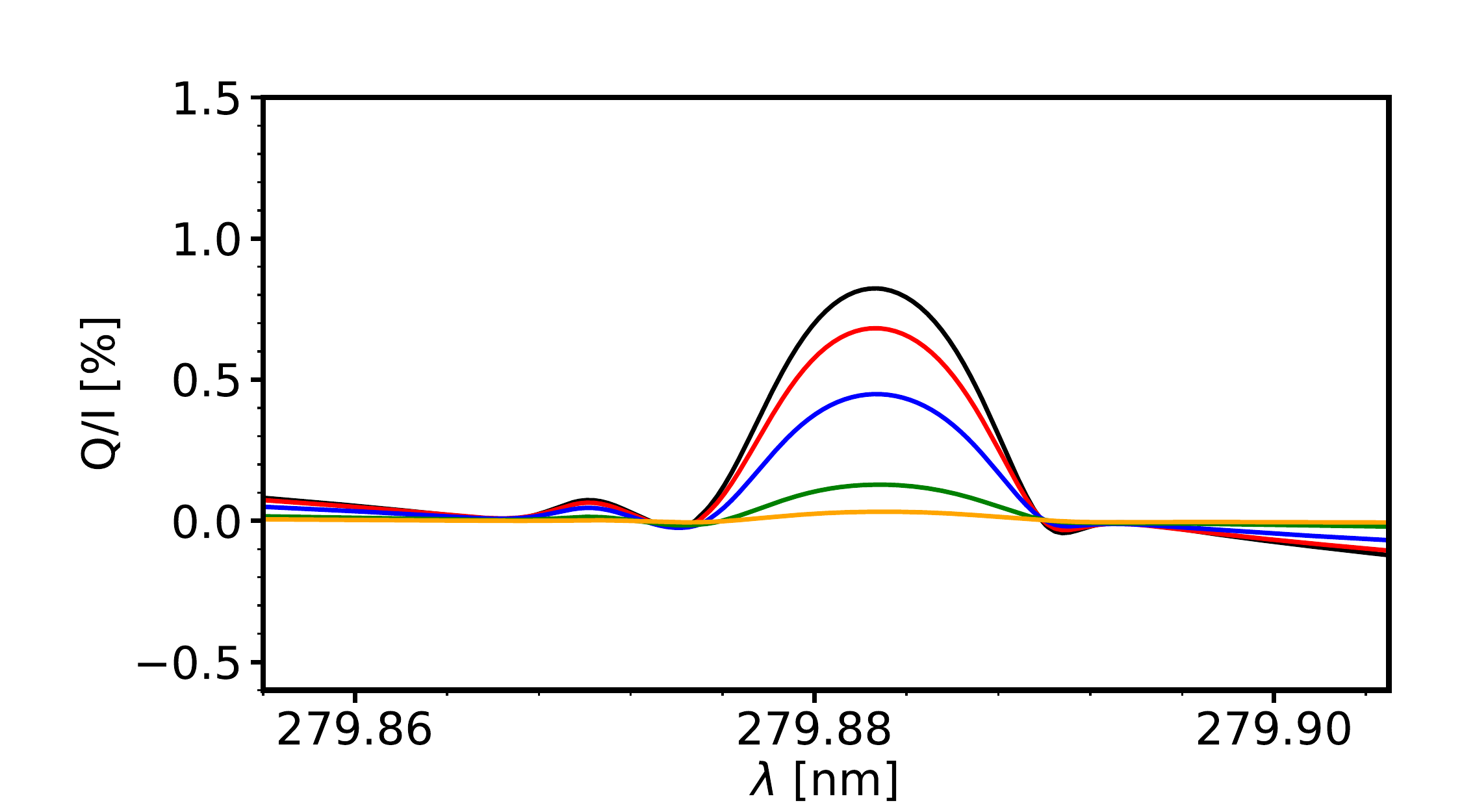}
\includegraphics[width=.4\hsize]{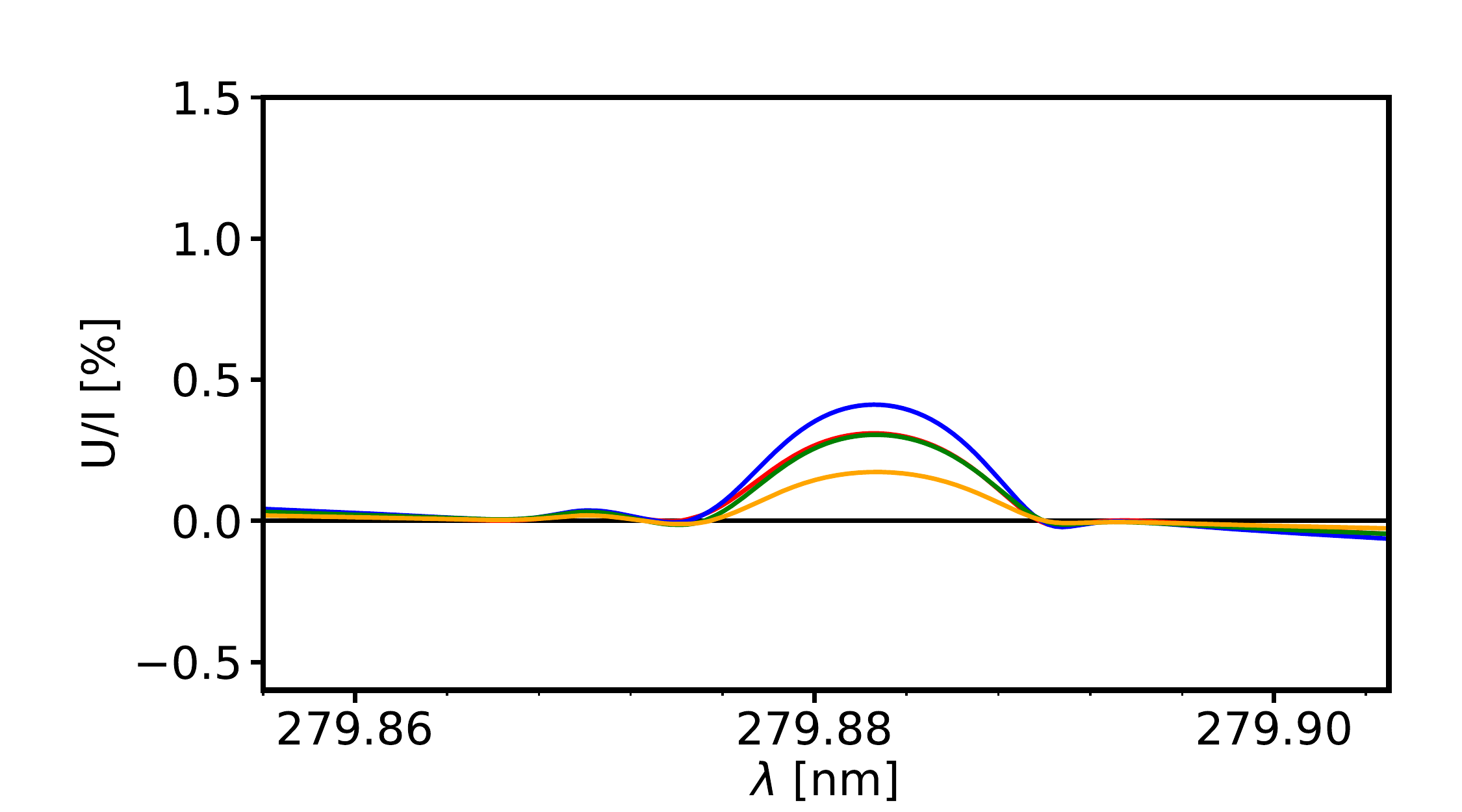}
\caption{Like in figure \ref{fig:MagnProfC}, but for the FAL-P model.}
\label{fig:MagnProfP}
\end{figure}

In this section we add a magnetic field of given strength and orientation
to the FAL-C and FAL-P atmospheric models in order to study the magnetic
sensitivity of the \ion{Mg}{2} h-k doublet and UV triplet.
Figures \ref{fig:MagnProfC} and \ref{fig:MagnProfP} show the emergent fractional
linear polarization along a LOS with $\mu = 0.1$ for a horizontal
magnetic field with strengths of 0, 10, 20, 50, 100, and 200\,G. For these
magnetic field strengths, we see two physical processes shaping the polarization
profiles. At the line center of the k and subordinate lines, the magnetic
field relaxes the coherence among the magnetic sublevels (Hanle effect) leading
to a depolarization of the signal with respect to the zero-field case.
Outside the atomic resonances, the depolarization of the $Q/I$ profile and the
corresponding appearance of a Stokes $U/I$ signal are due to M-O effects in
the line wings. These effects couple the $Q$ and $U$ Stokes parameters, and
transfer the broadband polarization, created by PRD in a magnetically split line,
from one Stokes parameter to the other. At the same time, the total linear
polarization $\sqrt{Q^2+U^2}$ is also reduced due to radiation transfer effects
(\citealt{delPinoetal2016,Alsinaetal2016}).

None of the levels of the h transition can carry atomic alignment and,
therefore, the line core does not show scattering polarization and is thus
unaffected by the Hanle effect. The transition $\rm s_{r_a}$ shows
a very small linear polarization signal, which is practically of
no diagnostic use. Regarding the M-O effects in the line wings, they
display a magnetic sensitivity similar to that of the k line (the broadband
pattern is due entirely to the combined action of PRD effects and quantum
interference in the upper term of the h-k doublet). Consequently,
the amplitude of the broadband polarization vanishes for magnetic fields in
the saturation regime of the k line (fields larger than $50$\,G in
Figs.~\ref{fig:MagnProfC} and \ref{fig:MagnProfP}).

\begin{figure}[htp!]
\centering
\includegraphics[width=.48\hsize]{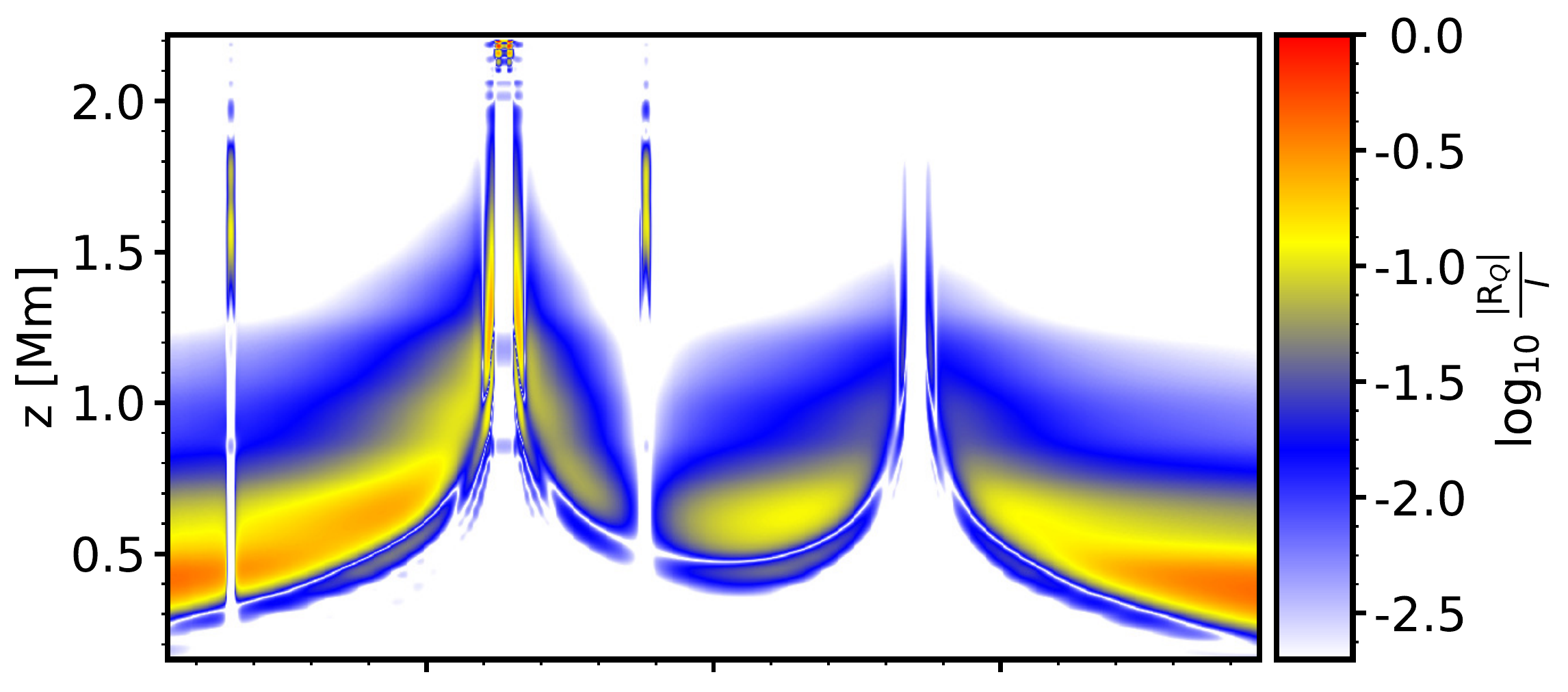}
\includegraphics[width=.48\hsize]{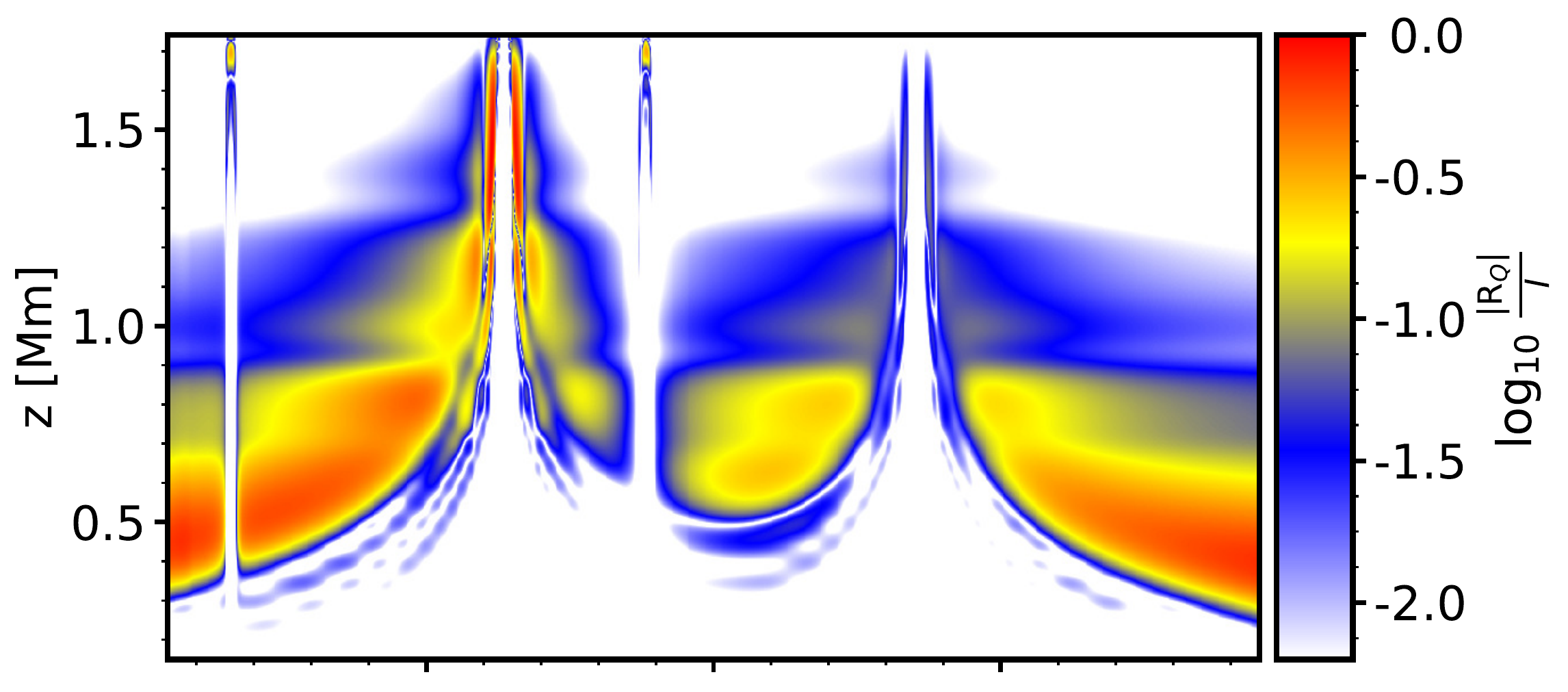} \\
\vspace{-5pt}
\includegraphics[width=.48\hsize]{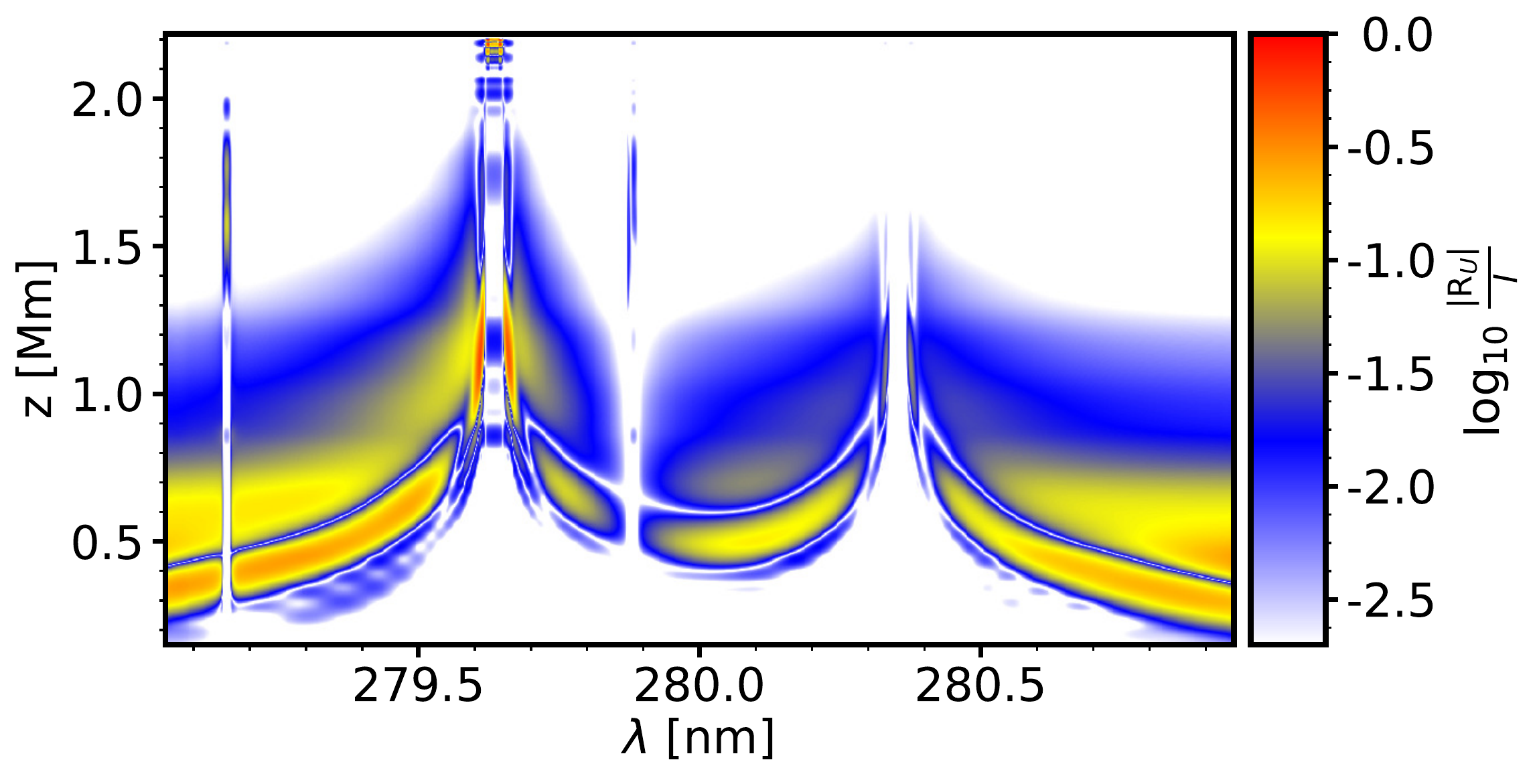}
\includegraphics[width=.48\hsize]{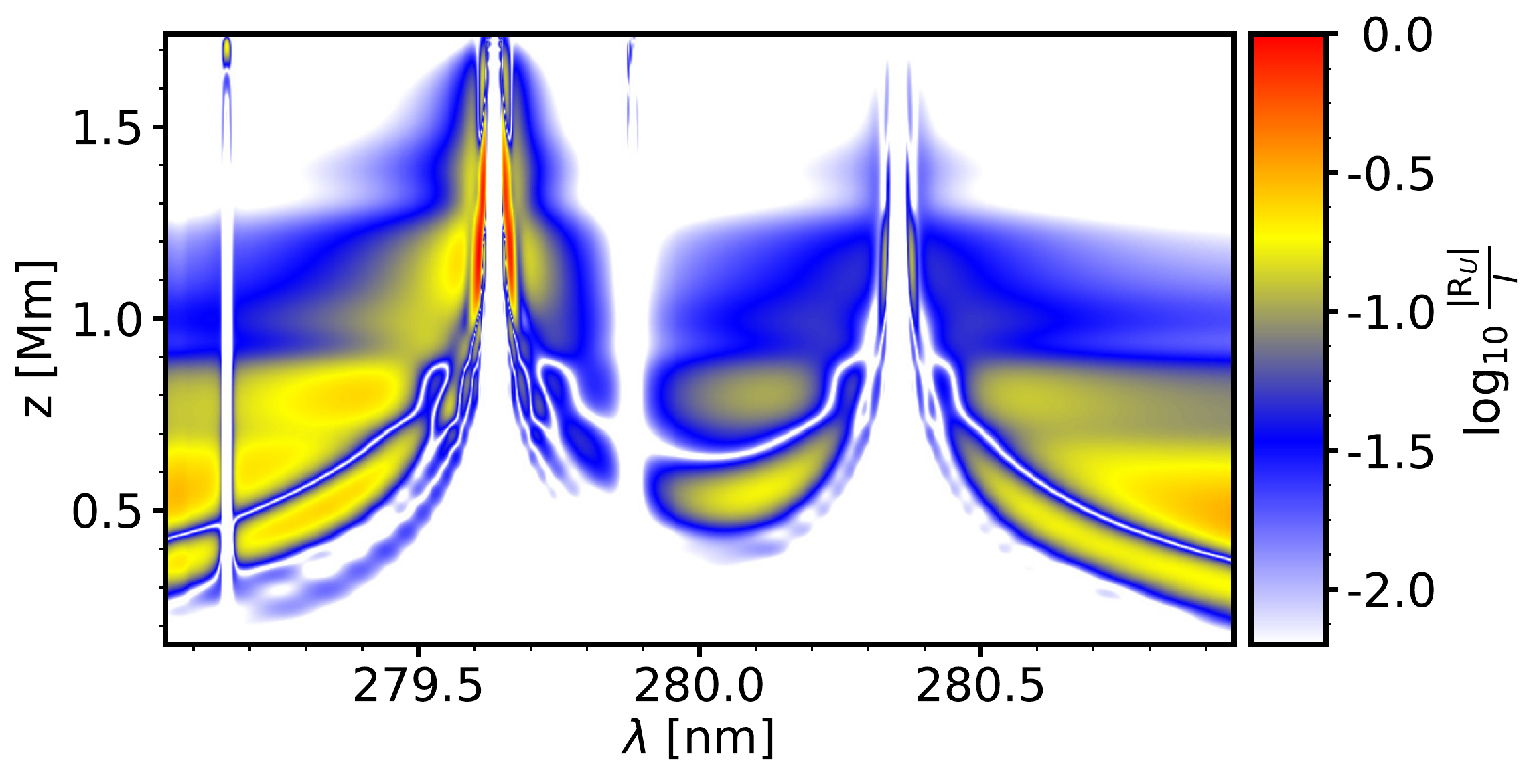}
\caption{Absolute value of the Stokes-Q (top row) and 
Stokes-U (bottom row) response functions for the \ion{Mg}{2} h-k doublet
and the UV triplet in the FAL-C (left) and FAL-P (right) models for a LOS
with $\mu=0.1$. The reference model has a horizontal magnetic field
of 20G directed towards the observer and it is perturbed with a horizontal
magnetic field of $1\,$G, pointing in the same direction.}
\label{fig:RespQU}
\end{figure}

The polarization signal at the core of the k line shows quantitatively a very
similar behavior in both the FAL-C (Fig.~\ref{fig:MagnProfC}) and
FAL-P (Fig.~\ref{fig:MagnProfP}) models. This happens because the core
of this transition forms in the top layers of the atmosphere, where the
two models are relatively similar. On the other hand, the polarization
of the broadband profile comes from an extended layer in the upper photosphere
(with its maximum response around $\sim 500$\,km above the visible surface),
while the cores of the subordinate lines form at chromospheric heights,
with the polarization signal being sensitive to the presence of magnetic
fields in a relatively extended layer above the line core height of
formation. Note that the FAL-P model shows a
significant compression in height in comparison to the FAL-C model, and thus the
subordinate and resonance lines form much closer in geometric scale in the
plage model than in the FAL-C average atmosphere. This can be more
easily seen in
Fig.~\ref{fig:RespQU}, which shows the response function of the Stokes-Q
and Stokes-$U$ parameters to perturbations in the magnetic field.
The response function tells us how
the emergent Stokes signal responds to a change in a physical parameter (e.g.,
the magnetic field) at each point within the model atmosphere and for each
wavelength (\citealt{LandiLandi1977}). Because the two atmospheric models are
different everywhere except in the lower photosphere and at the very top layers,
the quantitative behavior of the fractional linear polarization differs in the
whole spectral range except at the core of the k line.

The second column in Figs.~\ref{fig:MagnProfC} and \ref{fig:MagnProfP} shows the
fractional linear polarization profile $U/I$. Because in a 1D axially symmetric
atmosphere the $U/I$ polarization is zero in the absence of fields,
the introduction of a magnetic field inclined with respect to the local vertical
is necessary to generate Stokes $U$. At the same time, the field has
overall a depolarizing effect, because of the relaxation of the quantum coherence
of the atomic levels (Hanle effect). Therefore,
for magnetic fields below the critical Hanle field strength
($\sim20$, $60$, $10$, and $40\,$G for k, $\rm s_b$, $\rm s_{r_a}$,
and $\rm s_{r_b}$,
respectively), the $U/I$ polarization amplitude increases with the magnetic
strength (at the expense of the $Q/I$ polarization).
Above the critical Hanle field strength, the magnetic field reduces both
the $Q/I$ and $U/I$ polarization signals due to the Hanle effects. It is important
to emphasize that, in a 3D model atmosphere without axial-symmetry
constraint, a non-zero $U/I$ profile can emerge even in the absence
of magnetic fields (e.g., \citealt{StepanTrujillo2016}). In such a case,
the presence of a magnetic field typically results in a reduction of the
zero-field polarization signal (see also \citealt{delPinoetal2018}).

\begin{figure}[htp!]
\centering
\includegraphics[width=.32\hsize]{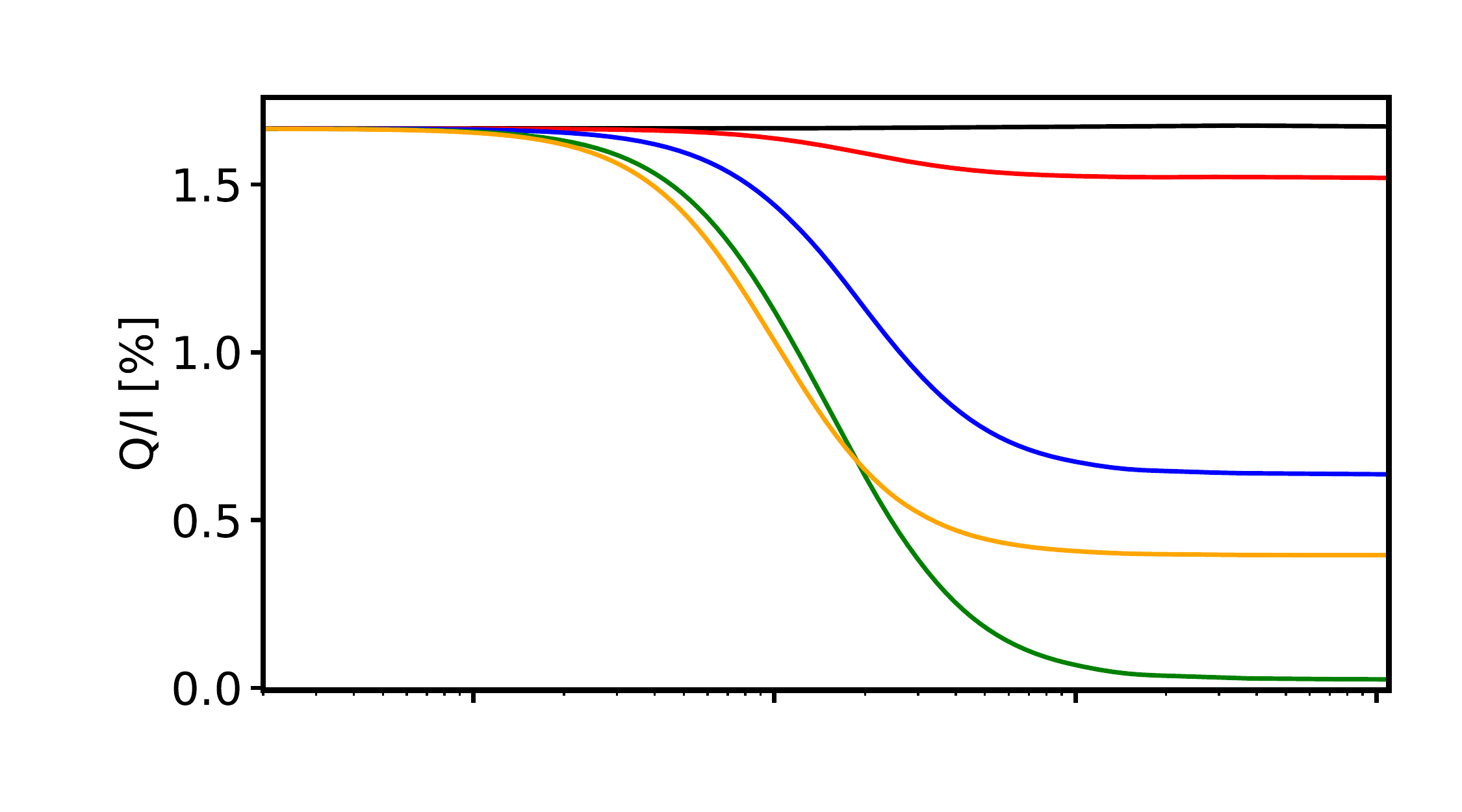} \hspace{-11pt}
\includegraphics[width=.32\hsize]{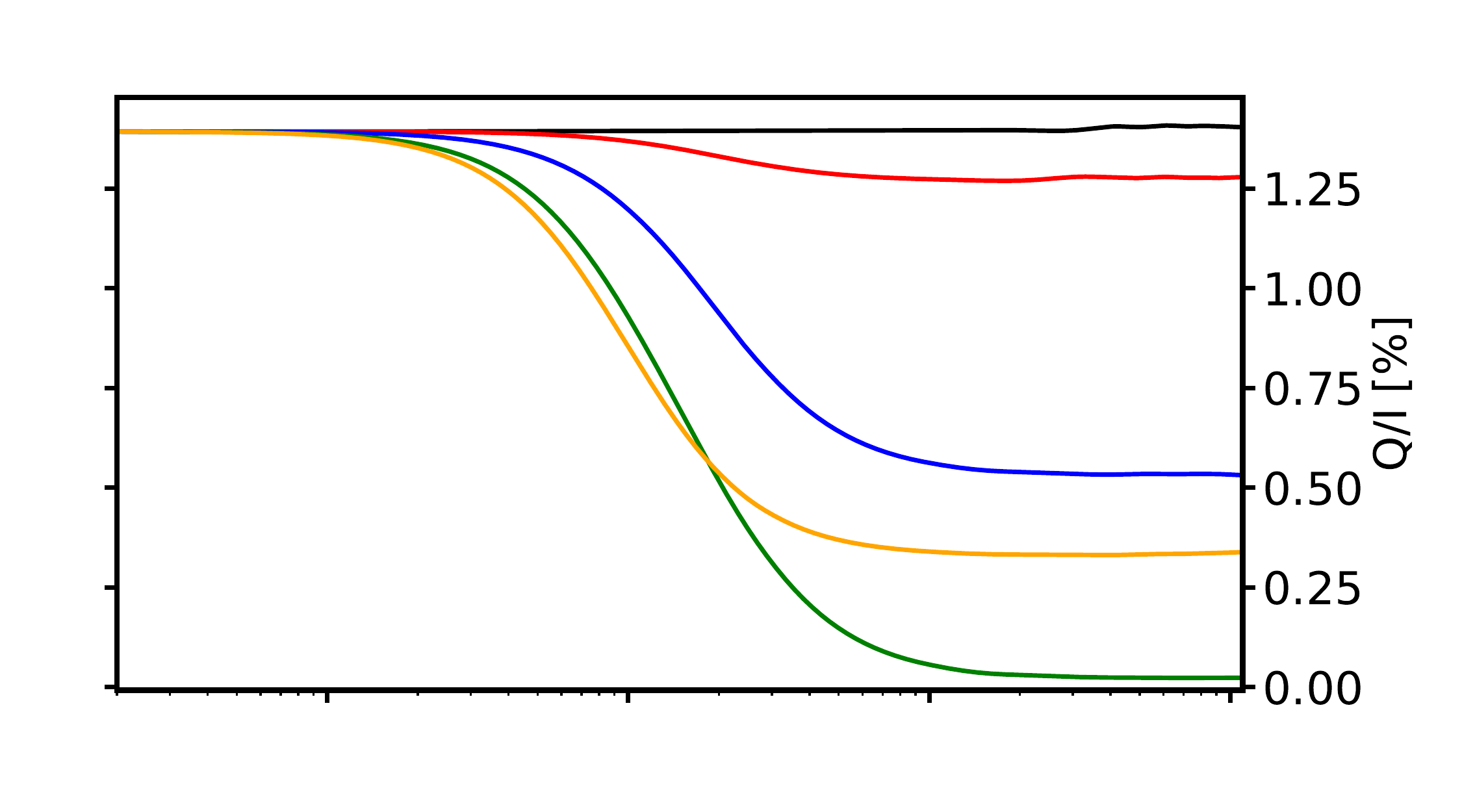} \\
\vspace{-13pt}
\includegraphics[width=.32\hsize]{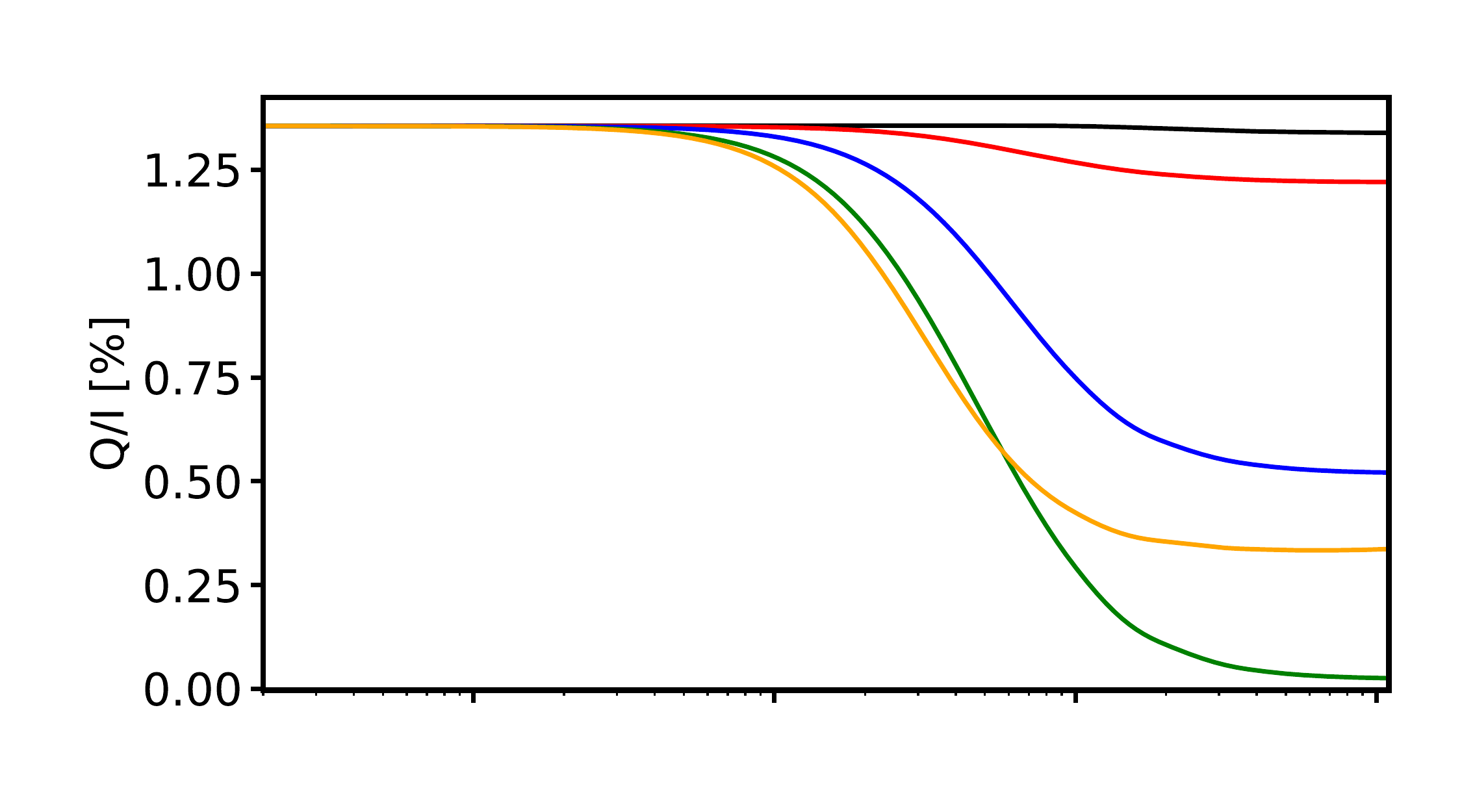} \hspace{-11pt}
\includegraphics[width=.32\hsize]{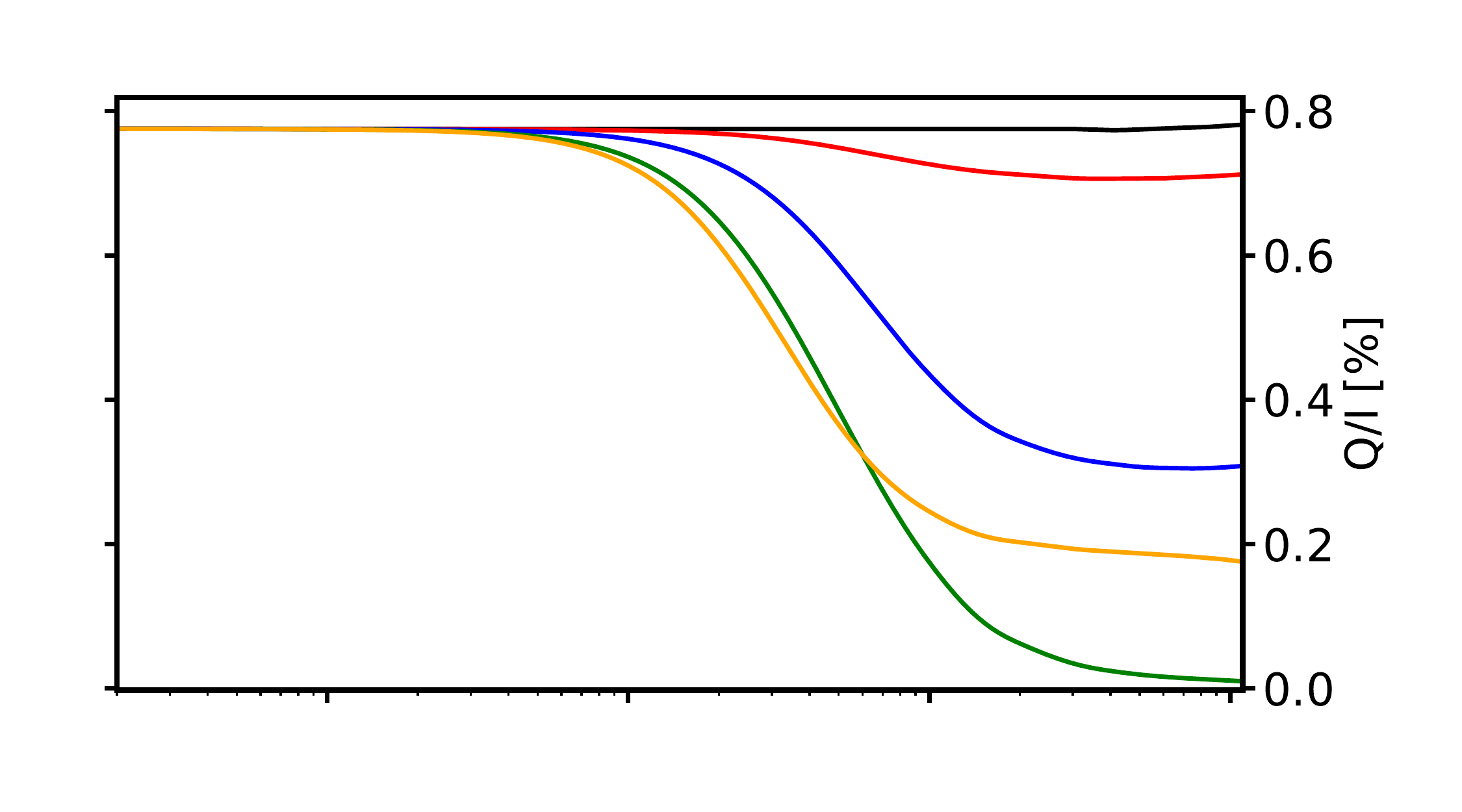} \\
\vspace{-13pt}
\includegraphics[width=.32\hsize]{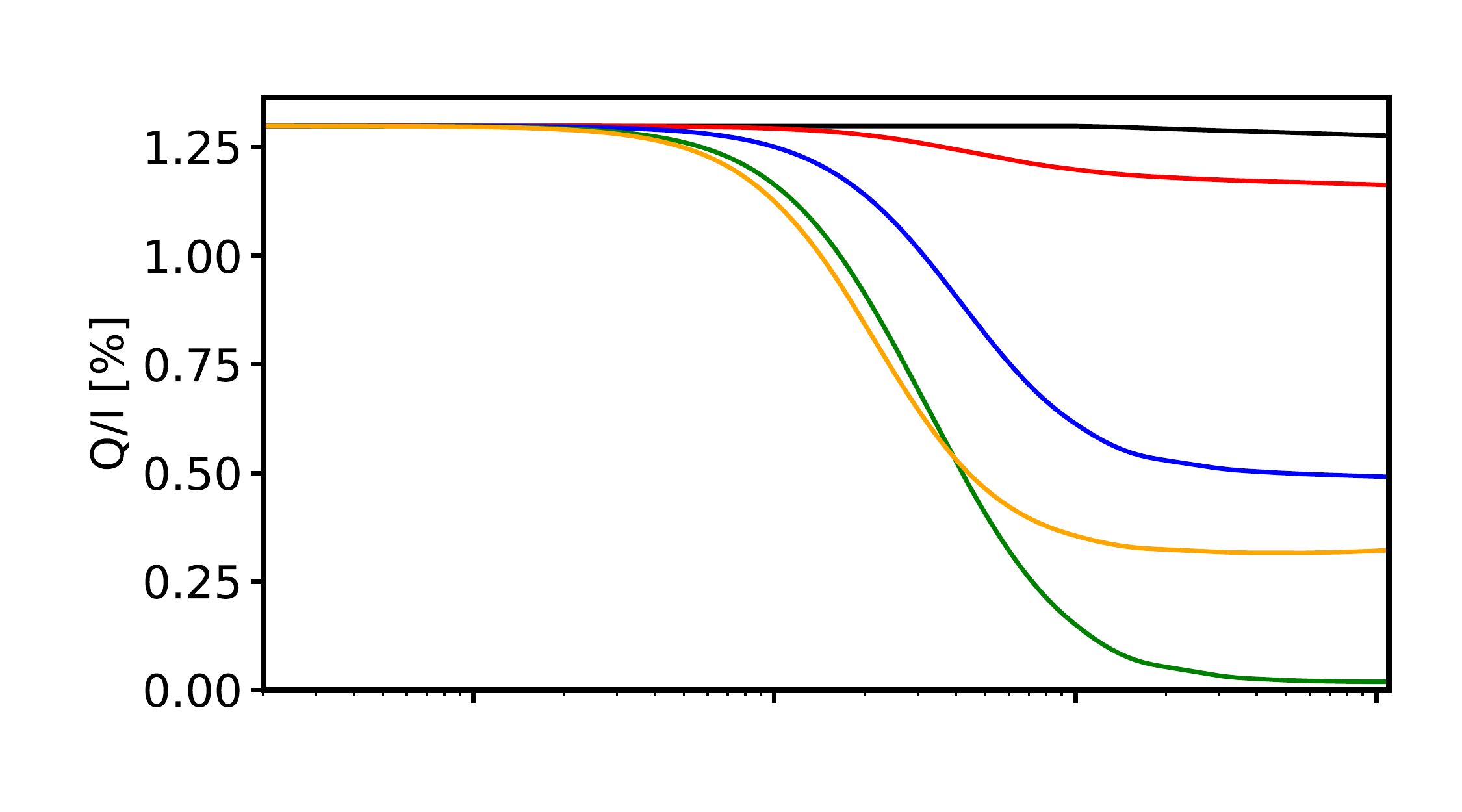} \hspace{-11pt}
\includegraphics[width=.32\hsize]{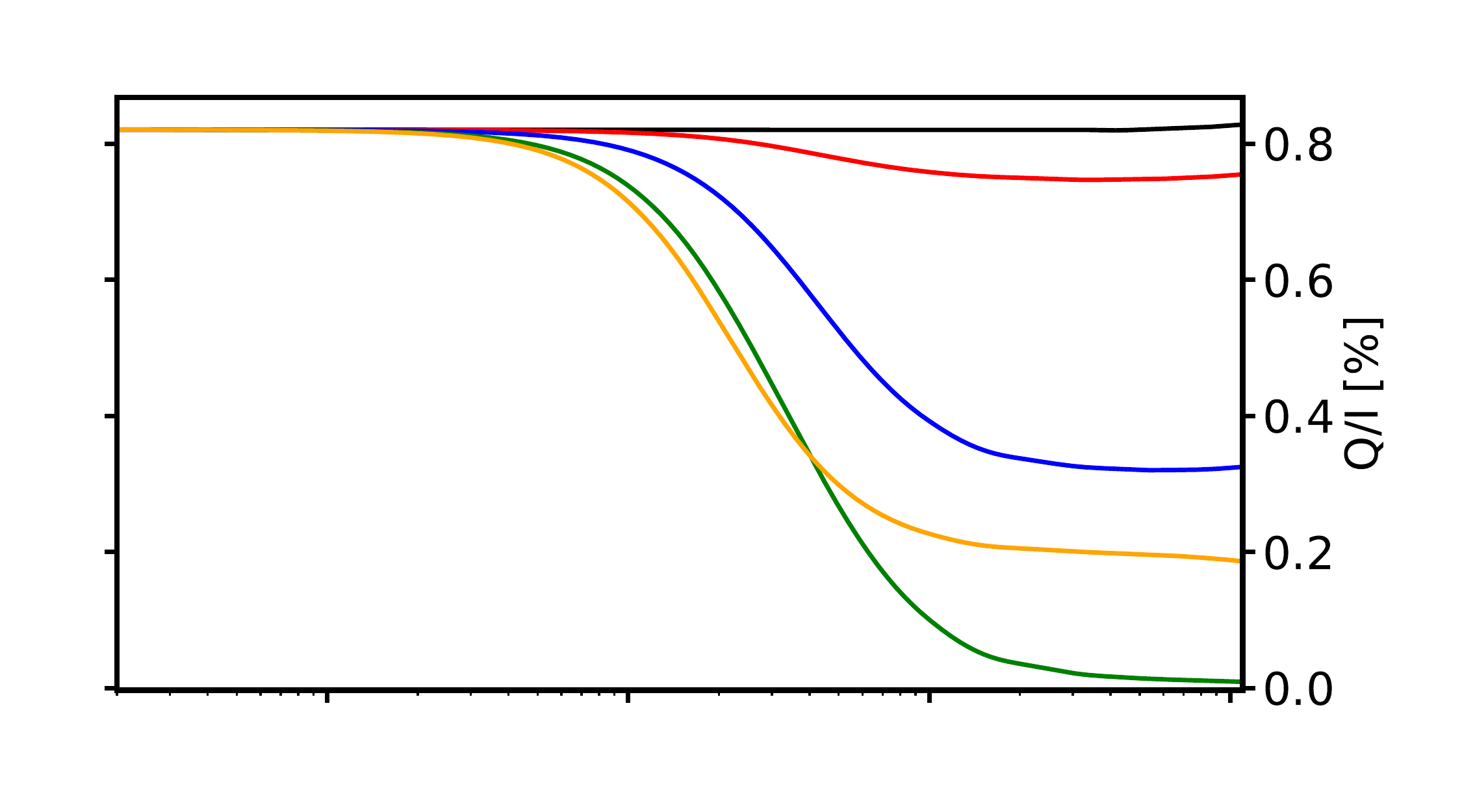} \\
\vspace{-13pt}
\includegraphics[width=.32\hsize]{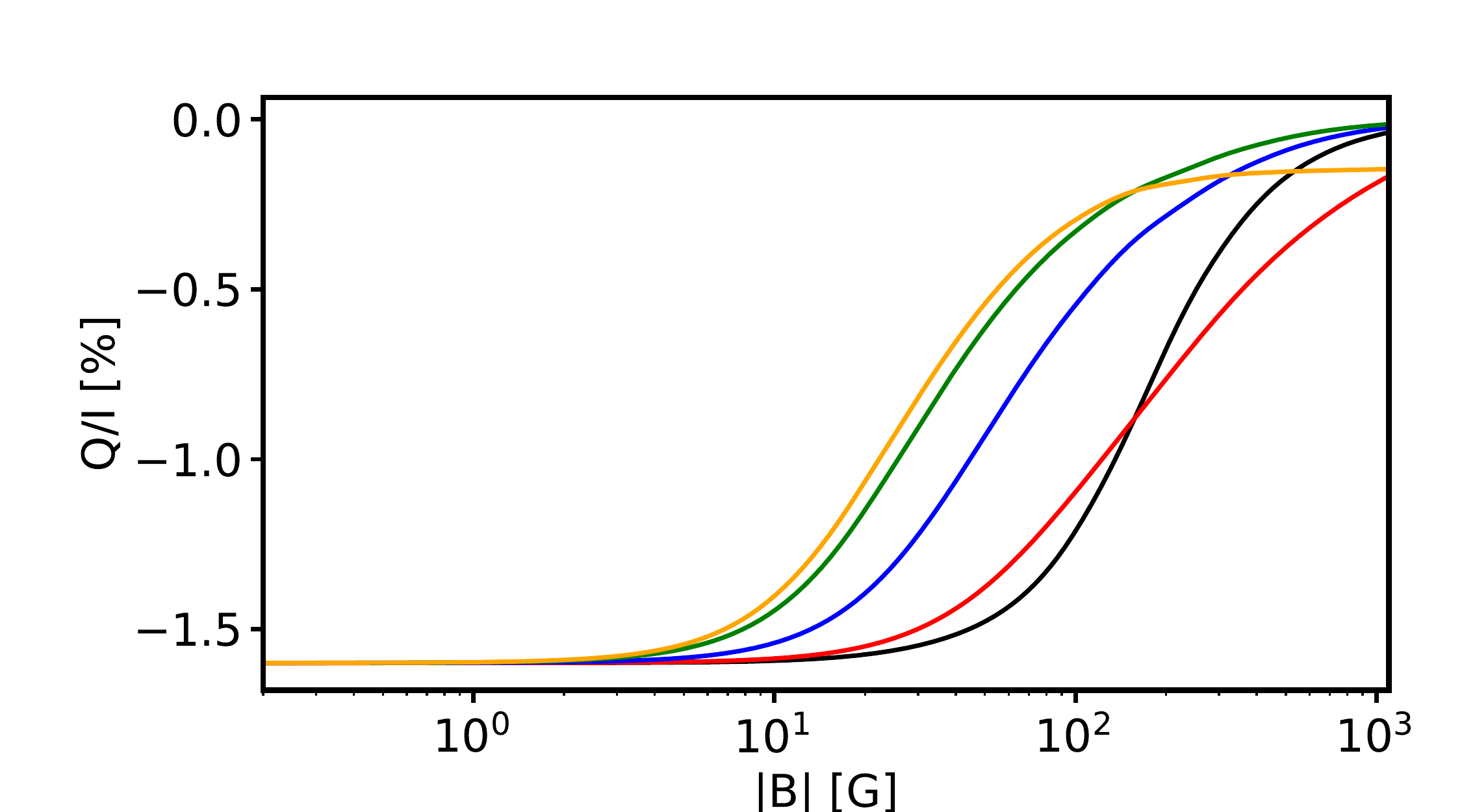} \hspace{-11pt}
\includegraphics[width=.32\hsize]{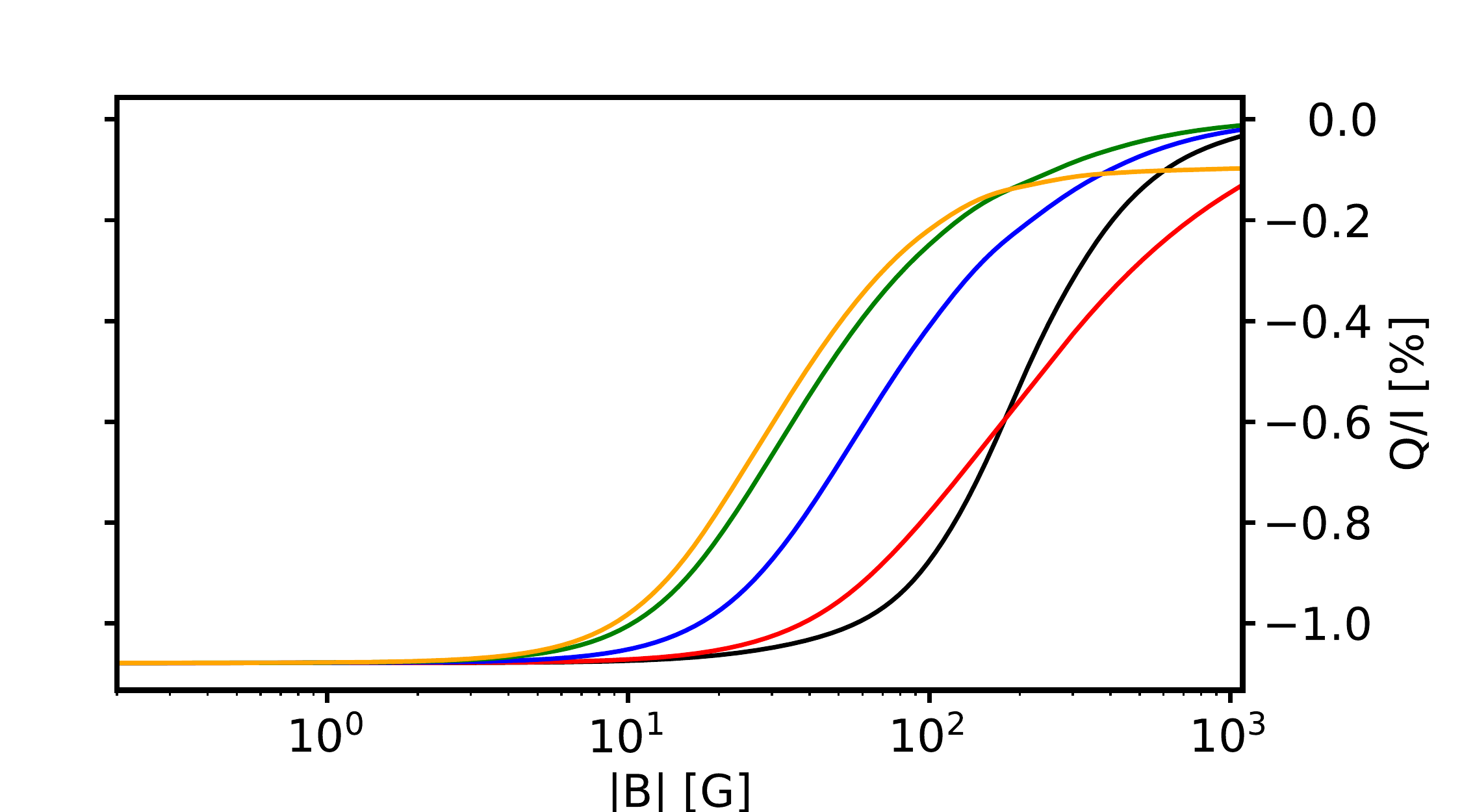}
\caption{Variation of the fractional linear polarization $Q/I$ with the magnetic
field strength for a homogeneous magnetic field with a fixed inclination,
averaged over all possible azimuths of the magnetic field, for a LOS with
$\mu=0.1$. The left (right) column shows calculations using the FAL-C (FAL-P) model.
From top to bottom, each row shows the polarization at the center of the k,
$\rm s_b$, $\rm s_{r_b}$ lines, and at a wavelength ($280.1\,$nm) close to the
minimum of the broadband pattern between h and k. The colors correspond to
different inclinations of the magnetic field: $0^\circ$ (black), $10^\circ$ (red),
$30^\circ$ (blue), $60^\circ$ (green), and $90^\circ$ (orange).}
\label{fig:MagnSensitQ}
\end{figure}

Figure \ref{fig:MagnSensitQ} shows the $Q/I$ polarization signal as a function
of magnetic strength, for a homogeneous magnetic field with different
inclinations, and after integrating the signal over all possible
magnetic field azimuths. This mimics the behavior of a ``canopy'' field
taking all possible azimuth orientations within the spatial resolution element.
From top to bottom, the various
panels give the polarization at the line center of the \ion{Mg}{2} transitions
(with the exception of h, which is unpolarized), and at the wavelength
$280.1\,$nm,
in the continuum between h and k, which shows the most negative polarization.
The first conclusion we can draw is that the behavior of the
polarization signal
with the magnetic field strength and inclination is qualitatively the same in both
the FAL-C (left panels) and FAL-P (right panels) models, with the only
difference
that the polarization emerging from the FAL-P model is consistently smaller.
Therefore, we expect smaller polarization signals in plage regions, even before
taking into account that the stronger magnetic fields of plages are already
more effective in depolarizing the scattered radiation than in the quiet Sun.
The only wavelength where the two models do not show the
same qualitative behavior is for the $\rm s_{r_a}$ line center. However,
its polarization signal is so small that this line is of no interest for magnetic
field diagnostics and thus we do not include a figure for this wavelength.

As it is expected from the theory, because the line center is dominated by
scattering polarization and the Hanle effect, the sensitivity to vertical (black
curves) or quasi-vertical ($10^\circ$ inclination, red curves) magnetic fields
is quite small, while the depolarization is maximum for inclinations close to
the Van Vleck angle ($\sim 54.7^\circ$; green curves). On the other hand, the
polarization outside the line cores (e.g., last row in Fig.~\ref{fig:MagnSensitQ})
is due to PRD effects, quantum interference, and to the M-O effects caused
by the Zeeman splitting. Therefore, the behavior of the polarization signal with
the magnetic field strength for different inclinations is also qualitatively
different than for the line centers.

\begin{figure}[htp!]
\centering
\includegraphics[width=.32\hsize]{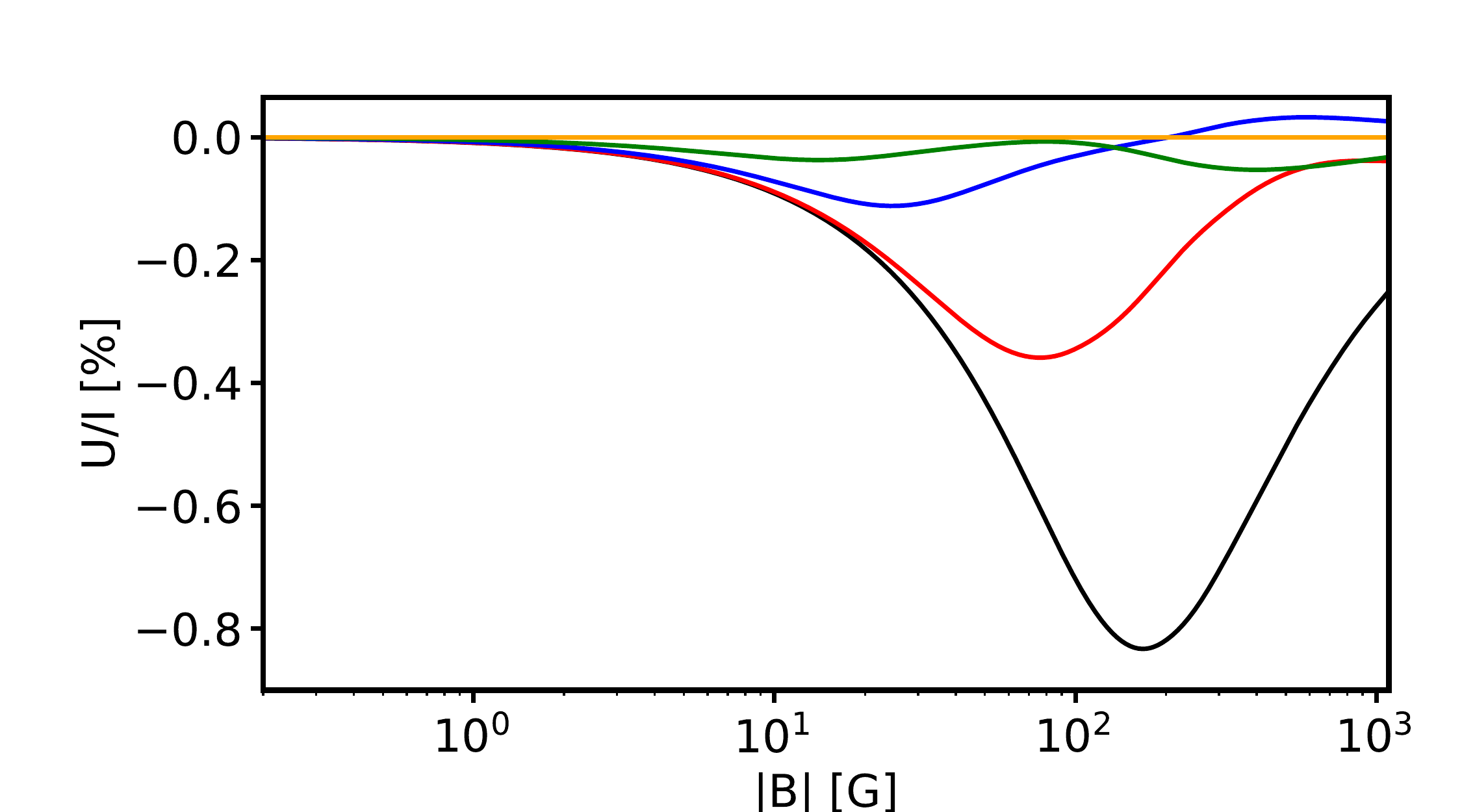} \hspace{-11pt}
\includegraphics[width=.32\hsize]{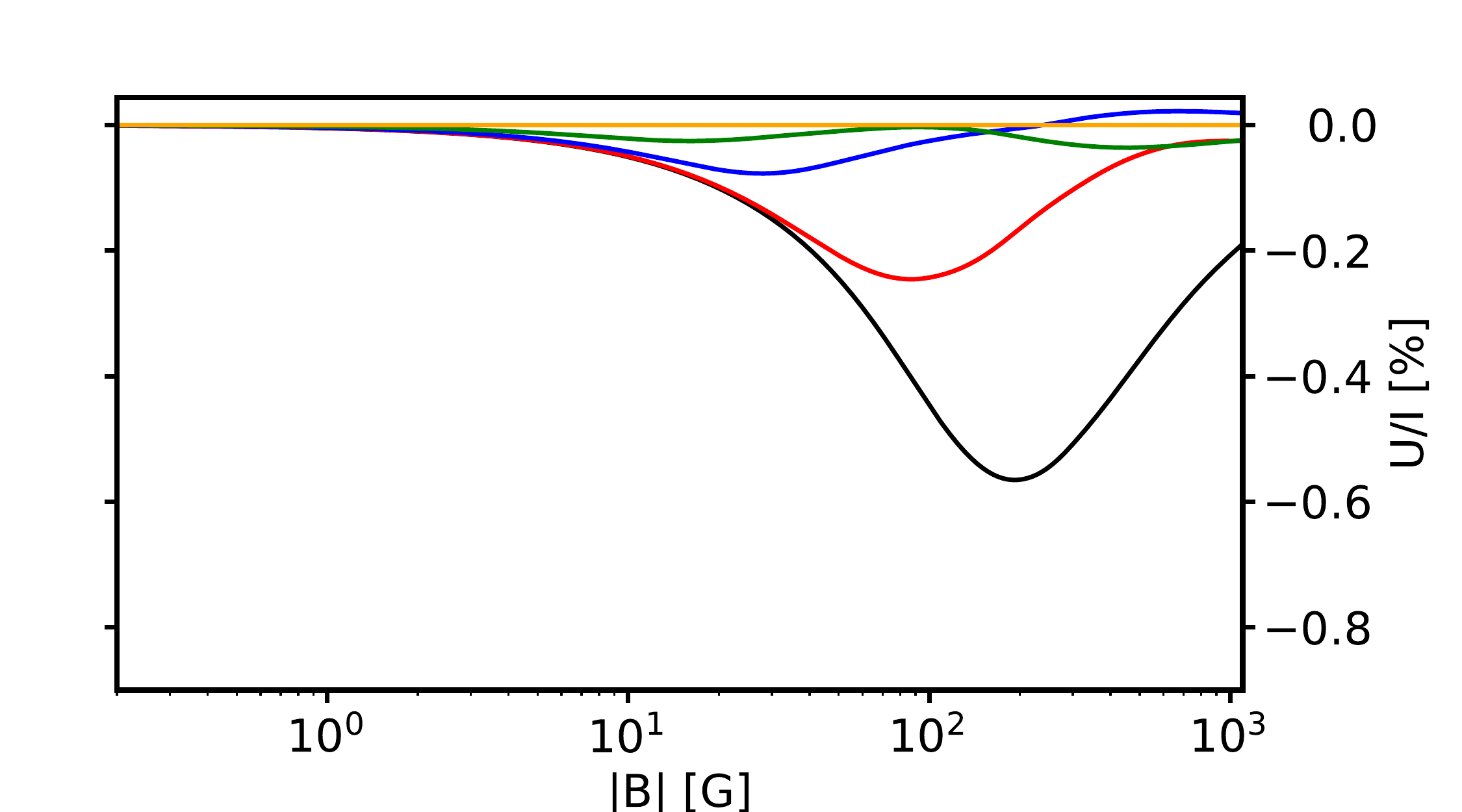}
\caption{Variation of the fractional linear polarization $U/I$ with the magnetic
field strength for a homogeneous magnetic field with a fixed inclination,
averaged over all possible azimuths, for a LOS with $\mu=0.1$, at
a wavelength ($280.1\,$nm) close to the minimum of the broadband pattern
between h and k. The left (right) column shows calculations using the FAL-C (FAL-P)
model. The colors correspond to different inclinations of the magnetic field:
$0$ (black), $10^\circ$ (red), $30^\circ$ (blue), $60^\circ$ (green), and
$90^\circ$ (orange).}
\label{fig:MagnSensitU}
\vspace{3.6ex}
\end{figure}

Because we averaged over all possible magnetic field azimuths, for
symmetry reasons the $U/I$ signals in the line cores due to the Hanle effect
cancel out. However, this is not the case for the $U/I$ broadband polarization.
Fig.~\ref{fig:MagnSensitU} shows the fractional linear polarization $U/I$ at the
continuum wavelength $280.1\,$nm for the same models of Fig.~\ref{fig:MagnSensitQ}.
Notably, vertical fields generate the largest $U/I$ signals. Initially, increasing
the magnetic field strength results in a larger $U/I$ polarization signal, as more
$Q/I$ polarization is being ``rotated'' into $U/I$. This trend of $U/I$ polarization
has a turning point (dependent on the field inclination), as a consequence of the
overall depolarization induced by the M-O effects in an optically thick atmosphere
(see also \citealt{Alsinaetal2018}).

Because the broadband pattern of polarization depends on both the magnetic field
\emph{and} the 3D thermal structure of the solar atmosphere, it can be difficult
to unequivocally demonstrate the manifestation of M-O effects through an
observing sequence of only a few minutes. Nevertheless, we know that a very strong
magnetic field should be able to completely destroy this broadband polarization.
Therefore, spectropolarimetric observations where the slit stretches over both the
quiet Sun and a strong active region would be helpful to establish the actual role
of M-O effects: if the broadband polarization pattern were visible in the quiet-Sun
region but disappeared in the strong-field region, this could only be due to
M-O effects, since it is the only mechanism that can completely destroy linear
polarization.\footnote{Other physical mechanisms, such as a significant increase
of collisional rates, can also completely destroy linear polarization,
but there is no physical reason why these rates should be much larger in an active
region than in the surrounding quiet-Sun atmosphere.}

\begin{figure}[htp!]
\centering
\includegraphics[width=.47\hsize]{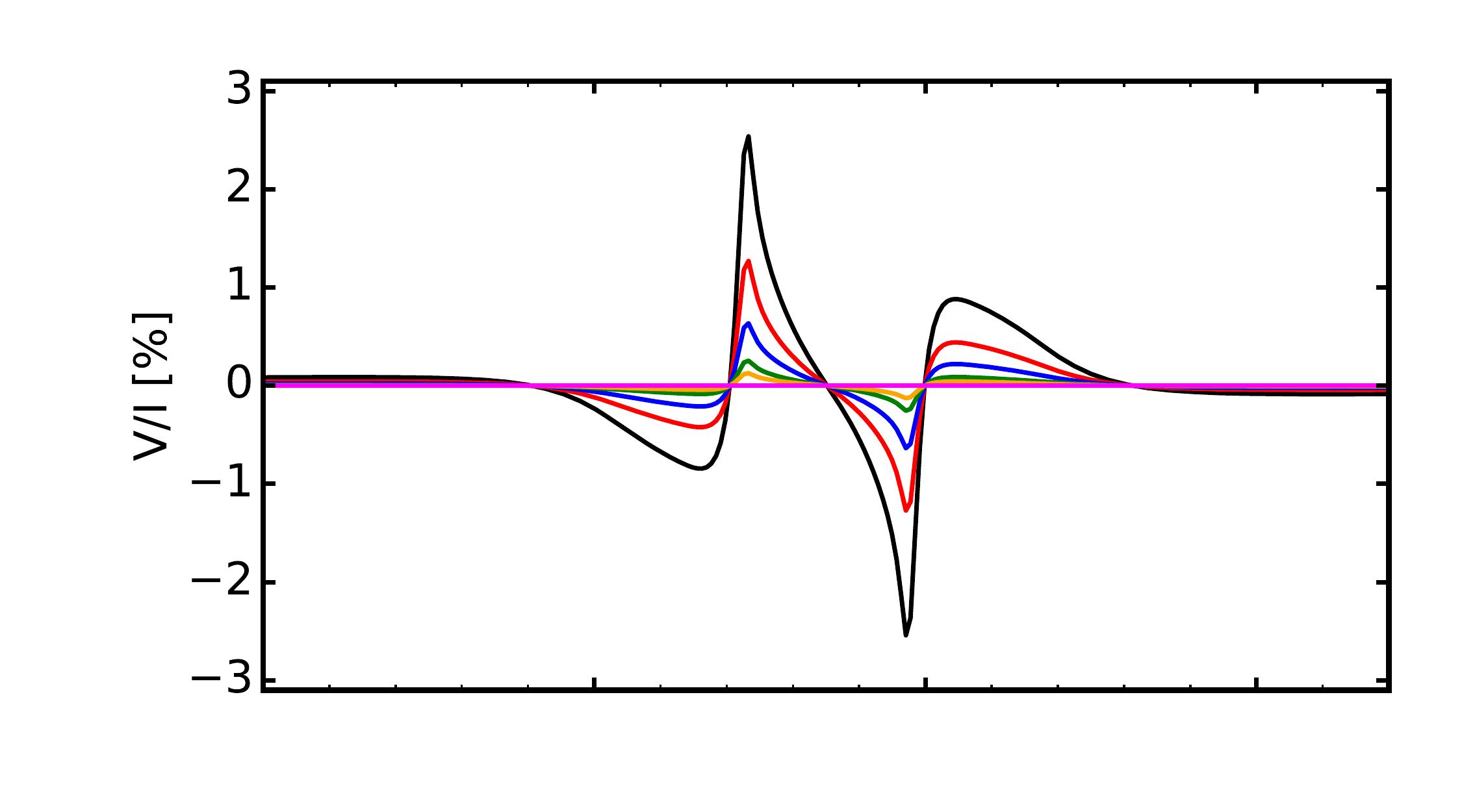}
\includegraphics[width=.47\hsize]{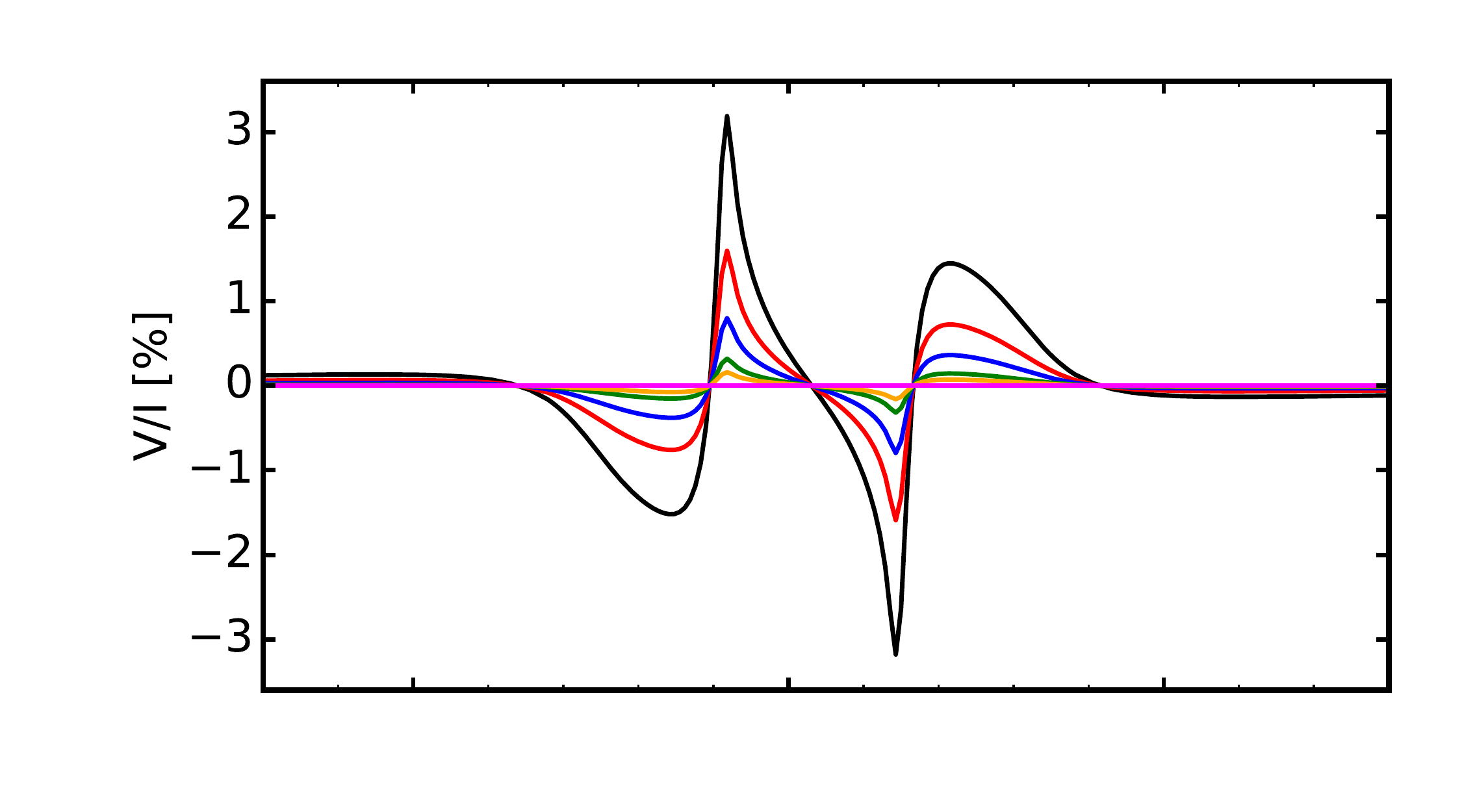} \\
\vspace{-20pt}
\includegraphics[width=.47\hsize]{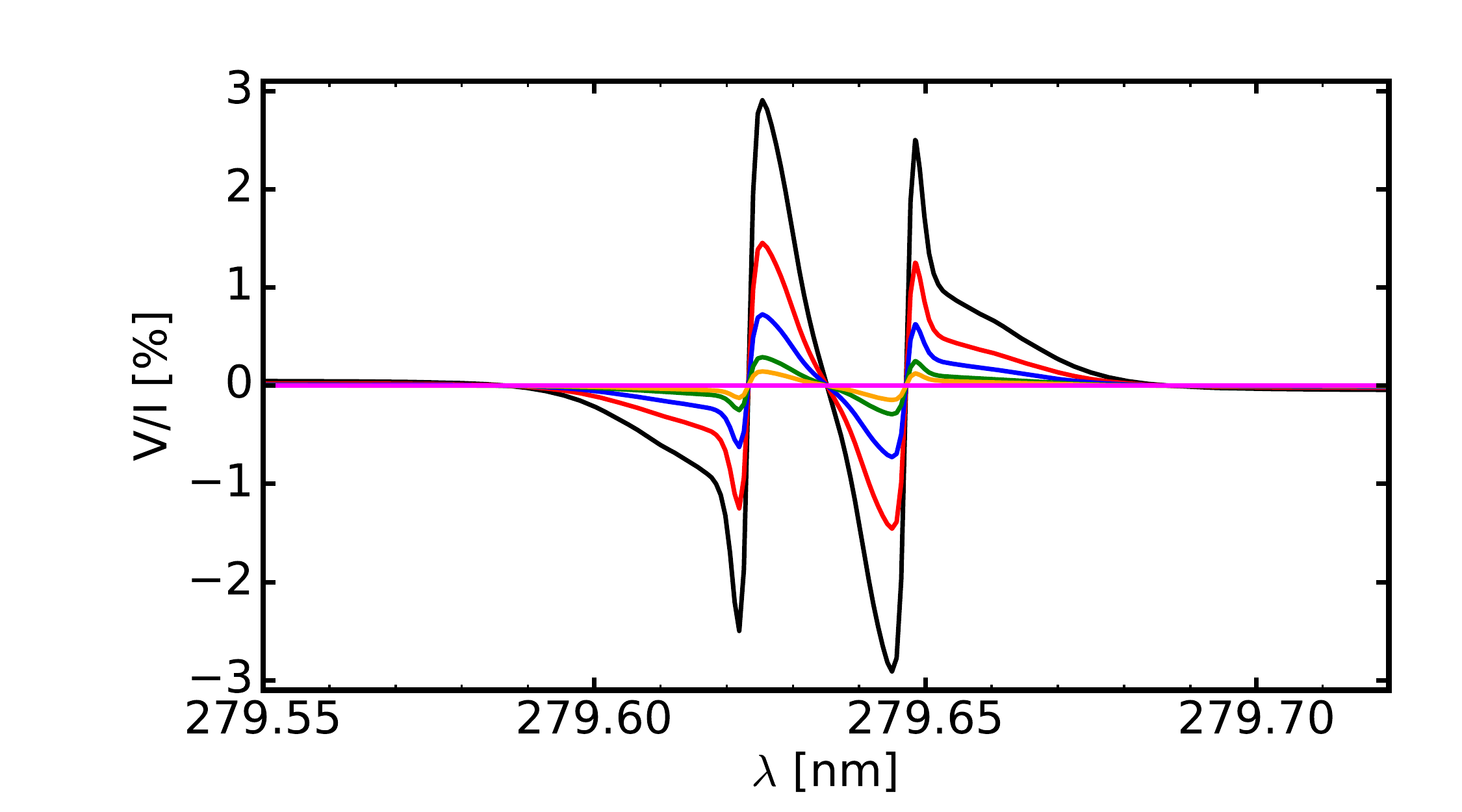} 
\includegraphics[width=.47\hsize]{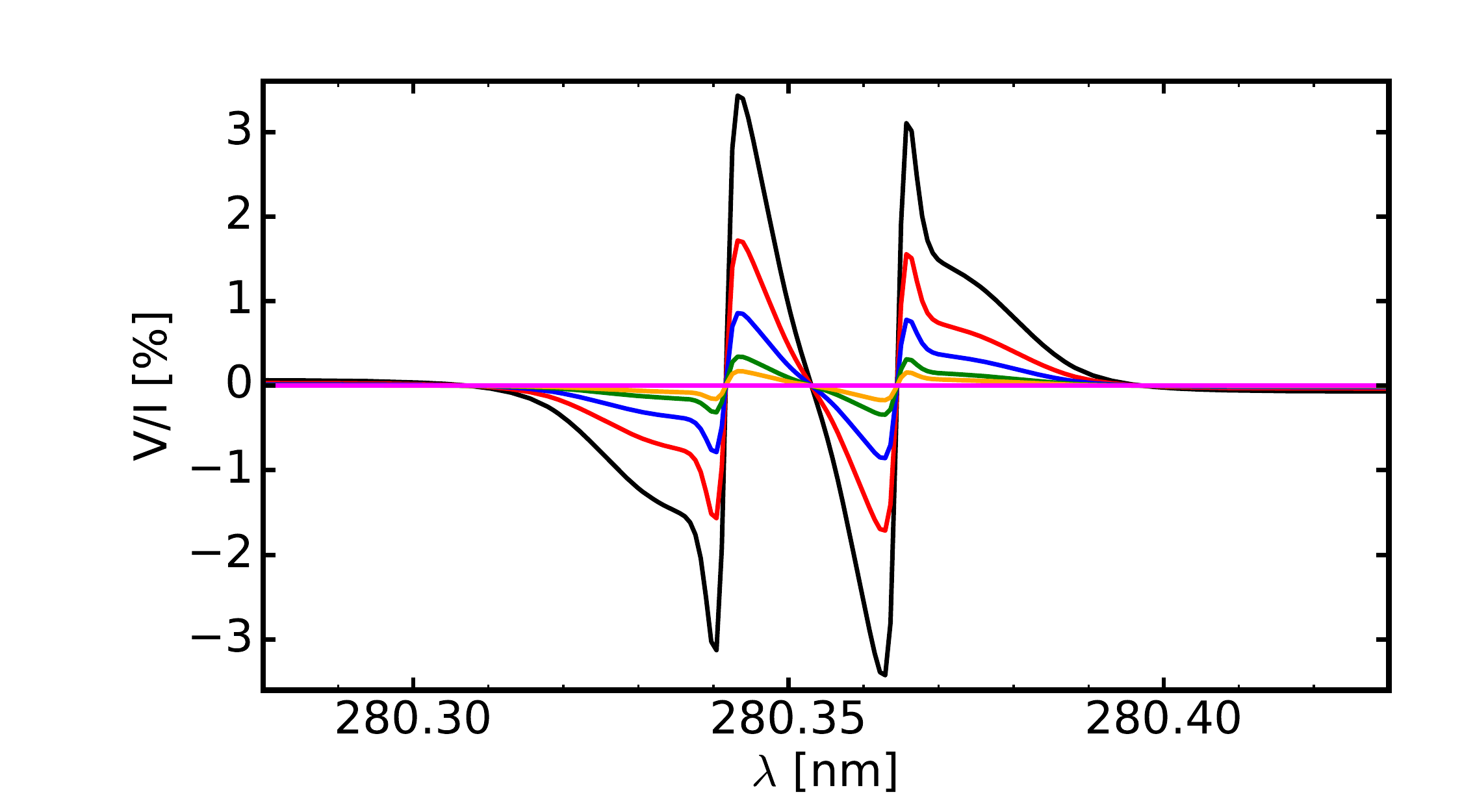}
\caption{Fractional circular polarization profile $V/I$ for the \ion{Mg}{2}
h (right column) and k (left column) doublet. The top (bottom) row shows
calculations using the FAL-C (FAL-P) model, for a LOS with
$\mu=1.0$. The color of the curves indicates the strength of the vertical
magnetic field: $0\,$G (magenta), $10\,$G (orange), $20\,$G (green),
$50\,$G (blue), $100\,$G (red), and $200\,$G (black).}
\label{fig:MagnProfVhk}
\end{figure}

\begin{figure}[htp!]
\centering
\includegraphics[width=.47\hsize]{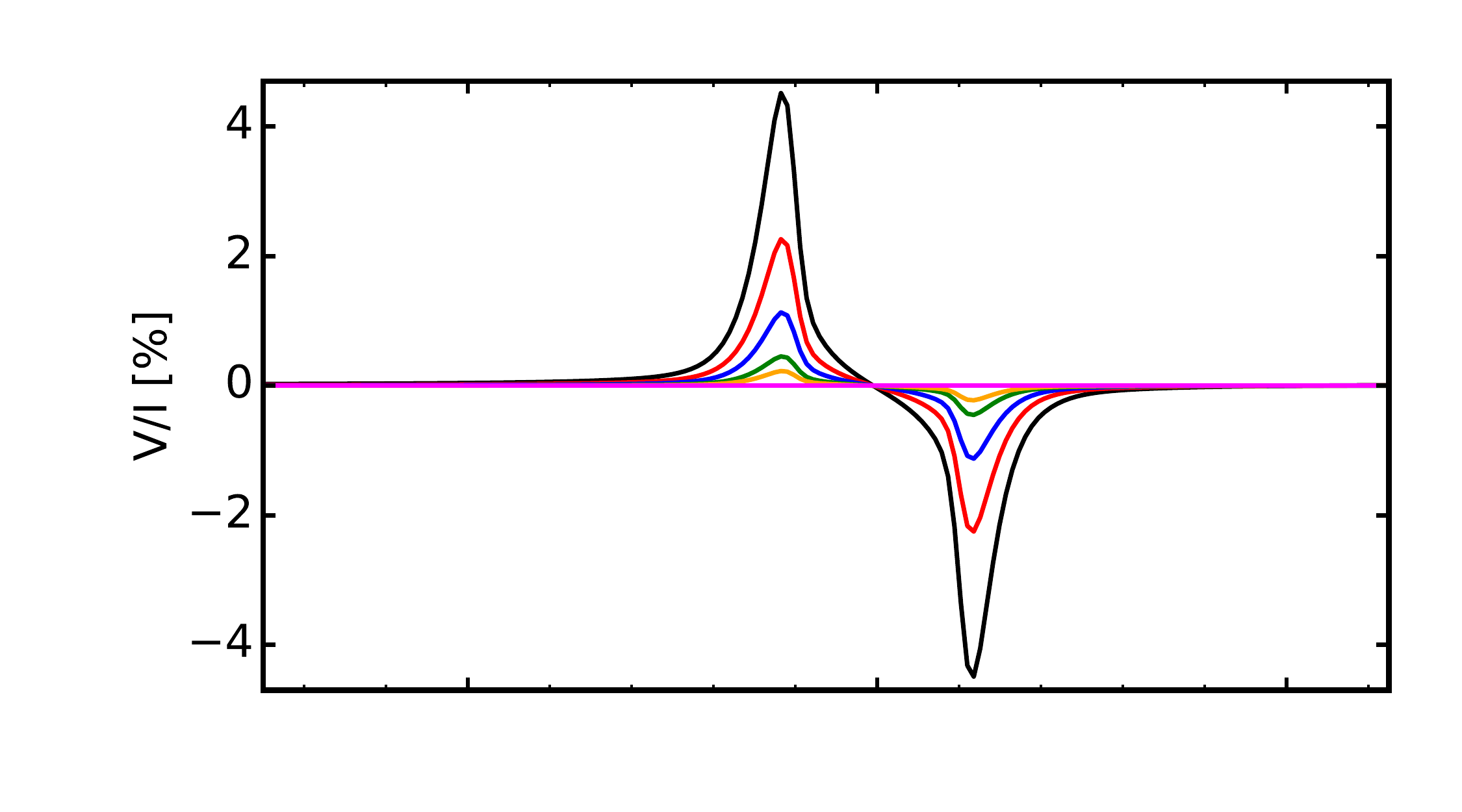}
\includegraphics[width=.47\hsize]{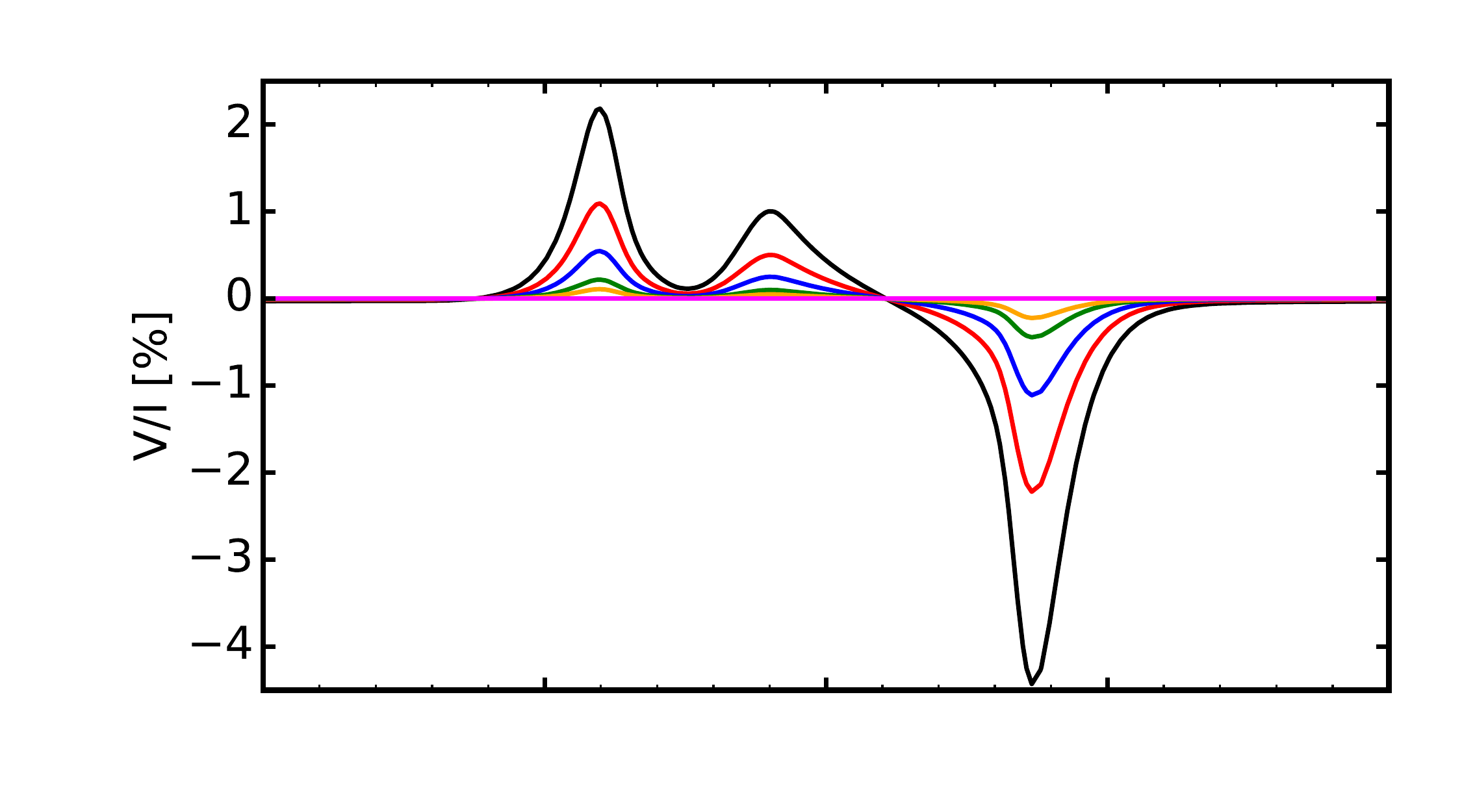} \\
\vspace{-20pt}
\includegraphics[width=.47\hsize]{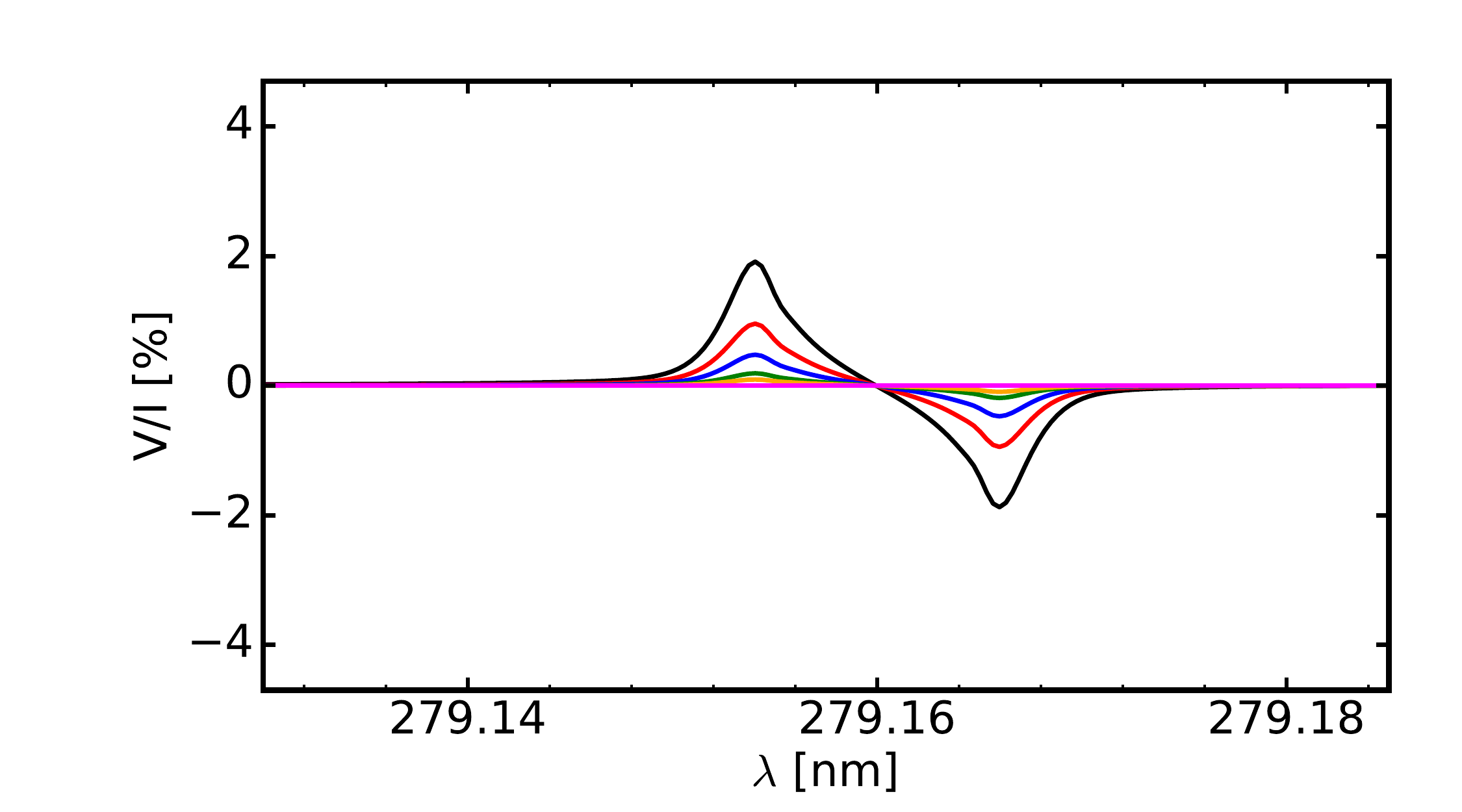}
\includegraphics[width=.47\hsize]{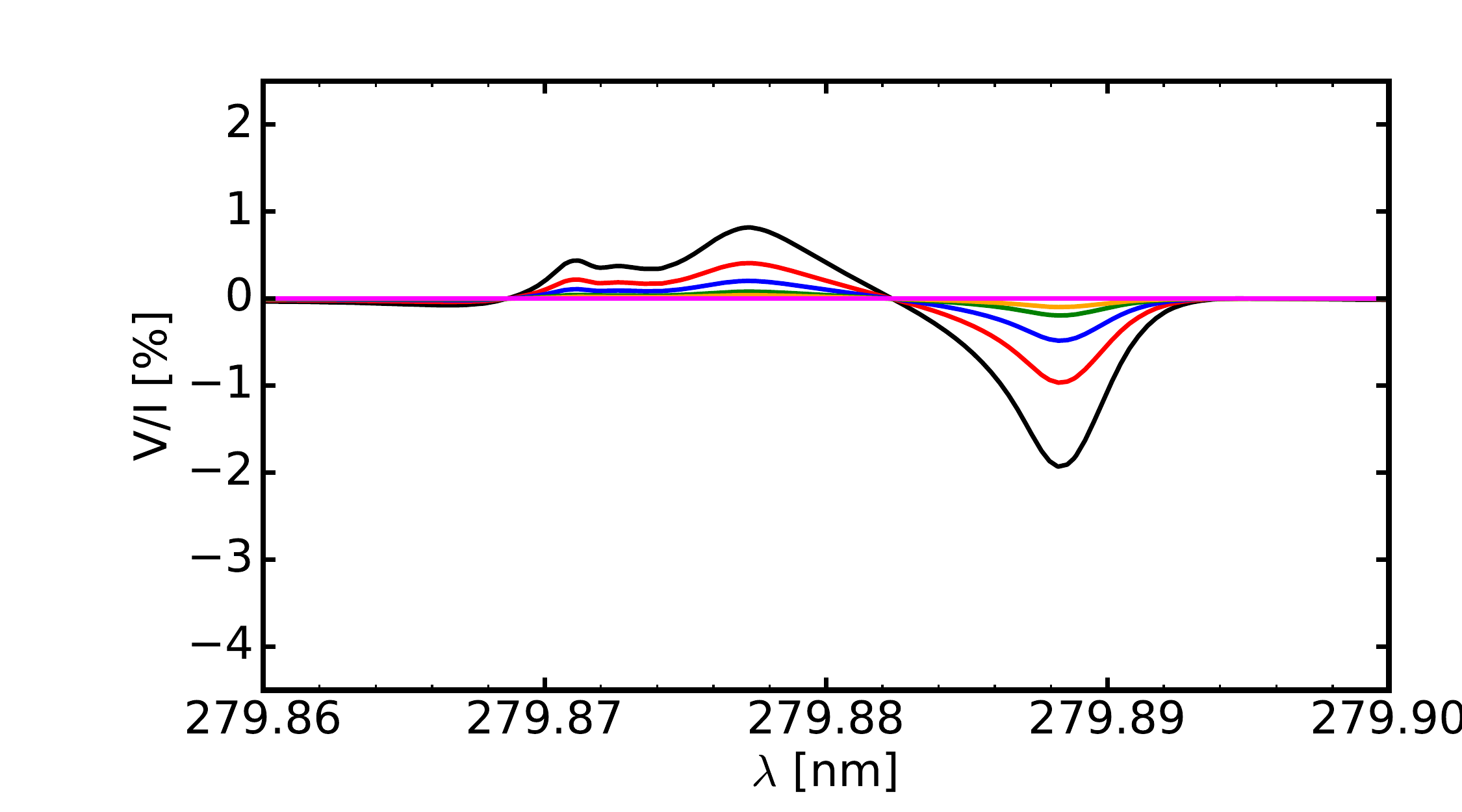}
\caption{Fractional circular polarization profile $V/I$ for the \ion{Mg}{2}
UV triplet. The left (right) column shows the spectral range around 
$\rm s_b$ ($\rm s_r$). The top (bottom) row shows
calculations using the FAL-C (FAL-P) model, for a LOS with
$\mu=1.0$. The color of the curves indicates the strength of the
vertical magnetic field: $0\,$G (magenta), $10\,$G (orange),
$20\,$G (green), $50\,$G (blue), $100\,$G (red), and $200\,$G (black).}
\label{fig:MagnProfVsub}
\end{figure}

\begin{figure}[htp!]
\centering
\includegraphics[width=.48\hsize]{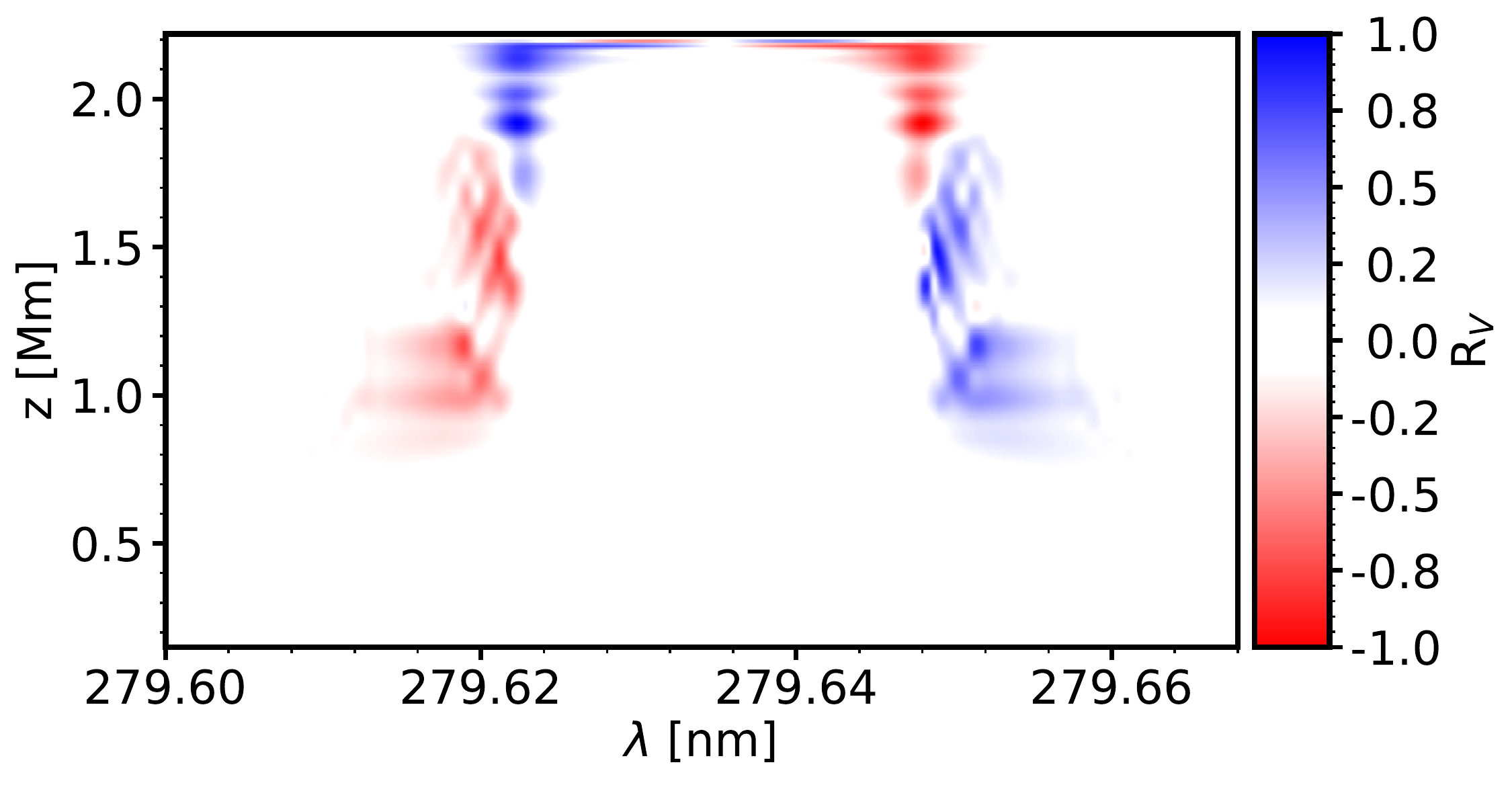}
\includegraphics[width=.48\hsize]{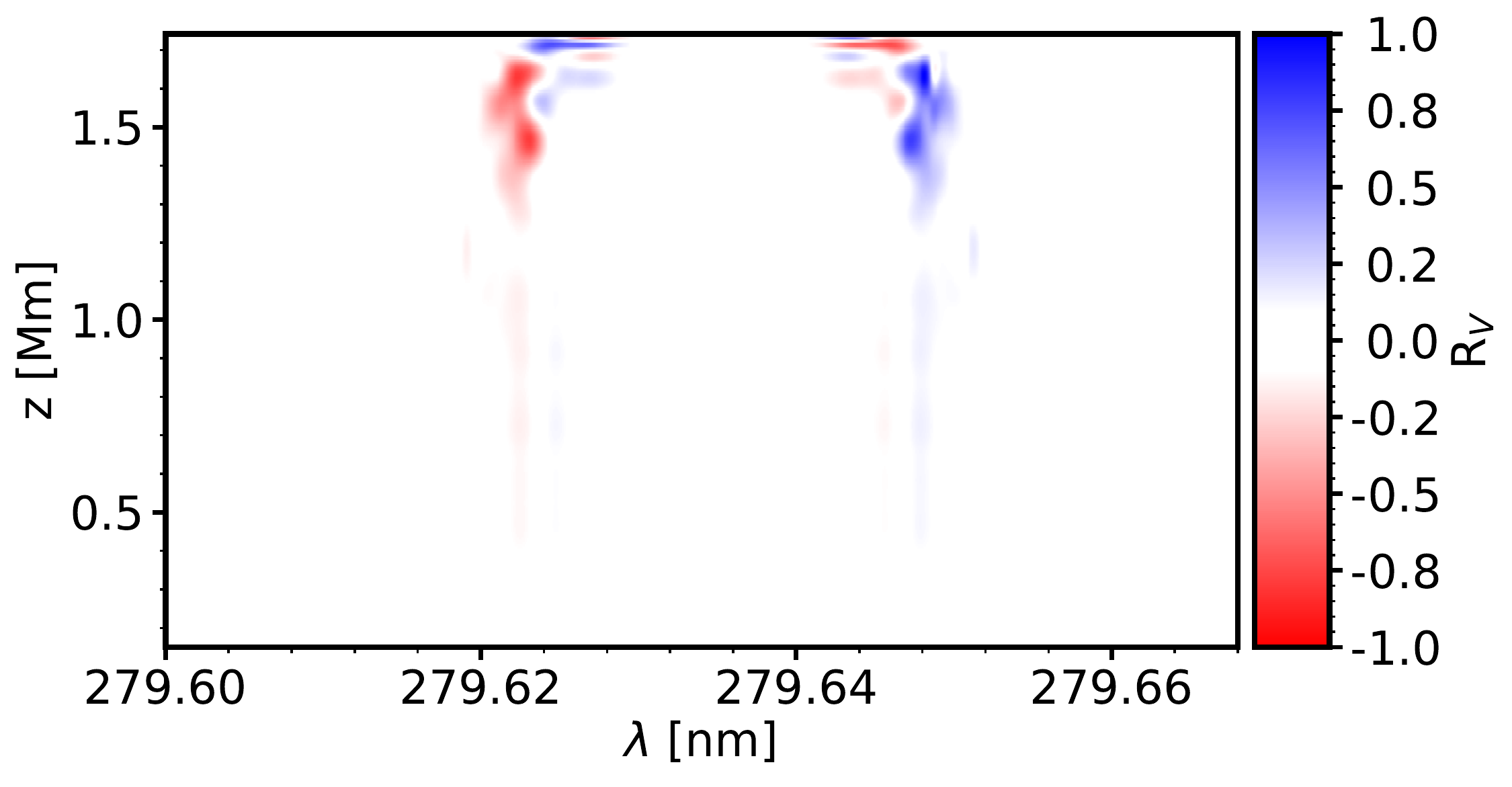}
\caption{Normalized value of the Stokes $V$ response function to magnetic
field perturbations for the \ion{Mg}{2} k line in the FAL-C (left) and FAL-P
(right) models, for the disk-center LOS. The calculation considers a
reference model atmosphere with a vertical magnetic field of $200$\,G, pointing
towards the observer, and a perturbation of $1\,$G.}
\label{fig:RespV}
\vspace{3.6ex}
\end{figure}

Finally, Figs. \ref{fig:MagnProfVhk} and \ref{fig:MagnProfVsub} show the
fractional circular polarization profile, Stokes $V/I$, for vertical magnetic
fields with different strengths and a LOS pointing at disk center. In this
geometry, only circular polarization is possible in a 1D axially symmetric model
atmosphere. While the subordinate lines are less polarized in the
plage model, the resonance lines show a similar degree of polarization in
the FAL-C and FAL-P models with regard to
the inner lobes of the profile (closer to line center).
The outer lobes of the $V/I$ polarization are due to PRD effects, and are
significantly larger in the plage model, likely a consequence of the
smaller geometrical extension of the model atmosphere. Due to this
''compression'' of the plage model, the outer peaks respond more strongly
to magnetic fields from $\sim1500\,$km upwards (see Fig.~\ref{fig:RespV},
right panel), quite close to the
region where the inner lobes form. The farther, smoother part of the profile forms
in a much more extended region. In the average FAL-C atmospheric model, instead,
the response is more extended in height, resulting in a smoother $V/I$ profile.

\subsection{Dynamic Sensitivity}\label{SSDynamics}

In this section we study the sensitivity of the \ion{Mg}{2} h-k and UV triplet
lines to vertical velocity gradients, and the effect of the time
integration on the observed Stokes profiles. In order to achieve this, we solve
the polarized radiation transfer problem at each time step in the CS
time-dependent hydrodynamic model
(\citealt{CarlssonStein1997}).
Our calculations of the angle-dependent PRD have been carried out with the
observer's frame method using a suficiently large number of frequency nodes
(of the order of $10^3$).
The presence of velocity fields introduces
changes in the radiation anisotropy because of the associated Doppler shifts of
the spectral line radiation. Hence, we must solve the problem using the more
general angle-dependent redistribution function for coherent scattering.
However, as the velocity in this 1D time series is always directed along the
vertical, we can still take advantage of the axial symmetry of the model in
order to simplify the calculation.\footnote{For general, non-axially symmetric
problems, we would need to take into account at least a factor 8 more directions,
and even more in order to attain sufficient accuracy in the polarization
profiles.}

\begin{figure}[htp!]
\centering
\includegraphics[width=.48\hsize]{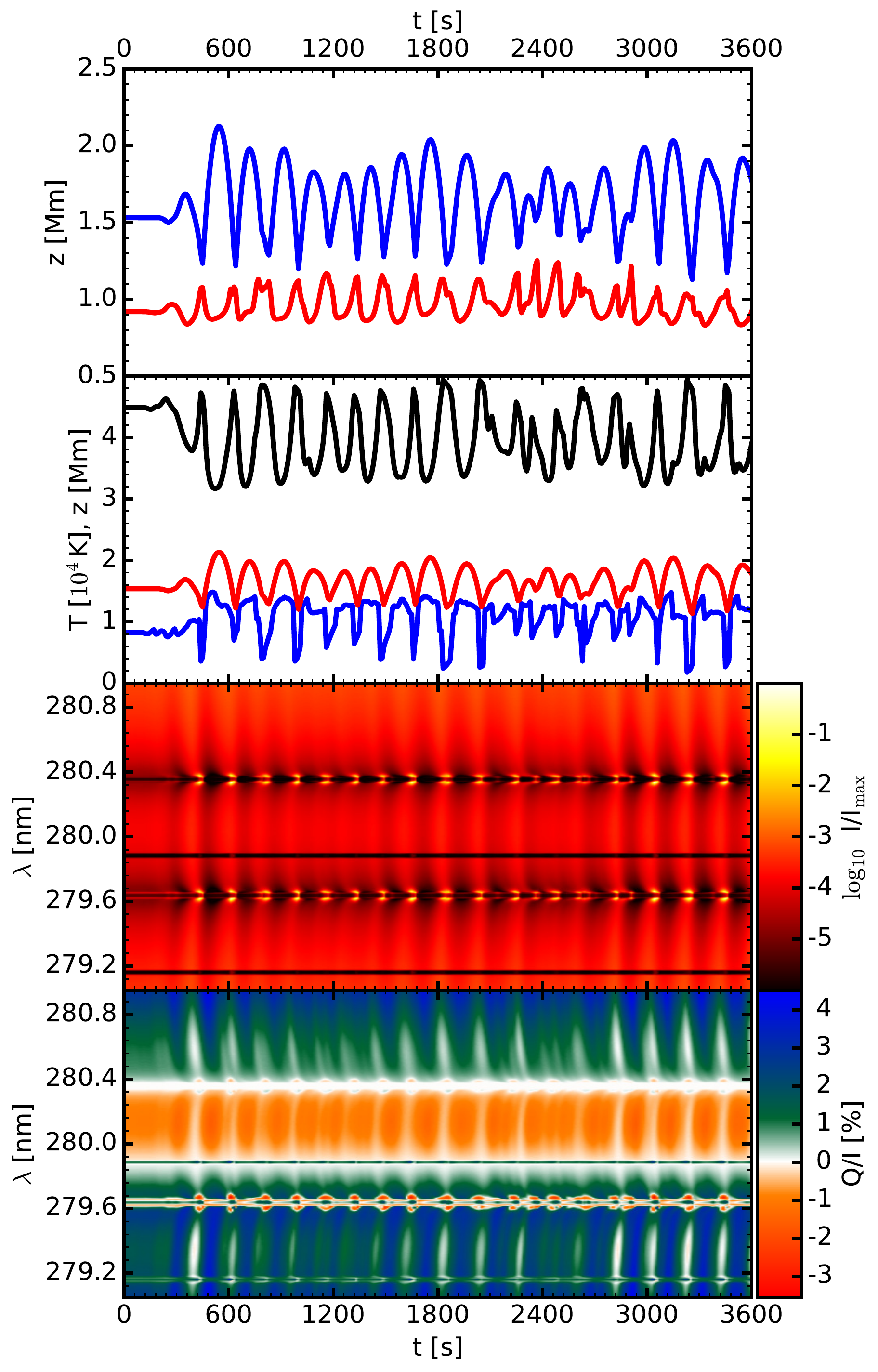}
\caption{Top panel: variation with time of the height where the optical depth
is equal unity for the line center of the k line at $279.64\,$ nm (blue) and
of the $\rm s_{r_b}$ transition at $279.88\,$nm (red). Second
panel: variation with time of the temperature minimum (black), the height
corresponding to the temperature minimum (blue), and the height with the maximum
temperature gradient (red). Third panel: variation with time of the intensity
profile around the \ion{Mg}{2} resonance lines, normalized to the maximum value
in the range. Bottom panel: variation with time of the fractional polarization
$Q/I$ profile around the resonance lines of \ion{Mg}{2}. The scale is saturated
to $|Q/I| = 2$ to facilitate the visualization, but the polarization in this
range reaches values of $4\%$.}
\label{fig:serie1}
\end{figure}

Figure \ref{fig:serie1} shows some of the main characteristics of the formation
of the \ion{Mg}{2} lines in the CS model. The heights where the
optical depth is unity (a rough estimation of the height of formation) for the
resonance and the subordinate lines are in opposition of phase (Fig.~
\ref{fig:serie1}, top panel). The lower-boundary piston used by
\cite{CarlssonStein1997} to drive the hydrodynamic simulation makes the
atmosphere oscillate in such a way that it goes through compression and
expansion phases. Given that the resonance lines form higher in the atmosphere
and the subordinate lines are formed at much lower heights, the regions of
formation get closer during the compression phase and separate instead during
the expansion phase.

During compression, a temperature shock rises through the
atmosphere (see \citealt{CarlssonStein1997}), producing a much hotter
temperature minimum (Fig.~\ref{fig:serie1} second panel, black curve). The
red curve in the second panel of Fig.~\ref{fig:serie1} shows the height with
the maximum temperature gradient, which traces the position of the transition
region. The blue curve in the same panel shows the height of the temperature
minimum. During a compression phase, the CS atmosphere has a different
stratification from a typical semi-empirical model atmosphere (color curves in
Fig.~\ref{fig:Atmos} top-left panel) due to the shock. During an expansion
phase, a ``valley'' with an evident temperature minimum is created in the space
left by the rising transition region and the falling photosphere.

\begin{figure}[htp!]
\centering
\includegraphics[width=.48\hsize]{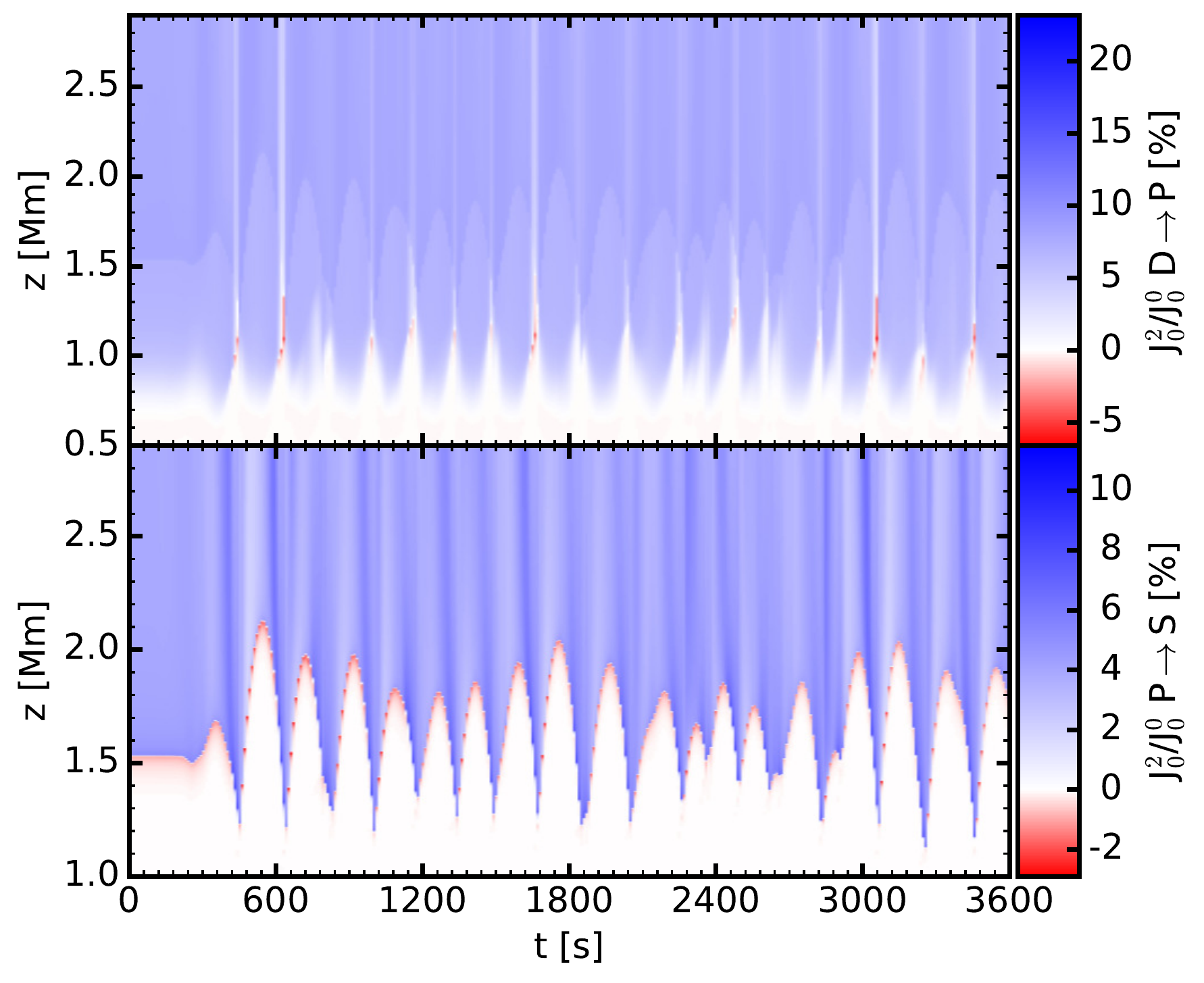}
\includegraphics[width=.48\hsize]{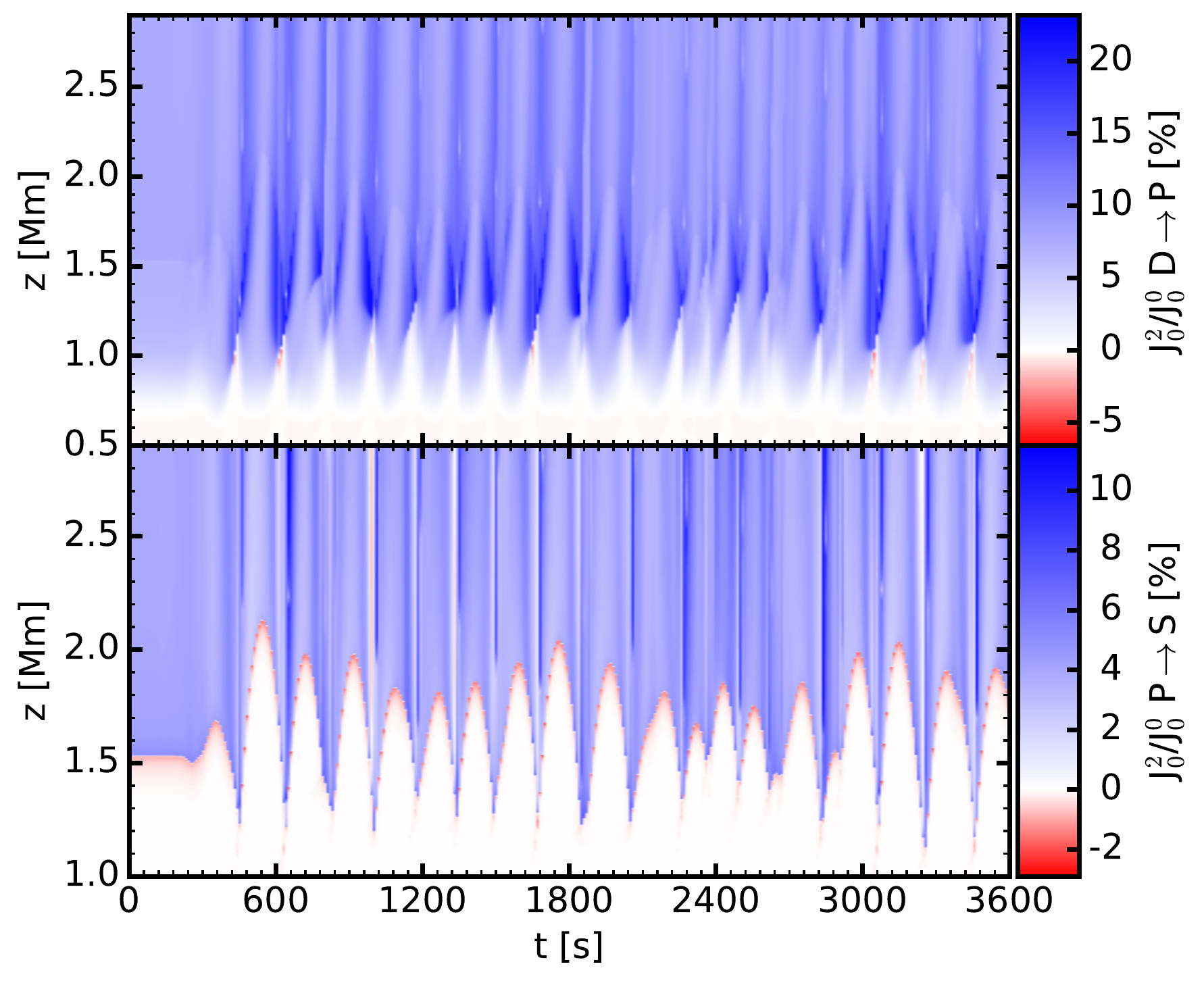}
\caption{Variation with time and height of the fractional anisotropy 
$J^2_0/J^0_0$ for the \ion{Mg}{2} UV triplet (top
panels) and the \ion{Mg}{2} h-k doublet (bottom panels) in the CS time
series. The left (right) panels correspond to the solution of the radiation
transfer problem when excluding (including) the presence of the
velocity fields of the model atmosphere.}
\label{fig:J20serie}
\end{figure}

\begin{figure}[htp!]
\centering
\includegraphics[width=.48\hsize]{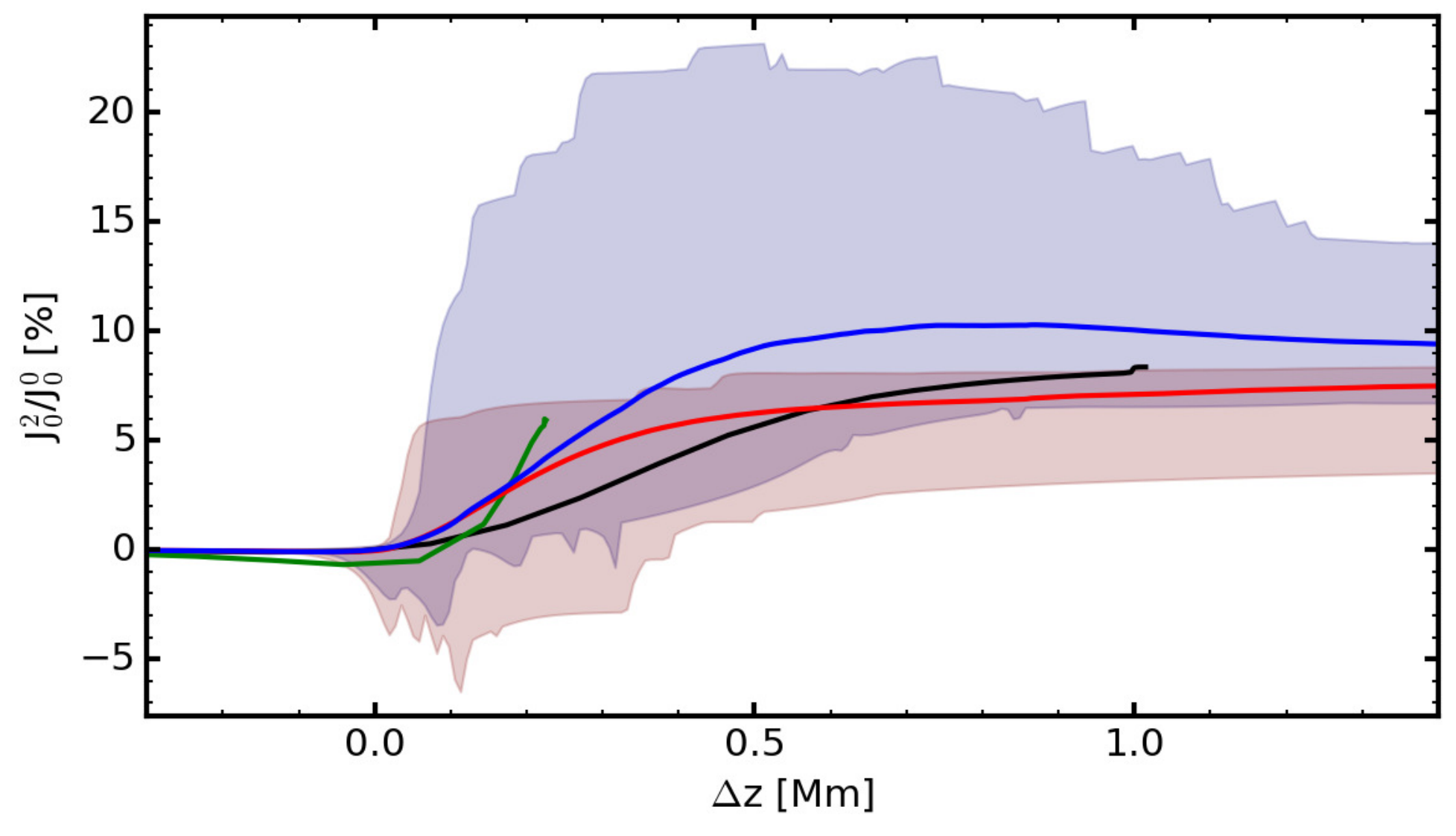}
\includegraphics[width=.48\hsize]{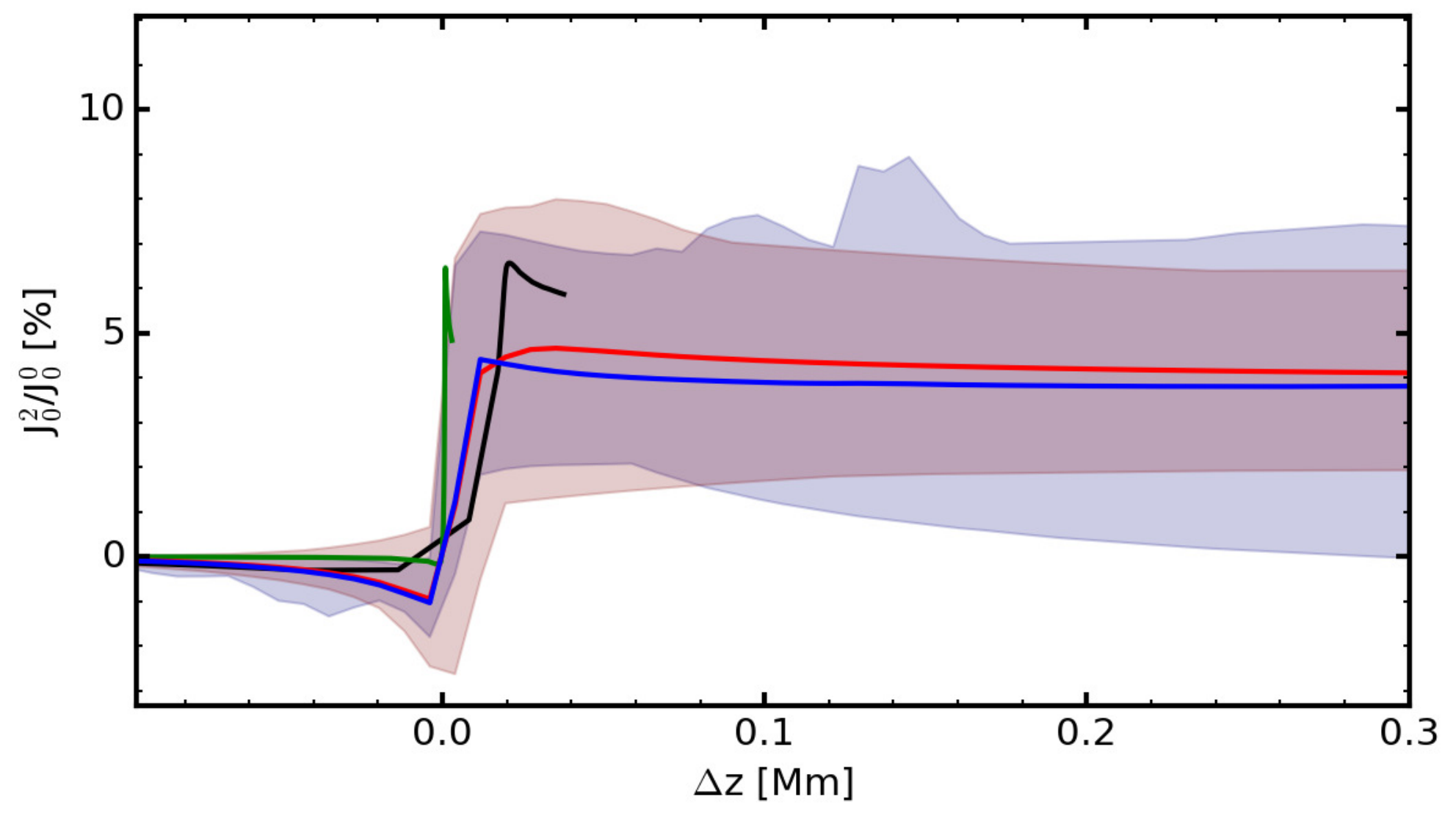}
\caption{Variation with height of the fractional anisotropy 
$J^2_0/J^0_0$ for the \ion{Mg}{2} UV triplet (left
panel) and the \ion{Mg}{2} h-k doublet (right panel). The shaded
areas demarcate the range of values of the anisotropy attained during
the CS time series, including (light blue) and excluding (light red)
velocities. The solid curves of the corresponding colors show the average
anisotropy over the time series for the two cases.
For comparison, the black (green) curve shows the anisotropy in the
FAL-C (FAL-P) atmospheric model.
We set the zero of the height axis at the maximum height
of optical depth unity, within the wavelengths of the set of transitions for each
model atmosphere.}
\label{fig:J20serieavg}
\end{figure}

The temporal change of the heights of formation is due to the change of the
thermal structure of the CS model, and not directly to the presence
of velocity gradients. In fact, when we solve the radiation transfer problem
excluding velocities, we obtain very similar results for those heights.
This is not the case for the radiation field and
the emergent Stokes parameters. Fig.~\ref{fig:J20serie} shows the variation
with height and time of the fractional anisotropy for the \ion{Mg}{2} h-k
doublet and UV triplet, excluding (left) and including (right)
the velocity field in the solution of the radiation transfer problem.
In agreement
with previous studies of different spectral lines (e.g.,
\citealt{Carlinetal2012,SampoornaNagendra2015}), the anisotropy is significantly
enhanced in the presence of a velocity field with gradients. In particular, at some
time steps, the anisotropies of the resonance and subordinate lines
appear to be
enhanced by a factor $\sim$2 and $\sim$4, respectively. This is more directly seen
in Fig.~\ref{fig:J20serieavg}, where we show the average anisotropy in the CS
time series including (blue curve) and excluding (red curve) velocities, together
with the corresponding full ranges of values (shaded areas), as a function of
height.
For comparison we also show in Fig.~\ref{fig:J20serieavg} the anisotropy
in the unperturbed FAL-C (black curve) and FAL-P (green curve) models.
Because the ``absolute'' height of formation is changing continuosly with time,
we set the zero of the height axis to the maximum height of optical depth unity,
separately for each set of transitions and atmospheric model.
The anisotropy reaches the largest values in the presence of velocity
gradients (blue shaded areas). This enhacement is much more significant in the
subordinate lines (left panel of Fig.~\ref{fig:J20serieavg}) and their anisotropy,
averaged over the whole series, is enhanced in the presence of velocity gradients
(the result is similar when taking the median). The anisotropy in the resonance lines
(right panel of Fig.~\ref{fig:J20serieavg}) also changes notably between
including and excluding velocity gradients during the time series, but the
differences in the average anisotropy is not significant.

\begin{figure}[htp!]
\centering
\includegraphics[width=.32\hsize]{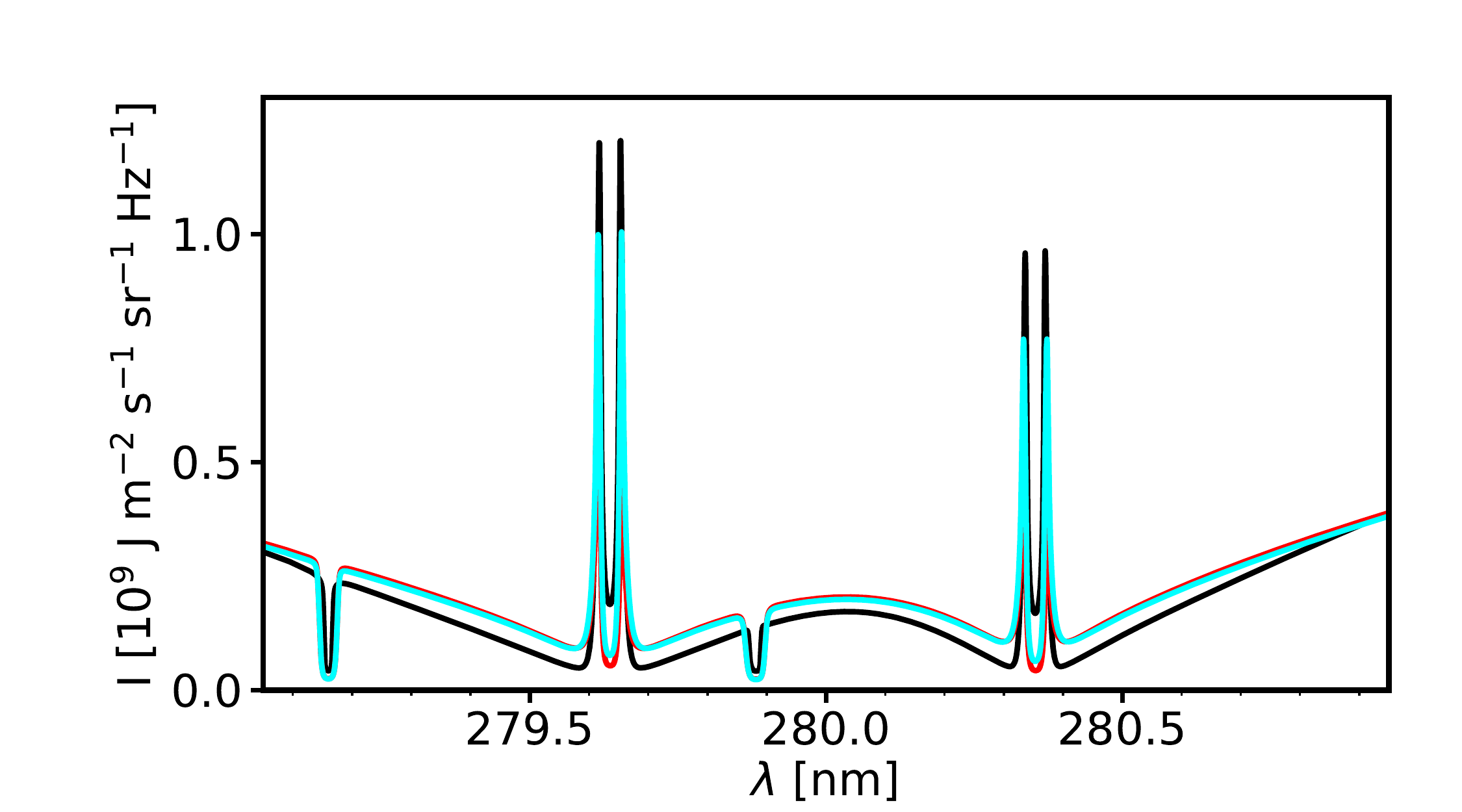}
\includegraphics[width=.32\hsize]{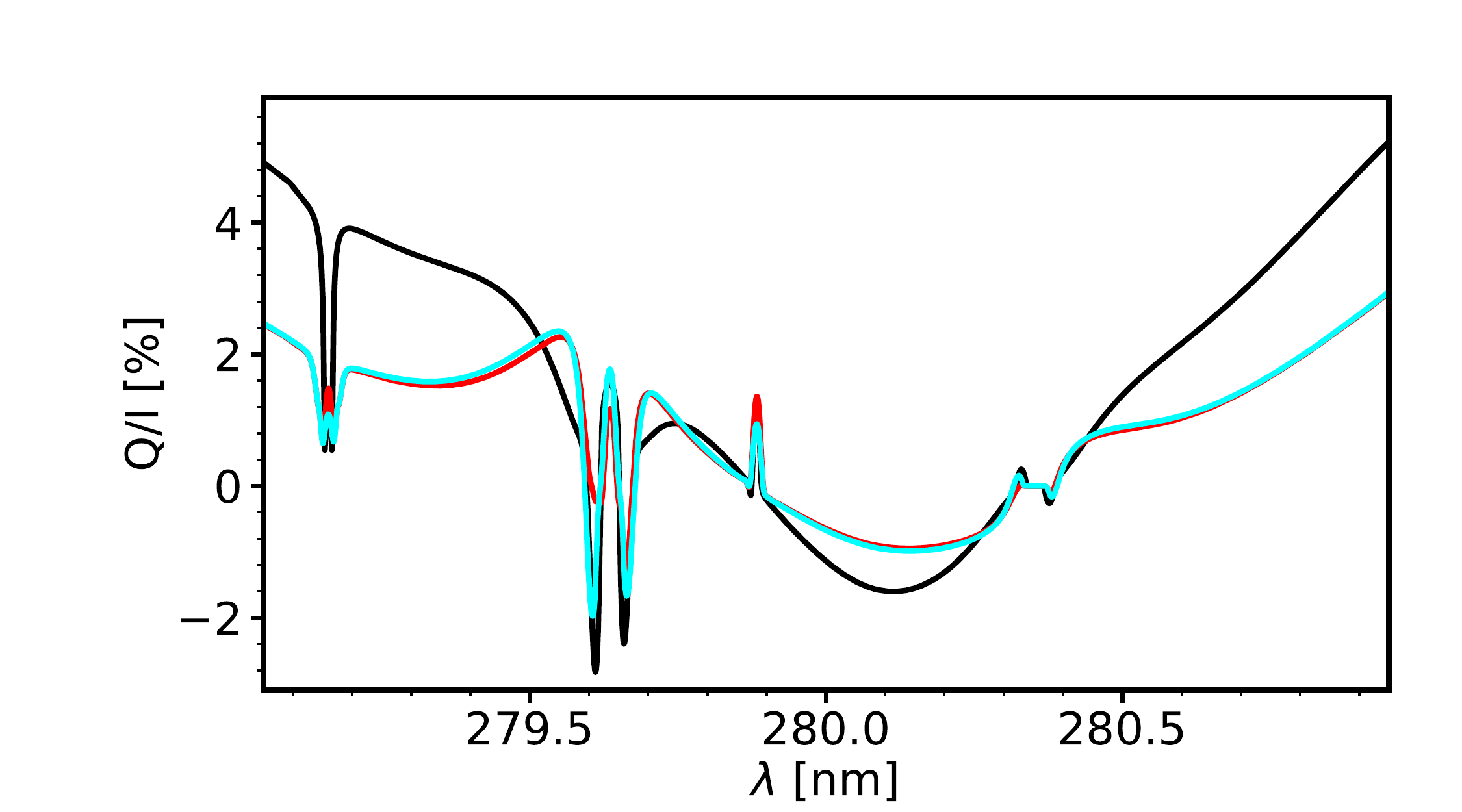} \\
\includegraphics[width=.32\hsize]{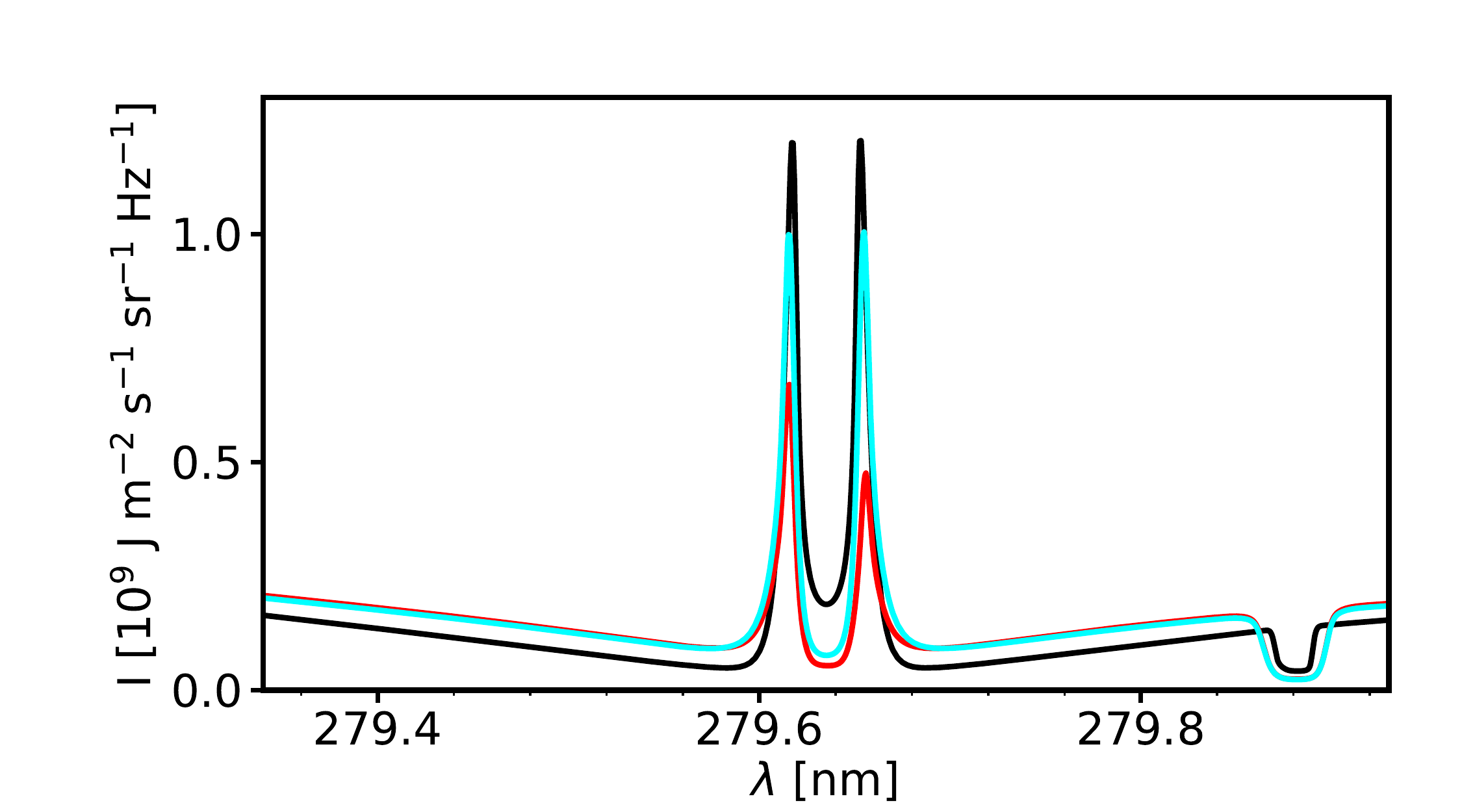}
\includegraphics[width=.32\hsize]{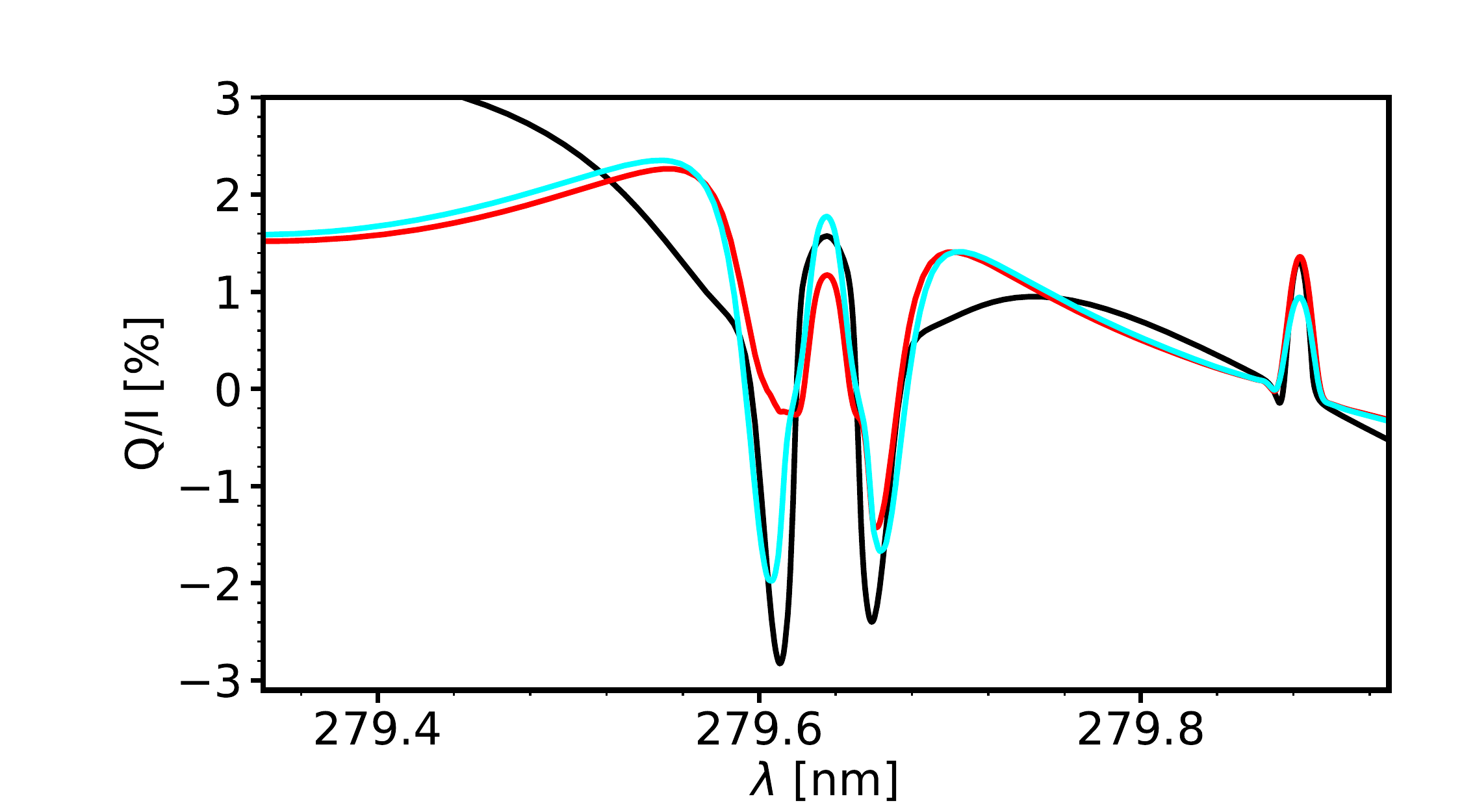} \\
\includegraphics[width=.32\hsize]{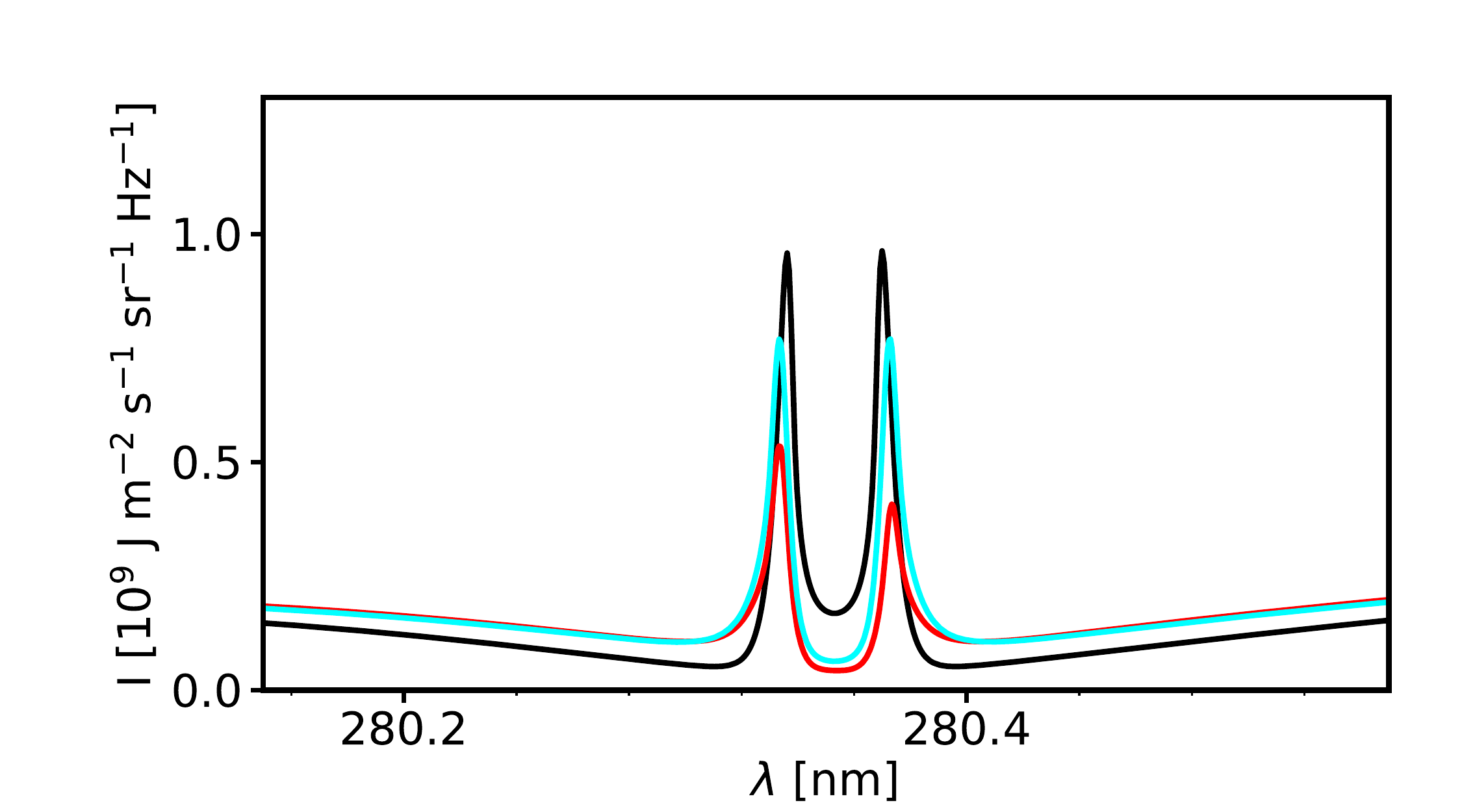}
\includegraphics[width=.32\hsize]{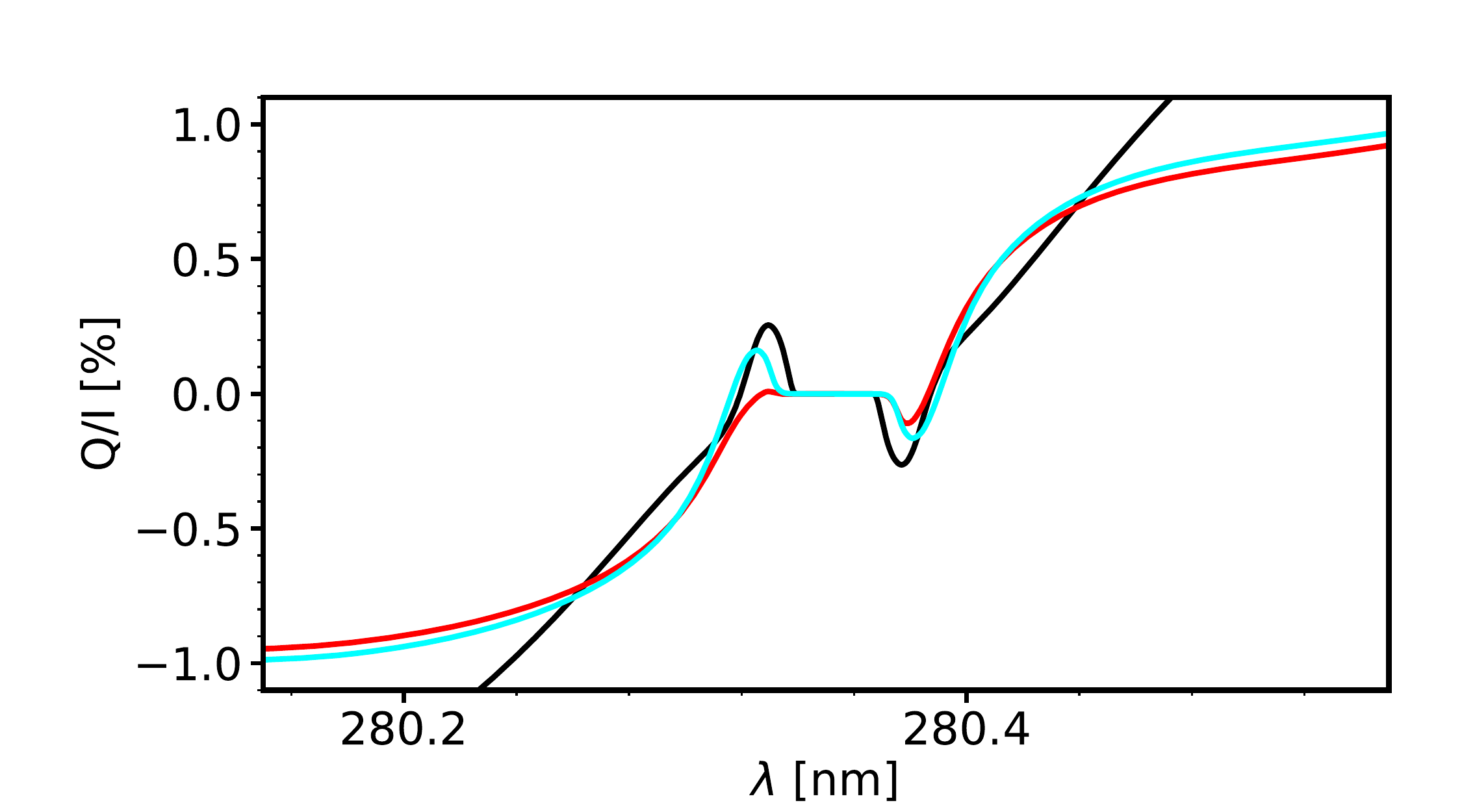} \\
\includegraphics[width=.32\hsize]{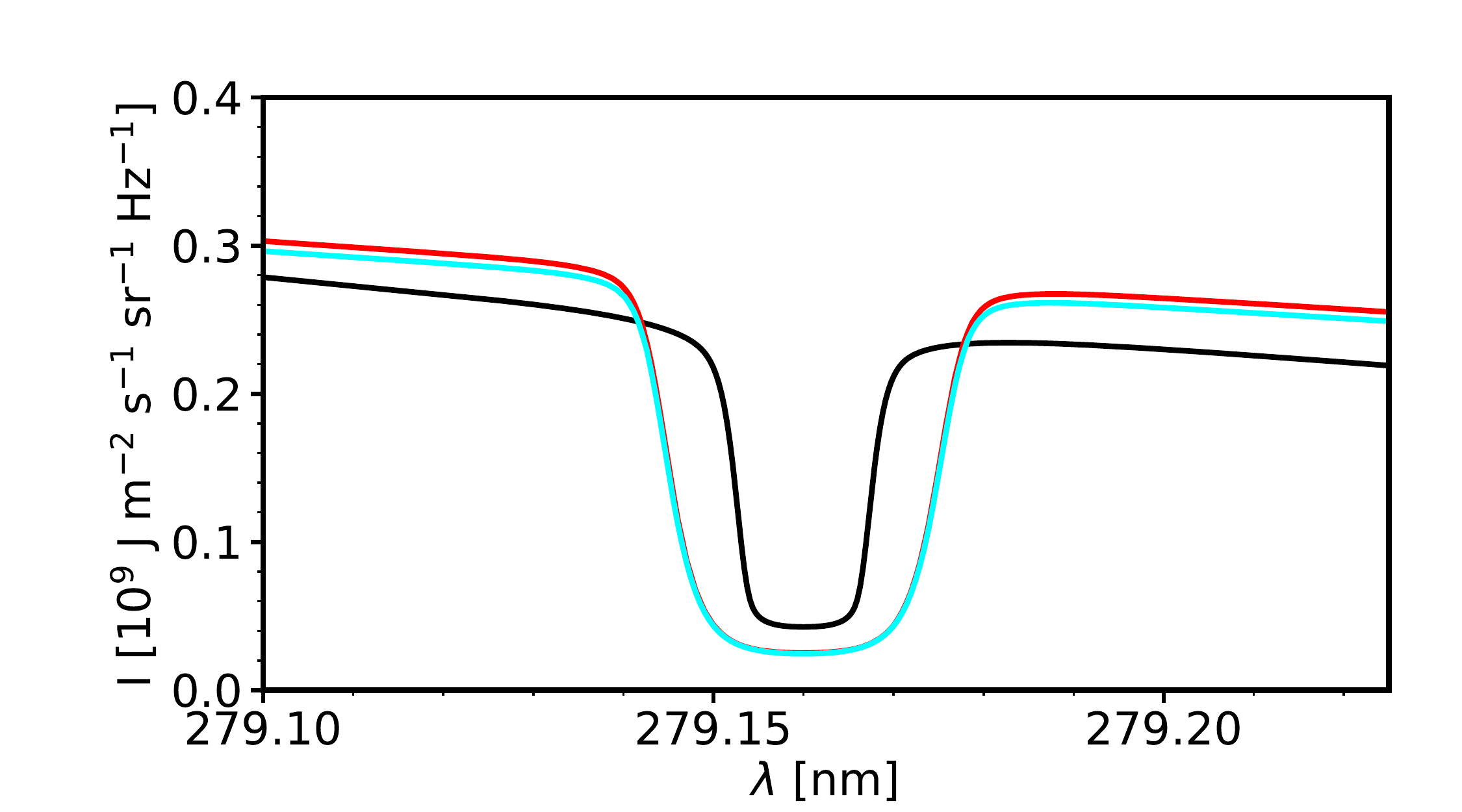}
\includegraphics[width=.32\hsize]{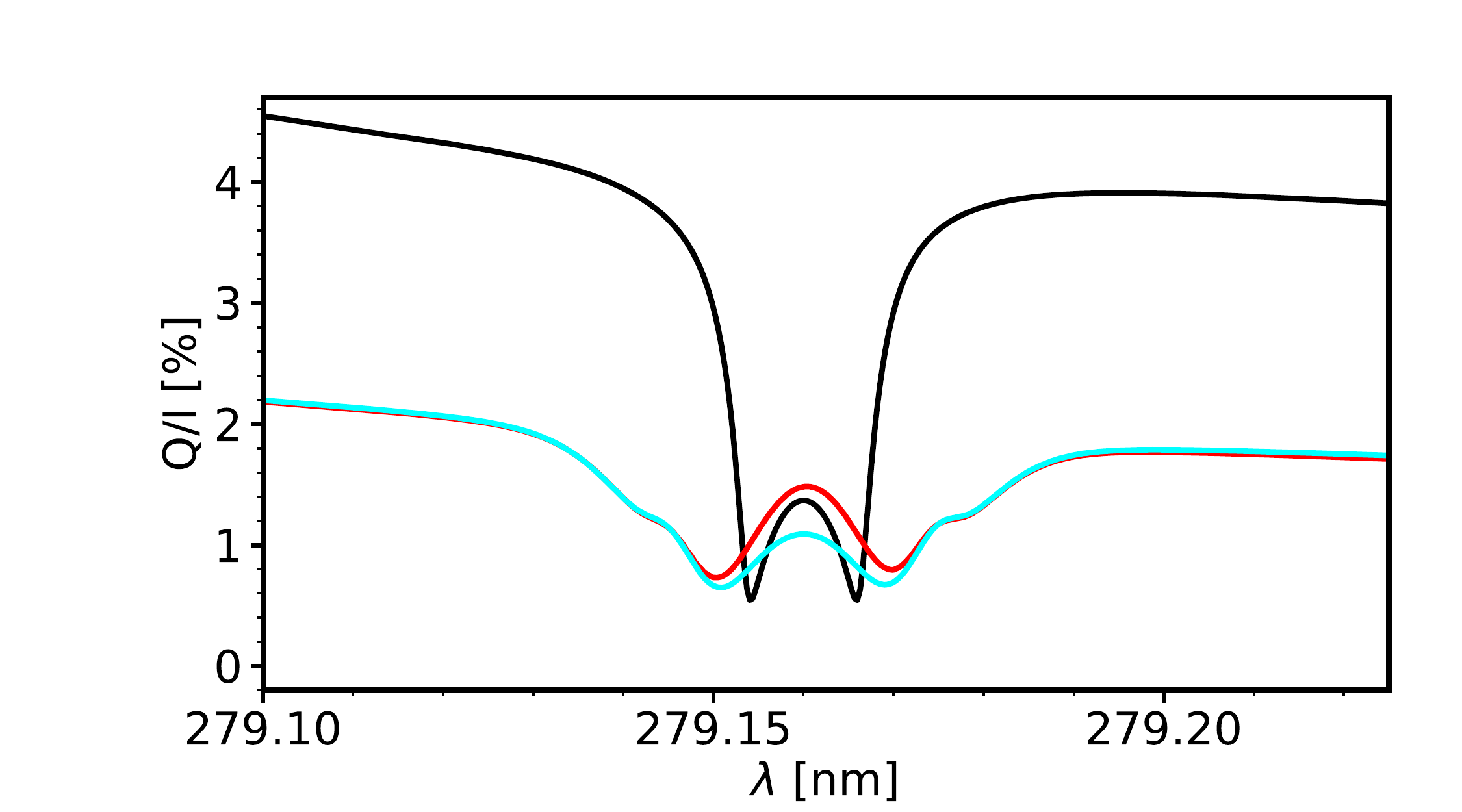} \\
\includegraphics[width=.32\hsize]{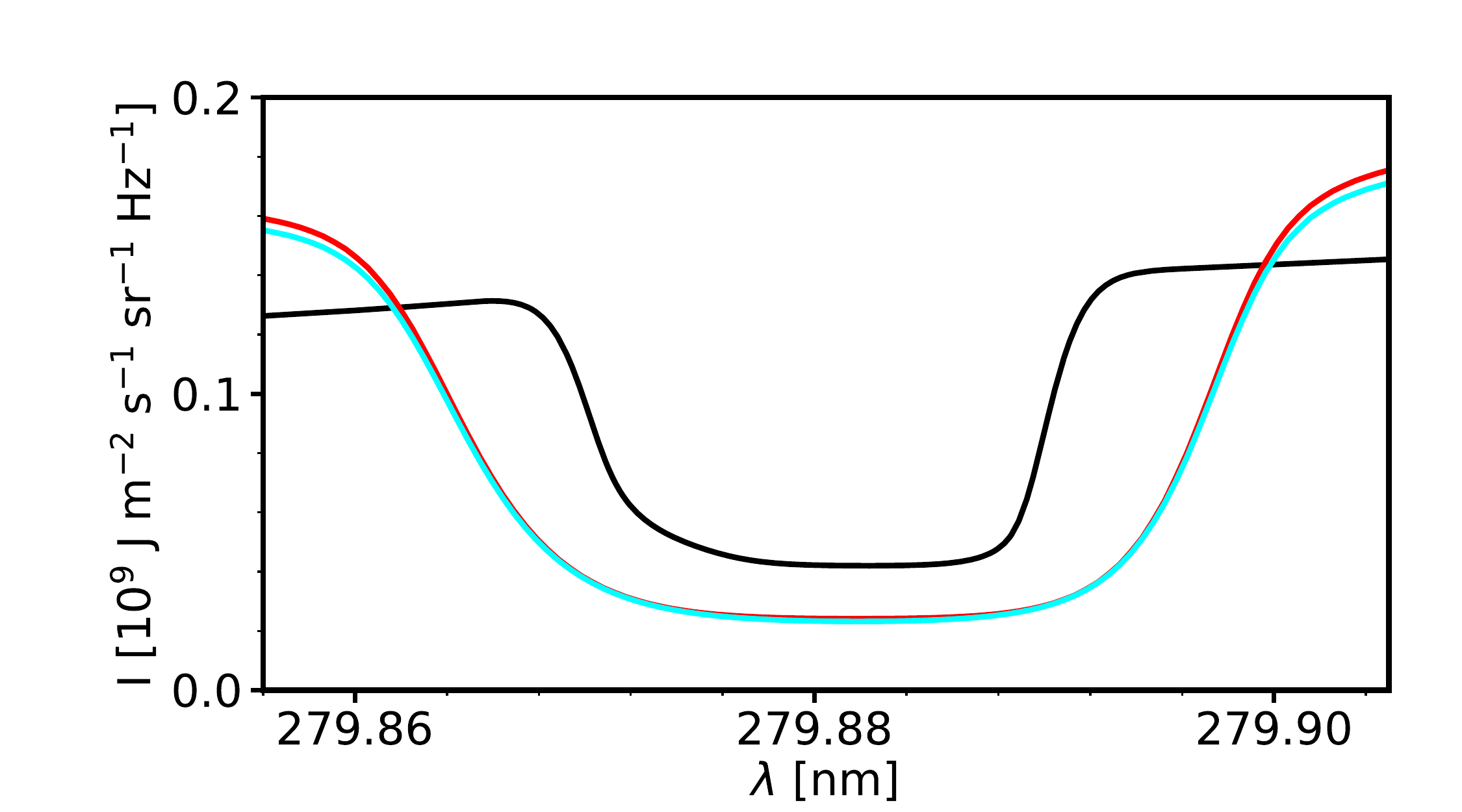}
\includegraphics[width=.32\hsize]{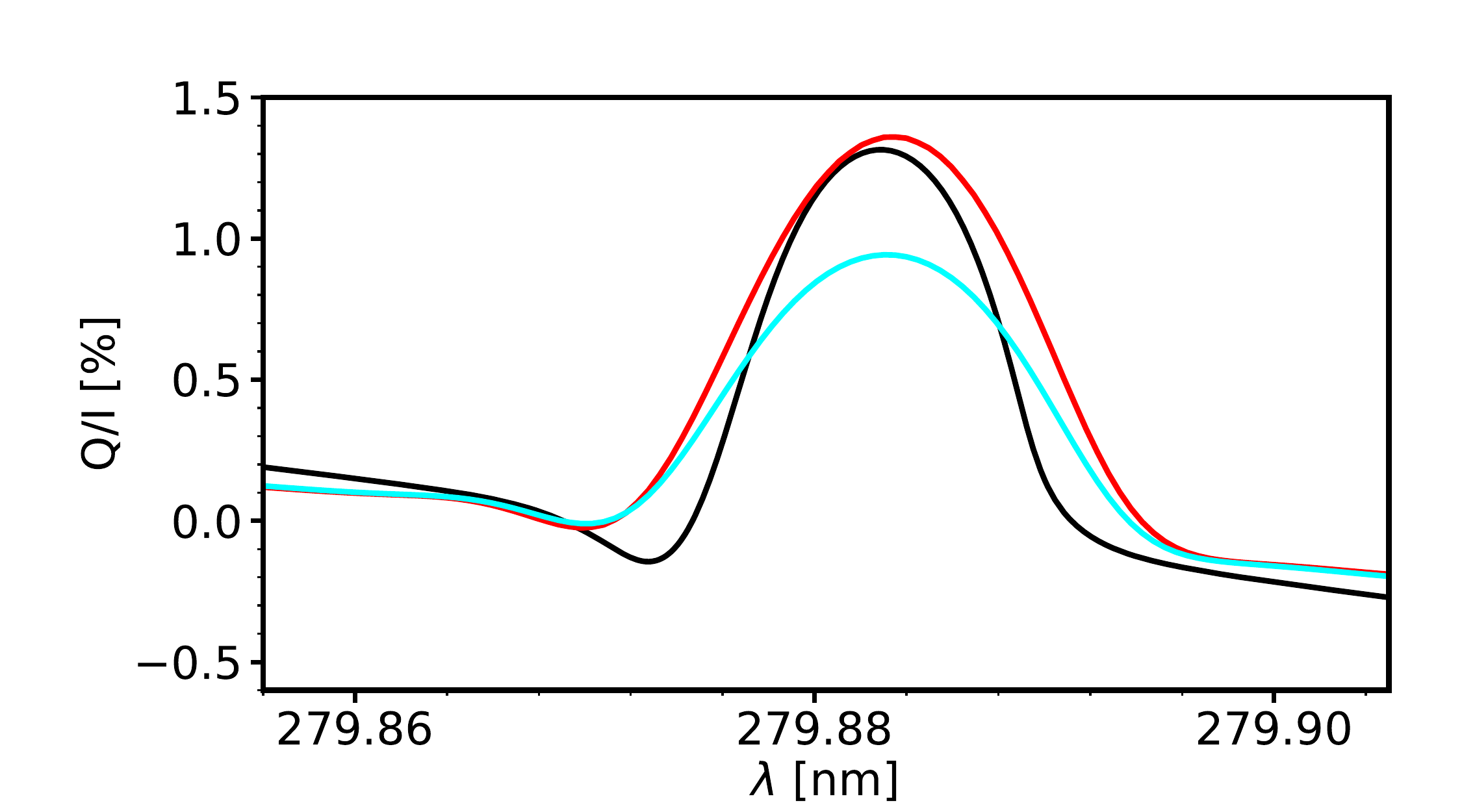}
\caption{Intensity $I$ profiles (left column), and fractional linear
polarization $Q/I$ profiles (right column) for the \ion{Mg}{2} h-k doublet and
UV triplet. The first row shows the full spectral range, while from
the second to the fifth rows show the spectral regions around the individual
transition lines. The black, light blue, and red curves correspond to the
FAL-C, CS (ignoring velocity fields), and CS models, respectively, for a LOS
with $\mu=0.1$.}
\label{fig:fullavg}
\end{figure}

Figure \ref{fig:fullavg} shows the intensity and fractional polarization
profiles corresponding to the FAL-C model and the time average over the full
CS time series, with and without velocities.
The subordinate lines are much wider in the CS series than in the FAL-C model.
This is mainly due to our choice of microturbulent velocity. While $7\,$km/s
appears to be adequate to produce a profile width for the h-k doublet similar
to that of the FAL-C model, it turns out to be excessive for the triplet.
However, since the objective of this section is to study the impact of the
velocity field, it is not necessary to determine the best microturbulence
stratification (a very computationally intensive task), and it is sufficient
to simply compare the emerging profiles in the CS series when velocities are
included or excluded. While the impact of the velocity fields is small on the
intensity profile of the subordinate lines, the fractional linear polarization
clearly shows the effect of the anisotropy enhacement, resulting in a significantly
larger polarization signal.

The situation changes dramatically for the h-k doublet. First, there is a clear
asymmetry in the emission peaks of Stokes $I$ for both lines of the doublet,
when velocities are included. The same asymmetry is also found in the fractional
polarization profile (see Fig.~\ref{fig:fullavg}). In fact, the degree of
polarization is reduced, and the asymmetry due to the intrinsic physical properties
of the k line (blue lobe larger than the red lobe) is reversed by the dynamics
(i.e., the red lobe is now larger than the blue lobe). The polarization of these
lines is clearly sensitive to the velocity field within the atmosphere.

\begin{figure}[htp!]
\centering
\includegraphics[width=.49\hsize]{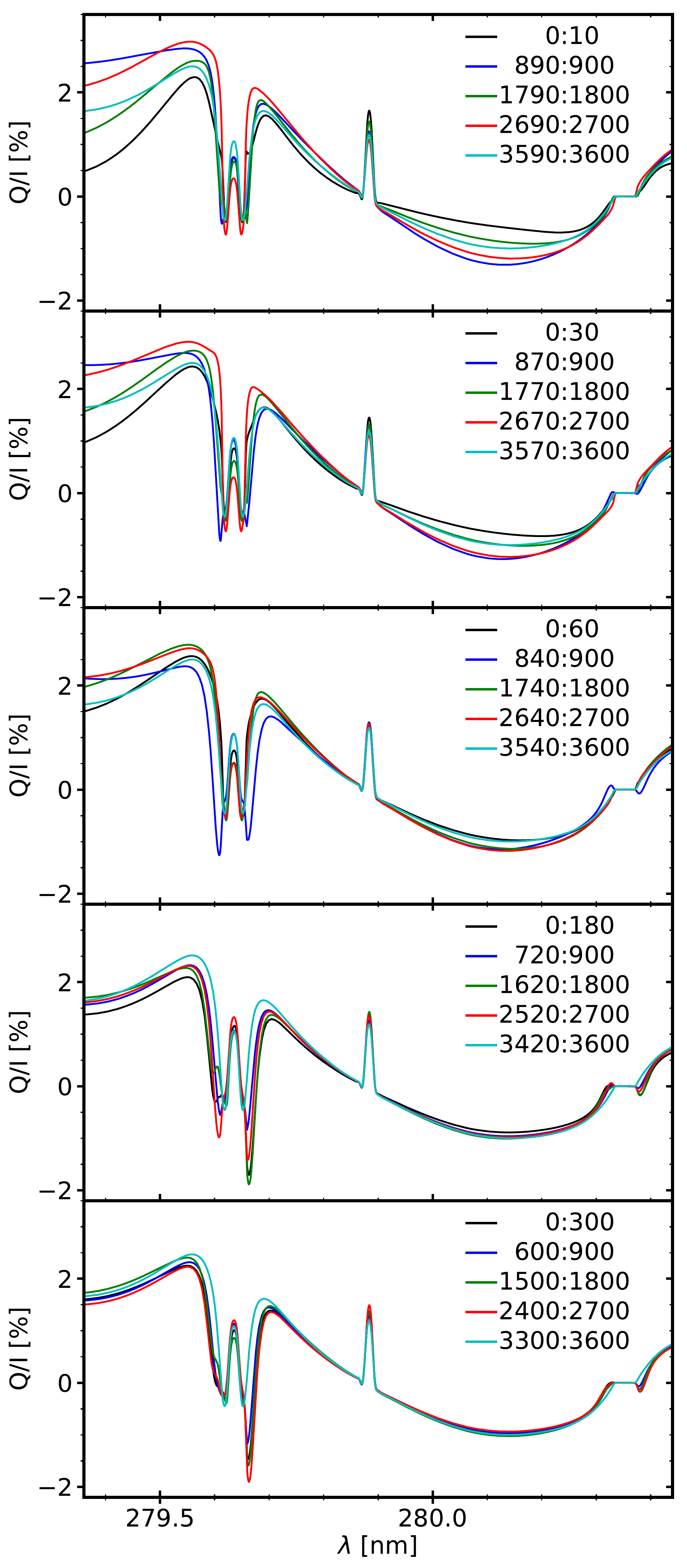}
\caption{Time integral of the fractional linear polarization $Q/I$ profiles for
the \ion{Mg}{2} h-k doublet and UV triplet in the CS model for a
LOS with $\mu=0.1$. Colors represent the starting and ending times of the
integration interval, indicated in the legend of the panels. In the different
panels, the profiles shown are the results of an integration of increasingly
larger times, from top to bottom: 10\,s, 30\,s, 60\,s, 180\,s, and 300\,s.}
\label{fig:stepavg}
\end{figure}

The profiles shown in Fig.~\ref{fig:fullavg} are the result of the integration
over the full series, i.e., 3600\,s. However, in a real observation the
integration time is significantly smaller. Fig.~\ref{fig:stepavg}
shows the fractional polarization profile, at different times, for different
exposure times. As expected, with small integration times (top panels), the
profiles are quite different depending on the interval of the series that is
being integrated (see \citealt{Carlinetal2013}). For larger integration times,
the different profiles converge to each other and to the total average
profile, as expected based on the behavior of the CS
model. However, it is important to note that the fractional polarization of the
spectral region between the $\rm s_r$ and h lines ``converges'' in amplitude
already for 60\,s of exposure time, and the integrated signal is around a
factor 2 smaller than the one expected from the calculations in the FAL-C
model. Therefore, the joint effect of velocity and magnetic fields could make
a clear detection of the negative fractional polarization in
this spectral region more difficult.

\section{Conclusions}\label{Sconclusions}

We carried out a detailed numerical investigation of the intensity and
polarization
of the \ion{Mg}{2} h-k doublet and UV triplet located around 280\,nm. We
used radiative transfer calculations in 1D models of quiet and plage regions
of the solar
atmosphere, in order to study the sensitivity of the emergent Stokes profiles
to the thermal, magnetic, and dynamic properties of those model
atmospheres. These
calculations take into account the combined action of anisotropic
irradiation of the atomic system (with partial frequency redistribution
for the h-k doublet) and the Hanle, Zeeman, and magneto-optical (M-O) effects,
allowing us to model the presence of magnetic fields of arbitrary field
strengths (i.e., in the general regime of the incomplete Paschen-Back effect).

We find that this UV spectral region is sensitive to the magnetic field over a wide
range of heights. The core of \ion{Mg}{2} k forms in the upper chromosphere,
the subordinate lines and the near wings of h and k form in the intermediate
chromosphere, whereas the broadband polarization pattern of the h-k wings is
sensitive to conditions in the upper photosphere and low chromosphere.

The magnetic sensitivity across the \ion{Mg}{2} h-k lines is caused by different
physical mechanisms. The Hanle effect operates at the center of the $Q/I$ and $U/I$
profiles of the k-line and the subordinate lines, whereas the M-O effects determine
the magnetic sensitivity of those polarization signals in the far wings.
In contrast, the Zeeman effect practically dominates the circular
polarization of all
the lines in this spectral region.

The Stokes $V$ profiles of the h-k doublet show two lobes in each
wing. The inner lobe is formed in the high chromosphere, while the outer one
is formed at lower chromospheric heights. Therefore, the relative amplitudes
of the two lobes can be used as a diagnostic of the relative strengths
of the magnetic fields in the corresponding layers of the solar atmosphere.

As expected, the dynamics of the time-series model atmosphere considered
in this work
introduces some variability in the shapes and amplitudes of the Stokes profiles, as
well as an enhancement of the radiation anisotropy. Nevertheless, when we take into
account the typical exposure times of spectropolarimetric observations, this
variability tends to be drastically reduced. In particular, for exposure times over
1 minute, the $Q/I$ polarization profile between the k and h transitions does not
change appreciably along the time series.

The polarization signals and the physical effects described in this paper are
within the observable range of instruments like CLASP-2. The modeling undertaken
for this work is very complex, but it misses the additional symmetry breaking
effects of 3D radiative transfer.
Nonetheless, we are confident that the
results presented in this work, and the effects discussed in this and previous
papers, can already be verified or falsified at this stage through
observations, without the need to further increase the realism and complexity
of the model.

Finally, we want to emphasize the importance of designing and deplying
space telescopes
equipped with UV spectropolarimeters in the near future, in order to
routinely attain the observational data necessary for the detailed study
of the magnetism of the upper solar chromosphere and transition
region. The complement of UV spectropolarimetric observations with the
visible and IR
observations already attainable from the ground will enormously increase
the opportunities for new discoveries in both solar and stellar physics.

\acknowledgements 
We acknowledge the funding received from the European Research Council (ERC)
under the European Union's Horizon 2020 research and innovation programme (ERC
Advanced Grant agreement No 742265).

\bibliographystyle{apj}
\bibliography{apj-jour,biblio}

\end{document}